\documentclass[aps,prb,notitlepage,showkeys,showpacs,superscriptaddress,nofootinbib,longbibliography,preprintnumbers]{revtex4-2}

\usepackage[utf8]{inputenc}

\usepackage{here}
\usepackage{ulem}

\usepackage[top=30truemm,bottom=30truemm,left=25truemm,right=25truemm]{geometry}

\usepackage{spectralsequences}
\usepackage{tikz}

\usepackage{xcolor}

\newcommand{\R}{\mathbb{R}}
\newcommand{\Z}{\mathbb{Z}}
\newcommand{\Zmod}[1]{\mathbb{Z}/#1\mathbb{Z}}
\newcommand{\zmod}[1]{\mathbb{Z}/#1\mathbb{Z}}
\newcommand{\rp}[1]{\mathbb{R}P^{#1}}
\newcommand{\tr}[1]{\mathrm{tr}\left(#1\right)}
\newcommand{\U}[1]{\mathrm{U}\left(#1\right)}

\newcommand{\PU}[1]{\mathrm{PU}\left(#1\right)}
\newcommand{\coho}[3]{\mathrm{H}^{#1}(#2;#3)}
\newcommand{\cohoZ}[2]{\mathrm{H}^{#1}(#2;\mathbb{Z})}

\newcommand{\cohoU}[2]{\mathrm{H}^{#1}(#2;\U{1})}

\newcommand{\abs}[1]{\left|#1\right|}

\usepackage[all]{xy}

\usepackage[colorlinks=true,linktoc=page,citecolor=red,linkcolor=blue]{hyperref}	
\usepackage[whole]{bxcjkjatype} 

\usepackage{slashed}

\usepackage{amsthm}
\usepackage{thmtools}
\usepackage{blindtext}
\usepackage{braket}

\declaretheoremstyle[
       shaded={bgcolor=\color{rgb}{0.9,0.9,0.9}}  
]{theorem}

\declaretheoremstyle[
       shaded={bgcolor=\color{rgb}{0.9,0.9,0.9}}
]{question}

\declaretheoremstyle[
       shaded={bgcolor=\color{rgb}{0.9,0.9,0.9}}  
]{remark}

\declaretheoremstyle[
       shaded={bgcolor=\color{rgb}{0.9,0.9,0.9}}  
]{proposition}

\declaretheoremstyle[
       shaded={bgcolor=\color{rgb}{0.9,0.9,0.9}}  
]{definition}

\declaretheoremstyle[
       shaded={bgcolor=\color{rgb}{0.9,0.9,0.9}}  
]{assumption}

\declaretheoremstyle[
       shaded={bgcolor=\color{rgb}{0.9,0.9,0.9}}  
]{conjecture}

\declaretheoremstyle[
       shaded={bgcolor=\color{rgb}{0.9,0.9,0.9}}  
]{corrorary}

\declaretheoremstyle[
       shaded={bgcolor=\color{rgb}{0.9,0.9,0.9}}  
]{axiom}

\declaretheoremstyle[
       shaded={bgcolor=\color{rgb}{0.9,0.9,0.9}}  
]{lemma}

\def\N{{\mathbb{N}}}
\def\Z{{\mathbb{Z}}}

\def\R{{\mathbb{R}}}
\def\C{{\mathbb{C}}}

\def\oc{{\mathcal{U}}}
\def\D{{\mathcal{D}}}

\def\a{{\alpha}}

\def\th{{\theta}}
\def\si{{\sigma}}

\def\>{{\geq }}
\def\<{{\leq }}

\def\os{\overset}

\def\ua{{\underline{A}}}

\def\hol{{\text{Hol}}}
\def\tt{{T}}

\usepackage{mathrsfs}
\usepackage{amsmath,amssymb}

\usepackage{esvect}

\begin{document}

\title{Discrete Higher Berry Phases and Matrix Product States}

\author{Shuhei Ohyama}
\email{shuhei.oyama@yukawa.kyoto-u.ac.jp}
 \affiliation{Center for Gravitational Physics and Quantum Information, Yukawa Institute for Theoretical Physics, Kyoto University, Kyoto 606-8502, Japan}
 
\author{Yuji Terashima}
\email{yujiterashima@tohoku.ac.jp}
\affiliation{Graduate School of Science, Tohoku University, Sendai 980-8578, Japan}

\author{Ken Shiozaki}
\email{ken.shiozaki@yukawa.kyoto-u.ac.jp}
\affiliation{Center for Gravitational Physics and Quantum Information, Yukawa Institute for Theoretical Physics, Kyoto University, Kyoto 606-8502, Japan}

\date{\today} 
\preprint{YITP-23-21}

\begin{abstract}
A $1$-parameter family of invertible states gives a topological transport phenomenon, similar to the Thouless pumping. As a natural generalization of this, we can consider a family of invertible states parametrized by some topological space $X$. This is called a higher pump. It is conjectured that $(1+1)$-dimensional bosonic invertible state parametrized by $X$ is classified by $\cohoZ{3}{X}$.  In this paper, we construct two higher pumping models parametrized by $X=\rp{2}\times S^1$ and $X=\mathrm{L}(3,1)\times S^1$ that corresponds to the torsion part of $\cohoZ{3}{X}$. As a consequence of the nontriviality as a family, we find that a quantum mechanical system with a nontrivial discrete Berry phase is pumped to the boundary of the $(1+1)$-dimensional system. We also study higher pump phenomena by using matrix product states (MPS), and construct a higher pump invariant which takes value in a torsion part of $\cohoZ{3}{X}$. This is a higher analog of the ordinary discrete Berry phase that takes value in the torsion part of $\cohoZ{2}{X}$. In order to define the higher pump invariant, we utilize the smooth Deligne cohomology and its integration theory. We confirm that the higher pump invariant of the model has a nontrivial value.
\end{abstract}

\maketitle

\setcounter{tocdepth}{3}
\tableofcontents

\section{Introduction}

\subsection{Invertible States and Higher Pump Phenomena}

An invertible state is a state which is realized as a ground state of a unique gapped Hamiltonian. It is known that a $1$-parameter family of $(1+1)$-dimensional $G$-symmetric invertible states gives a Thouless-like charge pumping phenomena\cite{Thouless83}, and classified by the group cohomology $\coho{1}{G}{\mathrm{U}(1)}$\cite{Shiozaki21,BDFJ22,OSS22}. This can be understood as a nontriviality of a family of invertible states with symmetry $G$ parametrized by $1$-dimensional circle $S^1$. Similarly, it is believed that a family of $(1+1)$-dimensional bosonic invertible states without any symmetry parametrized by some topological space $X$ is classified by $\cohoZ{3}{X}$\cite{KS20-1}. When $X=S^1$, this group is trivial, so no nontrivial classification arises. This means that if there is no symmetry, there are no nontrivial pump phenomena. On the other hand, when the dimension of $X$ is higher than $3$, this group can be nontrivial. This implies that, even if there is no symmetry, there is some kind of pumping phenomena\cite{KS20-1,KS20-2,XMAJMDAM21} (we called this the higher pump phenomenon), but its physical interpretation is still unclear.

\subsection{Summary of This Paper}

In this paper, we construct two models with nontrivial higher pump parametrized by $\rp{2}\times S^1$ and $\mathrm{L}(3,1)\times S^1$. We make a physical interpretation of higher pump phenomena: a pump of the ordinary discrete Berry phase and a boundary condition obstacle. 

In addition, we define a topological invariant of a higher pump by using an injective MPS bundle which takes value in the torsion part of $\cohoZ{3}{X}$. This invariant can be viewed as a higher analog of the discrete Berry phase, and this is the kind of nontriviality that cannot be detected by the higher Berry curvature proposed in \cite{KS20-1}. In the formulation of this invariant, the smooth Deligne cohomology and its integration theory are useful\cite{Gawedzki88,GT00,GT01,CJM04,GT09}.

\subsection{Outlook of This Paper}

The rest of this paper is organized as follows:

In Sec.\ref{sec:model}, we introduce models parametrized by $X=\rp{2}\times S^1$ and $\mathrm{L}(3,1) \times S^1$, and discuss the nontriviality of this model from the boundary perspective: we reveal that the ordinary discrete Berry phase is pumped to the boundary, and makes an effective $(0+1)$-dimensional model which is pumped to the boundary. We also show that there are no boundary terms that are parametrized by $X$ and open the gap over the whole $X$.

In Sec.\ref{sec:smooth Deligne}, we give a quick review of the smooth Deligne cohomology. This is a useful tool for describing generalizations of the Berry connection and the Berry curvature to higher dimensions. As an application example of the smooth Deligne cohomology, we write down the ordinary pump invariant of fermion parity\cite{OSS22} as an integration of the smooth Deligne cohomology class.

In Sec.\ref{sec:mps}, we define a higher pump invariant. To this end, we extract the Dixmier-Douady class\cite{DD63} of an injective MPS bundle, and construct a cocycle of the smooth Deligne cohomology. Then, we define the higher pump invariant as an integration of the smooth Deligne cocycle. This can be regarded as a higher analogue of the discrete Berry phase\footnote{This is not a common terminology, 
This is not a common term, but to avoid confusion with the holonomy we will refer to this quantity as the discrete Berry phase in this paper. See App.\ref{sec:complex line bundle} for definitions of terms.
}. As an example, we compute injective MPS bundles of the models introduced in Sec.\ref{sec:model}, and perform an integration of the smooth Deligne cohomology. As a result, we confirm that the higher pump invariants of these models are nontrivial.

\section{A Model of a Higher Pump}\label{sec:model}

In this section, we introduce a $(1+1)$-dimensional spin model with parameter $X=\rp{2}\times S^1$ and $X=\mathrm{L}(3,1)\times S^1$. In Sec.\ref{sec:model with rp2 times s1} and Sec.\ref{sec:Z/3 cluster model}, we define models parametrized by $X=\rp{2}\times S^1$ and $X=\mathrm{L}(3,1)\times S^1$ respectively, and construct their ground states. In Sec.\ref{sec:chern number pump} and \ref{sec:Z/3 chernnumber pump}, we argue the flow of the discrete Berry phase. Although the Berry connection in $(1+1)$-dimensional systems are known to diverge, parameter space allows us to define an effective Berry connection as the difference between divergent quantities, which is regarded as the ordinary discrete Berry phase of a quantum mechanical system pumped to the boundary. In Sec.\ref{sec:obstacle} and Sec.\ref{sec:Z/3 obstacle}, as another perspective, we examine the absence of a boundary condition that is parametrized by $X$ and that opens the gap of the system at all $x\in X$. This can be regarded as an obstacle to the boundary theory.

\subsection{$\rp{2}\times S^1$ model (or $\zmod{2}$ charge pump model)}

\subsubsection{Definition of the Model}\label{sec:model with rp2 times s1}

Let's consider a model on a $1$-dimensional lattice. We put labels on the lattice as $...,-\frac{3}{2},-1,-\frac{1}{2},0,\frac{1}{2},1,\frac{3}{2},...$ and so on. We will refer to integer sites as $\sigma$-sites and the others as $\tau$-sites. At $\tau$-site there is $3$-dimensional Hilbert space, and at $\sigma$-site there is $2$-dimensional Hilbert space. Let $\{\vec{z}=(z_1,z_2)\left|\right.\abs{z_1}^{2}+\abs{z_2}^{2}=1\}$ be a coordinate of $3$-dimensional sphere $S^3$ and  let $t\in [0,2\pi]$ be a coordinate of interval $I=[0,2\pi]$. At $\tau$-site, we take the following orthonormal basis depending on $\vec{z}$:
\begin{eqnarray}\label{eq:rp3 DDW basis 1}
\ket{u_{+}(\vec{z})}_{\tau}:=\begin{pmatrix}
1\\
0\\
0
\end{pmatrix},
\ket{u_{-}(\vec{z})}_{\tau}:=\begin{pmatrix}
0\\
z_1\\
z_2
\end{pmatrix},
\ket{u_{-}^{\perp}(\vec{z})}_{\tau}:=\begin{pmatrix}
0\\
-z^{\ast}_2\\
z_1^\ast
\end{pmatrix}.
\end{eqnarray}
At $\sigma$-site, we take the following orthonormal basis depending on $t$:
\begin{eqnarray}\label{eq:rp3 DDW basis 2}
\ket{\sigma_{\uparrow}(t)}_{\sigma}:=\begin{pmatrix}
\cos(\frac{t}{4})\\
-i\sin(\frac{t}{4})
\end{pmatrix},
\ket{\sigma_{\downarrow}(t)}_{\sigma}:=\begin{pmatrix}
-i\sin(\frac{t}{4})\\
\cos(\frac{t}{4})
\end{pmatrix}.
\end{eqnarray}
Note that $\ket{\sigma_{\uparrow}(t)}_\sigma$ and $\ket{\sigma_{\downarrow}(t)}_\sigma$ are not periodic but satisfy $\ket{\sigma_{\uparrow}(t+2\pi)}_\sigma = i\ket{\sigma_{\downarrow}(t)}_\sigma$ and $\ket{\sigma_{\downarrow}(t+2\pi)}_\sigma = -i\ket{\sigma_{\uparrow}(t)}_\sigma$. In the following, we omit the subscript $\sigma$ and $\tau$ of the basis.

Consider the operators on these Hilbert spaces:
\begin{eqnarray}
\tau^{x}(\vec{z})&:=&1_{3}-2\ket{u_-(\vec{z})}\bra{u_-(\vec{z})}-\ket{u_{-}^{\perp}(\vec{z})}\bra{u_{-}^{\perp}(\vec{z})},\\
&=&\ket{u_{+}(\vec{z})}\bra{u_{+}(\vec{z})}-\ket{u_{-}(\vec{z})}\bra{u_{-}(\vec{z})},\\
\tau^{z}(\vec{z})&:=&\ket{u_{+}(\vec{z})}\bra{u_{-}(\vec{z})}+\ket{u_{-}(\vec{z})}\bra{u_{+}(\vec{z})},\label{eq:tauzz_def}\\
\sigma^{x}(t)&:=&\ket{\sigma_{\uparrow}(t)}\bra{\sigma_{\downarrow}(t)}+\ket{\sigma_{\downarrow}(t)}\bra{\sigma_{\uparrow}(t)},\\
\sigma^{z}(t)&:=&\ket{\sigma_{\uparrow}(t)}\bra{\sigma_{\uparrow}(t)}-\ket{\sigma_{\downarrow}(t)}\bra{\sigma_{\downarrow}(t)}.\label{eq:sigmazt_def}
\end{eqnarray}
Together with $\sigma^y(t):=-i\sigma^z(t) \sigma^x(t)$, $\sigma^\mu(t)$ with $\mu=x,y,z$ are the usual Pauli matrices for at site $\sigma$. 
On the one hand, $\tau^x(\vec z)$ and $\tau^z(\vec z)$ satisfy only the anticommutation relation $\{ \tau^z(\vec z),\tau^x(\vec z)\} = 0$ for $\vec{z}\in S^3$.
On the subspace spanned by $\ket{u_+(\vec{z})}$ and $\ket{u_-(\vec{z})}$, the operators $\tau^x(\vec z), \tau^z(\vec{z})$ and $\tau^y(\vec{z}):= -i \tau^z(\vec{z}) \tau^x(\vec{z})$ behave as the usual Pauli matrices.

By using the above operators, we consider the following model:
\begin{eqnarray}\label{eq:model}
H(\vec{z},t)=-\sum_{j\in\mathbb{Z}}\tau^{z}_{j-\frac{1}{2}}(\vec{z})\sigma^{x}_{j}(t)\tau^{z}_{j+\frac{1}{2}}(\vec{z})-\sum_{j\in\mathbb{Z}}\sigma^{z}_{j}(t)\tau^{x}_{j+\frac{1}{2}}(\vec{z})\sigma^{z}_{j+1}(t).
\end{eqnarray}
At $(\vec{z}=(1,0), t=0)$, this model resembles the cluster model\cite{BR01}. Since $\tau^{z}(\vec{z})$ and $\tau^{x}(\vec{z})$ satisfy
\begin{eqnarray}
\tau^{z}(-\vec{z})=-\tau^{z}(\vec{z}),\tau^{x}(-\vec{z})&=&\tau^{x}(\vec{z}),
\end{eqnarray}
the Hamiltonian Eq.(\ref{eq:model}) coincide at $\vec{z}$ and $-\vec{z}$:
\begin{eqnarray}
H(-\vec{z},t)=H(\vec{z},t).
\end{eqnarray}
Also, since $\sigma^{z}(t)$ and $\sigma^{x}(t)$ satisfy
\begin{eqnarray}
\sigma^{z}(t+2\pi)=-\sigma^{z}(t), \sigma^{x}(t+2\pi)=\sigma^{x}(t),
\end{eqnarray}
the Hamiltonian Eq.(\ref{eq:model}) coincide at $t$ and $t+2\pi$:
\begin{eqnarray}
H(\vec{z},t+2\pi)=H(\vec{z},t).
\end{eqnarray}
Therefore, the operators are parametrized by $S^3 \times I$, but the Hamiltonian is parametrized by $\rp{3} \times S^1$. 

In order to write down the ground state of $H(\vec{z},t)$, we introduce the decorated domain wall state~\cite{CYV14} with respect to $\ket{u_{\pm}(\vec{z})}$ and $\ket{\sigma_{\uparrow/\downarrow}(t)}$. A typical decorated domain wall state is
\begin{eqnarray}\label{eq:typical DDW}
\ket{\cdots u_{+}(\vec{z})\sigma_{\uparrow}(t)u_{-}(\vec{z})\sigma_{\downarrow}(t)u_{+}(\vec{z})\sigma_{\downarrow}(t)u_{-}(\vec{z})\sigma_{\uparrow}(t)\cdots},
\end{eqnarray}
i.e., put $u_{-}(\vec{z})$ where the $\sigma$-arrow reverses and $u_{+}(\vec{z})$ otherwise. The place where the $\sigma$-arrow reverses is called the domain wall of $\sigma$-arrows. Since we ``decorate" $u_{-}(\vec{z})$ on the domain wall, the state in Eq.(\ref{eq:typical DDW}) is called a decorated domain wall state. We will denote the set of decorated domain wall states as ${\rm DDW}_{2}$. Here, the subscript $2$  means that it is a domain wall for $\zmod{2}$ with up and down arrows, and the purpose is to distinguish it from the domain wall for $\zmod{3}$ in Sec.\ref{sec:Z/3 cluster model}. The ground state of $H(\vec{z},t)$ is found to be the equal weight superposition of decorated domain wall states:
\begin{eqnarray}
\sum_{\{i_k,j_l\}\in\mathrm{DDW}_2}\ket{\cdots u_{i_1}(\vec{n})\sigma_{j_1}(t)\cdots u_{i_L}(\vec{n})\sigma_{j_L}(t)\cdots}.
\end{eqnarray}
Another useful representation of the ground state is the way to use the fluctuation term, which is defined by
\begin{eqnarray}\label{eq:fluctuation}
f_{j}(\vec{z},t):=1+\tau^{z}_{j-\frac{1}{2}}(\vec{z})\sigma^{x}_{j}(t)\tau^{z}_{j+\frac{1}{2}}(\vec{z}),
\end{eqnarray} 
for all $j\in\mathbb{Z}$. By using this term, the normalized ground state is given by
\begin{eqnarray}\label{eq:GS of the Z/2 cluster}
\ket{\mathrm{G.S.}(\vec{z},t)}:=\prod_{j\in\mathbb{Z}}\frac{f_{j}(\vec{z},t)}{\sqrt{2}}\ket{\mathrm{Ref}(\vec{z},t)},
\end{eqnarray}
where $\ket{\mathrm{Ref}(\vec{z},t)}$ is a decorated domain wall state\footnote{Remark that $f_j/2$ is not a projection on the whole Hilbert space as $(f_j/2)^2 \neq f_j/2$, but is a projection on decorated domain wall states. Thus, replacing $f_j$ by $p(\vec{z})-\tau^{z}_{j-\frac{1}{2}}(\vec{z})\sigma^{x}_{j}(t)\tau^{z}_{j+\frac{1}{2}}(\vec{z})$ in Eq.(\ref{eq:fluctuation}) gives the same state, where $p(\vec{z})$ is a projection onto the space orthogonal to all $\ket{u_{-}(\vec{z})}$. In this way, in considering the decorated domain wall states, we can handle $f_j/2$ as a projection.}. We call $\ket{\mathrm{Ref}(\vec{z},t)}$ a reference state of this representation. Note that $\ket{\mathrm{Ref}(\vec{z},t)}$ is not unique but the ground state Eq.(\ref{eq:GS of the Z/2 cluster}) is independent of this choice. 

In particular, by taking $z_1\in\mathbb{R}$, the parameter space becomes $\rp{2}\times S^1$. We define $n_{3}:=z_1$, $n_{1}:=\mathrm{Re}(z_{2})$ and $n_{2}:=\mathrm{Im}(z_{2})$, and use $\vec{n}=(n_{1},n_{2},n_{3})^{\rm T}$ as a coordinate of $\rp{2}$.  In the following, we consider a model
\begin{eqnarray}\label{eq:model with rp2}
H(\vec{n},t):=H(\vec{z},t)\left.\right|_{z_{1}\in\mathbb{R}},
\end{eqnarray}
and verify the nontriviality of this model\footnote{Since $\cohoZ{3}{\rp{2}\times S^1}\simeq\zmod{2}$, it can be nontrivial as a family of invertible states.} as a family of invertible states over $\rp{2}\times S^{1}$. 

\subsubsection{Physical Interpretation I: Discrete Berry Phase Pumping}\label{sec:chern number pump}

In the ordinary pump phenomenon of $(1+1)$-dimensional systems, a $(0+1)$-dimensional invertible state is pumped from one edge of the system to the other\cite{Kitaev13}. As a generalization of this, in higher pump phenomenon of $(1+1)$-dimensional systems, families of $(0+1)$-dimensional invertible states are pumped from one edge to the other. In particular, it is believed that when the parameter space $X$ is $S^1\times M^n$ for some $n$-dimensional topological space $M^n$, a $(0+1)$-dimensional system with parameter $M^n$ is pumped to the edge\cite{XMAJMDAM21}\footnote{In \cite{XMAJMDAM21}, they argue that the flow of the ordinary Berry curvature in the case of $X=S^3$ for $(1+1)$-dimensional systems, based on the Kapustin-Spodyneiko invariant\cite{KS20-1}.}. Let's check that this picture holds for the model Eq.(\ref{eq:model}).

In order to show a physical interpretation of the higher pump, we cut the system between site $0$ and site $\frac{1}{2}$, and create a boundary such that $\tau$ appears at the left edge:
\begin{eqnarray}\label{eq:model_with_boundary}
 H(\Vec{n},t)=-\sum_{j\in\mathbb{N}}\tau^{z}_{j-\frac{1}{2}}(\Vec{n})\sigma^{x}_{j}(t)\tau^{z}_{j+\frac{1}{2}}(\Vec{n})-\sum_{j\in\mathbb{N}}\sigma^{z}_{j}(t)\tau^{x}_{j+\frac{1}{2}}(\Vec{n})\sigma^{z}_{j+1}(t).
 \end{eqnarray}
The Hamiltonian Eq.(\ref{eq:model_with_boundary}) has doubly degenerated ground states
\begin{eqnarray}
\sum_{\{i_k,j_l\}\in\mathrm{DDW}_{2}}\ket{u_{i_1}(\vec{n})\sigma_{j_1}(t)\cdots u_{i_L}(\vec{n})\sigma_{j_L}(t)},
\end{eqnarray}
and 
\begin{eqnarray}
\sum_{\{i_k,j_l\}\in\mathrm{DDW}_{2}}\tau^{z}_{\frac{1}{2}}(\vec{n})\ket{u_{i_1}(\vec{n})\sigma_{j_1}(t)\cdots u_{i_L}(\vec{n})\sigma_{j_L}(t)}.
\end{eqnarray}
This can be seen from the fact that these states are eigenstates of all terms of the Hamiltonian Eq.(\ref{eq:model_with_boundary}), and the Hamiltonian Eq.(\ref{eq:model_with_boundary}) commute with $\tau^{z}_{\frac{1}{2}}(\vec{n})$. For simplicity, we fix the parameters as $(\vec{n}=(0,0,1)^{\rm T},t=0)$, and represent this state as follows:
\begin{eqnarray}
\ket{\pm\uparrow+\uparrow+\uparrow+\uparrow\cdots}+\ket{\mp\downarrow+\downarrow+\downarrow+\downarrow\cdots}+\ket{\mp\downarrow-\uparrow-\downarrow-\uparrow\cdots}+\ket{\pm\uparrow-\downarrow-\uparrow-\downarrow\cdots}+\cdots,
\end{eqnarray}
Here, we denote $\ket{u_{\pm}(\vec{n}=(0,0,1)^{\rm T})}$ as $\ket{\pm}$ and $\ket{\sigma_{\uparrow/\downarrow}(t=0)}$ as $\ket{\uparrow/\downarrow}$.
In this case, there is a degeneracy for the sign of the edge. 
By imposing appropriate boundary conditions, we choose the upper sign of $\pm$ for the initial state:
\begin{eqnarray}\label{eq:cannonical_pump_at_0}
\ket{+\uparrow+\uparrow+\uparrow+\uparrow\cdots}+\ket{-\downarrow+\downarrow+\downarrow+\downarrow\cdots}+\ket{-\downarrow-\uparrow-\downarrow-\uparrow\cdots}+\ket{+\uparrow-\downarrow-\uparrow-\downarrow\cdots}+\cdots.
\end{eqnarray}
Now we rotate all $\sigma$-spin by $\pi$, this is accomplished by varying $t$ from $0$ to $2\pi$: 
\begin{eqnarray}\label{eq:cannonical_pump_at_1}
\ket{+\downarrow+\downarrow+\downarrow+\downarrow\cdots}+\ket{-\uparrow+\uparrow+\uparrow+\uparrow\cdots}+\ket{-\uparrow-\downarrow-\uparrow-\downarrow\cdots}+\ket{+\downarrow-\uparrow-\downarrow-\uparrow\cdots}+\cdots.
\end{eqnarray}
By comparing the initial state Eq.(\ref{eq:cannonical_pump_at_0}) and the final state Eq.(\ref{eq:cannonical_pump_at_1}), we can see that only the sign of the edge is flipped. Intuitively, this sign flipping indicates that the ground state of the quantum mechanical system at the boundary changes between the initial state and the final state. In the above process, we considered a fixed parameter of $\rp{2}$, but by running it, we can see that the quantum mechanical system parametrized by $\rp{2}$ is pumped to the edge, as seen below. 

Let's implement this process using the Hamiltonian. We need to add a boundary term to remove the degeneracy. To realize state Eq.(\ref{eq:cannonical_pump_at_0}), simply add $-\tau^{x}_{\frac{1}{2}}(\vec{n})\sigma^{z}_{1}(t)$ to the boundary\footnote{Remark that this term break the $2\pi$-periodicity of the Hamiltonian. We will discuss this point in Sec.\ref{sec:obstacle}.} of the Hamiltonian $H(\vec{n},t)$.\footnote{Of course, there are other choices as boundary terms. We discuss this point in Appendix.\ref{sec:boundary condition}} 
Set $\vec{z}_{0}=(0,0,1)^{\rm T}$.
In the following, we use the notations $\tau^{\mu}_{j-\frac{1}{2}}:= \tau^{\mu}_{j-\frac{1}{2}}(\vec{z}=\vec{z}_0)$ and $\sigma^{\mu}_{j}:=\sigma^{\mu}_j(t=0)$ for $\mu=x,z$. 
First, consider the initial Hamiltonian, i.e., $t=0$:
\begin{eqnarray}\label{eq:model_with_boundary_0}
H_{\rm in.}(\Vec{n}):= -\tau^{x}_{\frac{1}{2}}(\vec{n})\sigma^{z}_{1}-\sum_{j=1,2,...}\tau^{z}_{j-\frac{1}{2}}(\Vec{n})\sigma^{x}_{j}\tau^{z}_{j+\frac{1}{2}}(\Vec{n})-\sum_{j=1,2,...}\sigma^{z}_{j}\tau^{x}_{j+\frac{1}{2}}(\Vec{n})\sigma^{z}_{j+1}.
\end{eqnarray}
This is a unique gapped Hamiltonian for all $\vec{n}\in\rp{2}$. The normalized ground state when $\vec{n}=\vec{z}_{0}$ is given by
\begin{eqnarray}\label{eq:GS_in}
\ket{\mathrm{G.S.}_{\rm in.}(\vec{n}=\vec{z}_{0})}:=\prod_{j=1,2,...}\frac{f_{j}}{\sqrt{2}}\ket{\mathrm{Ref}_{\rm in.}},
\end{eqnarray}
where $f_{j}$ is a fluctuation term defined by
\begin{eqnarray}
f_{j}:=1+\tau^{z}_{j-\frac{1}{2}}\sigma^{x}_{j}\tau^{z}_{j+\frac{1}{2}},
\end{eqnarray} 
and $\ket{\mathrm{Ref}_{\rm in.}}$ is a decorated domain wall state whose eigenvalue of $\tau^{x}_{\frac{1}{2}}\sigma^{z}_{1}$ is $1$, for example, $\ket{\mathrm{Ref}_{\rm in.}}=\ket{+\uparrow+\uparrow+\cdots}$.
Let $(\theta,\phi)$ be the spherical coordinate of $\vec{n}$:
\begin{eqnarray}
n_1&=&\sin(\theta)\cos(\phi),\\
n_2&=&\sin(\theta)\sin(\phi),\\
n_3&=&\cos(\theta).
\end{eqnarray}
For generic $\vec{n}$, noticing $H_{\rm in.}(\Vec{n})$ is given by the unitary transformation 
\begin{eqnarray}
    H_{\rm in.}(\Vec{n})=\left[\prod_{j=1}^{\infty}V_{\tau}(\vec{n})_{j-\frac{1}{2}}\right]H_{\rm in.}(\vec{n}=\vec{z}_0)\left[\prod_{j=1}^{\infty}V_{\tau}(\vec{n})_{j-\frac{1}{2}}\right]^\dag
\end{eqnarray}
with unitary matrices
\begin{align}
    V_{\tau}(\vec{n})_{j}:=\begin{pmatrix}
1&&\\
&\cos(\theta)&-e^{-i\phi}\sin(\theta)\\
&e^{i\phi}\sin(\theta)&\cos(\theta)
\end{pmatrix}
\end{align}
acting on the site $j$, and the ground state is given by 
\begin{eqnarray}
\ket{\mathrm{G.S.}_{\rm in.}(\vec{n})}=\prod_{j=1}^{\infty}V_{\tau}(\vec{n})_{j-\frac{1}{2}}\ket{\mathrm{G.S.}_{\rm in.}(\vec{n}=\vec{z}_{0})}.
\end{eqnarray}
Then the Berry connection $A_{\rm in.}(\vec{n})$ is formally given by
\begin{eqnarray}\label{eq:Berry connection at t=0}
A_{\rm in.}(\vec{n}):=\bra{\mathrm{G.S.}_{\rm in.}(\vec{n})}d\ket{\mathrm{G.S.}_{\rm in.}(\vec{n})}=\sum_{j=1,2,\cdots}\bra{\mathrm{G.S.}_{\rm in.}(\vec{n}=\vec{z}_{0})}V_{\tau}(\vec{n})_{j-\frac{1}{2}}^{\dagger}dV_{\tau}(\vec{n})_{j-\frac{1}{2}}\ket{\mathrm{G.S.}_{\rm in.}(\vec{n}=\vec{z}_{0})}.
\end{eqnarray}
Remark that the Berry connection $A_{\rm in.}(\vec{n})$ is ill-defined as a convergent quantity as it is an infinite sum. 
We will carefully extract only the contribution from the left boundary. We define  
\begin{eqnarray}
h_{j-\frac{1}{2}}(\vec{n}):=V_{\tau}(\vec{n})_{j-\frac{1}{2}}^{\dagger}dV_{\tau}(\vec{n})_{j-\frac{1}{2}}.
\end{eqnarray}
Since the support of $h_{j-\frac{1}{2}}(\vec{n})$ is $\{j-\frac{1}{2}\}$ and the support of $f_{j}$ is $\{j\pm\frac{1}{2},j\}$, $f_{j}$ is commute with $h_{k}(\vec{n})$ when $j\neq k,k-1$. Thus, each term of the Eq.(\ref{eq:Berry connection at t=0}) is recast into
\begin{eqnarray}
\bra{\mathrm{G.S.}_{\rm in.}(\vec{n}=\vec{z}_{0})}h_{j-\frac{1}{2}}(\vec{n})\ket{\mathrm{G.S.}_{\rm in.}(\vec{n}=\vec{z}_{0})}=\frac{1}{4}\bra{\mathrm{Ref}_{\rm in.}}f_{j}f_{j-1}h_{j-\frac{1}{2}}(\vec{n})f_{j-1}f_{j}\prod_{\substack{k=1,2,\cdots\\k\neq j-1,j}}f_{k}\ket{\mathrm{Ref}_{\rm in.}},
\end{eqnarray}
where $f_{0}:=\sqrt{2}$. Moreover, fluctuation terms in the product can be replaced by $1$. This is because
$f_{k}$ is the only operator which acts on the site $k$ among the operators sandwiched between states $\ket{\mathrm{Ref}_{\rm in.}}$, so the fluctuated part is projected out by $\bra{\mathrm{Ref}_{\rm in.}}$\footnote{Note that this argument is incorrect if the reference state is defined as a superposition of DDW states.}. Therefore, 
\begin{eqnarray}
A_{\rm in.}(\vec{n})=\sum_{j=1,2,\cdots}\frac{1}{4}\bra{\mathrm{Ref}_{\rm in.}}f_{j}f_{j-1}h_{j-\frac{1}{2}}(\vec{n})f_{j-1}f_{j}\ket{\mathrm{Ref}_{\rm in.}}. 
\end{eqnarray}
Let $\gamma: [0,2\pi] \to \rp2$ be a loop whose homotopy class is nontrivial. The discrete Berry phase\footnote{This is not a common terminology, 
This is not a common term, but to avoid confusion with the holonomy we will refer to this quantity as the discrete Berry phase in this paper. See App.\ref{sec:complex line bundle} for definitions of terms.
}  $n_{\rm in.}(\gamma)$ along a path $\gamma$ is 
\begin{eqnarray}\label{eq:holonomy at t=0}
n_{\rm in.}(\gamma):=\exp(\int_{\gamma}A_{\rm in.}-\frac{1}{2}\int_{\Sigma}dA_{\rm in.})\braket{\mathrm{G.S.}_{\rm in.}(\gamma(2\pi))|\mathrm{G.S.}_{\rm in.}(\gamma(0))},
\end{eqnarray}
where $\Sigma$ is a bounding surface of $2\cdot\gamma$, i.e.,  $\partial\Sigma=2\cdot\gamma$.

Similarly, the final Hamiltonian, i.e., $t=2\pi$ is given by
\begin{eqnarray}\label{eq:model_with_boundary_2pi}
 H_{\rm fin.}(\Vec{n}):=\tau^{x}_{\frac{1}{2}}(\vec{n})\sigma^{z}_{1}-\sum_{j=1,2,...}\tau^{z}_{j-\frac{1}{2}}(\Vec{n})\sigma^{x}_{j}\tau^{z}_{j+\frac{1}{2}}(\Vec{n})-\sum_{j=1,2,...}\sigma^{z}_{j}\tau^{x}_{j+\frac{1}{2}}(\Vec{n})\sigma^{z}_{j+1},
\end{eqnarray}
and the ground state is 
\begin{eqnarray}
\ket{\mathrm{G.S.}_{\rm fin.}(\vec{n}=\vec{z}_{0})}:=\prod_{j=1,2,...}\frac{f_{j}}{\sqrt{2}}\ket{\mathrm{Ref}_{\rm fin.}},
\end{eqnarray}
where $\ket{\mathrm{Ref}_{\rm fin.}}$ is a decorated domain wall state whose eigenvalue of $-\tau^{x}_{\frac{1}{2}}\sigma^{z}_{1}$ is $1$, for example, $\ket{\mathrm{Ref}_{\rm in.}}=\ket{-\uparrow+\uparrow+\cdots}$.
By a similar calculation to that of $A_{\rm in.}(\vec{n})$,  the Berry connection of the final Hamiltonian is 
\begin{eqnarray}
A_{\rm fin.}(\vec{n})&:=&\bra{\mathrm{G.S.}_{\rm fin.}(\vec{n})}d\ket{\mathrm{G.S.}_{\rm fin.}(\vec{n})},\\
&=&\sum_{j=1,2,\cdots}\frac{1}{4}\bra{\mathrm{Ref}_{\rm fin.}}f_{j}f_{j-1}h_{j-\frac{1}{2}}(\vec{n})f_{j-1}f_{j}\ket{\mathrm{Ref}_{\rm fin.}},
\end{eqnarray}
and the discrete Berry phase $n_{\rm fin.}(\gamma)$ along a path $\gamma$ is 
\begin{eqnarray}\label{eq:holonomy at t=2pi}
n_{\rm fin.}(\gamma):=\exp(\int_{\gamma}A_{\rm fin.}-\frac{1}{2}\int_{\Sigma}dA_{\rm fin.})\braket{\mathrm{G.S.}_{\rm fin.}(\gamma(2\pi))|\mathrm{G.S.}_{\rm fin.}(\gamma(0))}.
\end{eqnarray}

\begin{figure}[H]
 \begin{center}
  \includegraphics[width=55mm]{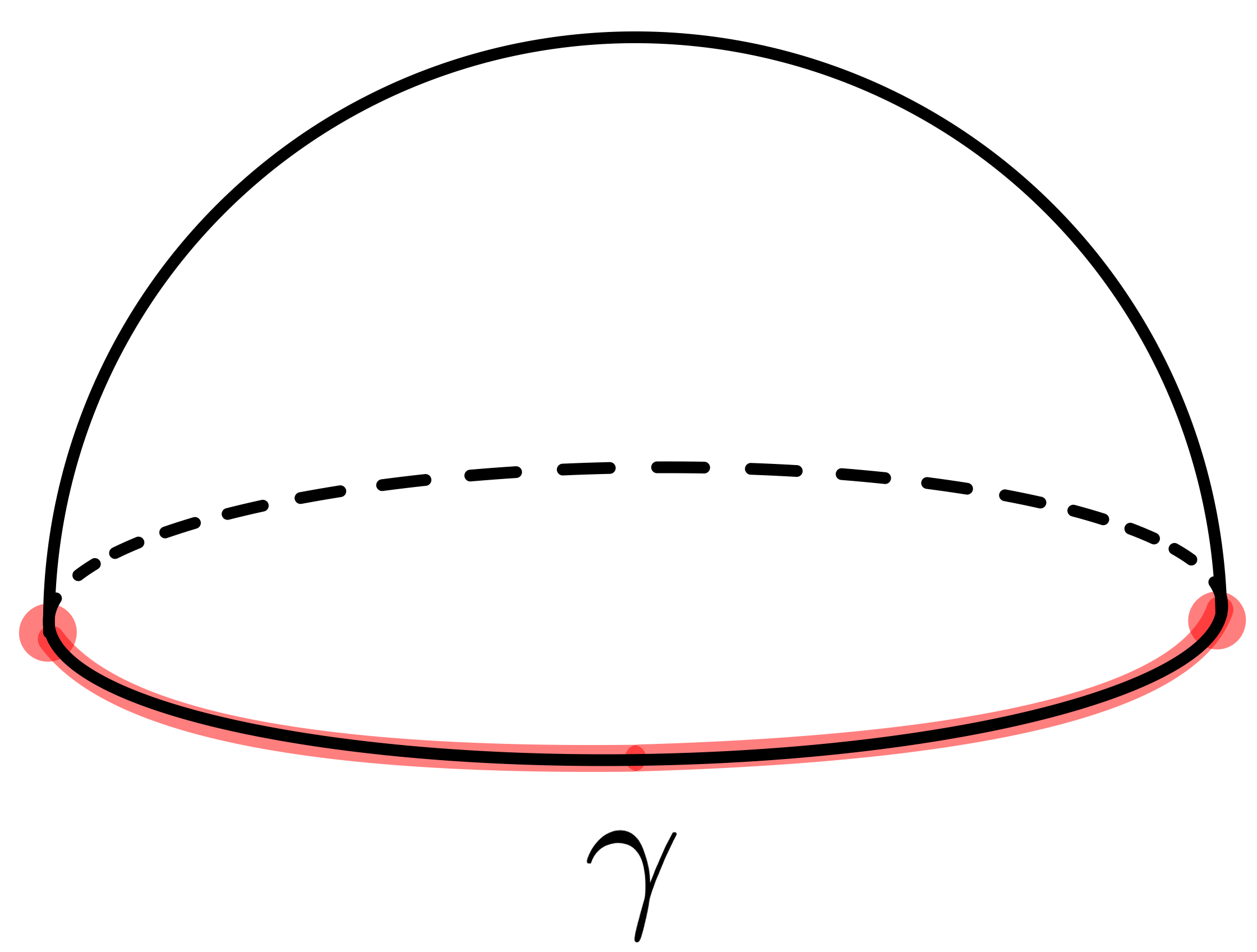}
 \end{center}
  \caption[]{$\gamma$ is a path defined by $\theta=\frac{\pi}{2}$. This is a nontrivial path in $\rp{2}$.
}
\label{fig:path}
\end{figure}

Being a semi-infinite system, the values of each discrete Berry phase Eq.(\ref{eq:holonomy at t=0}) and Eq.(\ref{eq:holonomy at t=2pi}) do not necessarily converge. However, in the pump model, the bulk states coincide at $t=0$ and $2\pi$, so we can choose reference states at $t=0$ and $2\pi$ with the same bulk configuration. Then, only the edge contribution remains in the ratio of the discrete Berry phases. We choose $\ket{\mathrm{Ref}_{\rm in.}}$ and $\ket{\mathrm{Ref}_{\rm fin.}}$ as
\begin{eqnarray}
\ket{\mathrm{Ref}_{\rm in.}}&=&\ket{+\uparrow+\uparrow+\cdots},\\
\ket{\mathrm{Ref}_{\rm fin.}}&=&\ket{-\uparrow+\uparrow+\cdots},
\end{eqnarray}
and let's compute the ratio $r$ of the discrete Berry phases for a nontrivial path $\gamma$ in $\rp{2}$:
\begin{eqnarray}\label{eq:ratio}
r:=\frac{n_{\rm in.}(\gamma)}{n_{\rm fin.}(\gamma)}=\exp(\int_{\gamma}(A_{\rm in.}-A_{\rm fin.})-\frac{1}{2}\int_{\Sigma}(dA_{\rm in.}-dA_{\rm fin.}))\frac{\braket{\mathrm{G.S.}_{\rm in.}(\gamma_{0})|\mathrm{G.S.}_{\rm in.}(\gamma_{1})}}{\braket{\mathrm{G.S.}_{\rm fin.}(\gamma_{0})|\mathrm{G.S.}_{\rm fin.}(\gamma_{1})}}.
\end{eqnarray}
We illustrate the path $\gamma$ in Fig.\ref{fig:path}. Since the only difference between $\ket{\mathrm{Ref}_{\rm in.}}$ and $\ket{\mathrm{Ref}_{\rm fin.}}$ is the site $\frac{1}{2}$, the expectation value of an operator not acting on site $\frac{1}{2}$ is the same. Thus we obtain 
\begin{eqnarray}
A_{\rm in.}(\vec{n})-A_{\rm fin.}(\vec{n})&=&\frac{1}{4}(2\bra{\mathrm{Ref}_{\rm in.}}f_{1}h_{\frac{1}{2}}(\vec{n})f_{1}\ket{\mathrm{Ref}_{\rm in.}}
+\bra{\mathrm{Ref}_{\rm in.}}f_{2}f_{1}h_{\frac{3}{2}}(\vec{n})f_{1}f_{2}\ket{\mathrm{Ref}_{\rm in.}})\nonumber\\
&-&\frac{1}{4}(2\bra{\mathrm{Ref}_{\rm fin.}}f_{1}h_{\frac{1}{2}}(\vec{n})f_{1}\ket{\mathrm{Ref}_{\rm fin.}}
+\bra{\mathrm{Ref}_{\rm fin.}}f_{2}f_{1}h_{\frac{3}{2}}(\vec{n})f_{1}f_{2}\ket{\mathrm{Ref}_{\rm fin.}}).
\end{eqnarray}
The second term can be nonzero only if one chooses $1$ twice or $\tau^{z}_{\frac{1}{2}}\sigma^{z}_{1}\tau^{z}_{\frac{3}{2}}$ twice from the two $f_{1}=1+\tau^{z}_{\frac{1}{2}}\sigma^{z}_{1}\tau^{z}_{\frac{3}{2}}$. By using this observation, the second term and the forth term cancel. Therefore, 
\begin{eqnarray}\label{eq:effective Berry phase}
A_{\rm in.}(\vec{n})-A_{\rm fin.}(\vec{n})&=&\frac{1}{2}(\bra{\mathrm{Ref}_{\rm in.}}f_{1}h_{\frac{1}{2}}(\vec{n})f_{1}\ket{\mathrm{Ref}_{\rm in.}}
-\bra{\mathrm{Ref}_{\rm fin.}}f_{1}h_{\frac{1}{2}}(\vec{n})f_{1}\ket{\mathrm{Ref}_{\rm fin.}}).
\end{eqnarray}
After a simple calculation, we obtain
\begin{eqnarray}
A_{\rm in.}(\vec{n})-A_{\rm fin.}(\vec{n})=\frac{1}{4}\bra{+}(1+\tau^{z}_{\frac{1}{2}})h_{\frac{1}{2}}(\vec{n})(1+\tau^{z}_{\frac{1}{2}})\ket{+}-\frac{1}{4}\bra{-}(1+\tau^{z}_{\frac{1}{2}})h_{\frac{1}{2}}(\vec{n})(1+\tau^{z}_{\frac{1}{2}})\ket{-}=0.
\end{eqnarray}
Since $\ket{\mathrm{G.S.}_{\rm fin.}(\vec{n})}\propto\tau^{z}_{\frac{1}{2}}(\vec{n})\ket{\mathrm{G.S.}_{\rm in.}(\vec{n})}$, 
\begin{eqnarray}
\braket{\mathrm{G.S.}_{\rm fin.}(\vec{n})|\mathrm{G.S.}_{\rm fin.}(-\vec{n})}&=&\bra{\mathrm{G.S.}_{\rm in.}(\vec{n})}\tau^{z}_{\frac{1}{2}}(\vec{n})\tau^{z}_{\frac{1}{2}}(-\vec{n})\ket{\mathrm{G.S.}_{\rm in.}(-\vec{n})},\\
&=&-\braket{\mathrm{G.S.}_{\rm in.}(\vec{n})|\mathrm{G.S.}_{\rm in.}(-\vec{n})}.
\end{eqnarray}
Thus the ratio $r$ is $-1$:
\begin{eqnarray}
r=\frac{n_{\rm in.}(\gamma)}{n_{\rm fin.}(\gamma)}=-1.
\end{eqnarray}
It is worth mentioning that for each $t$, the discrete Berry phase is ill-defined because it is a semi-infinite system, but the ratio of it at $t=0$ and $t=2\pi$ is well-defined because the bulk state returns to itself when the system goes around in the $S^1$ direction. In this sense, this quantity $r$ essentially measures the nontriviality as a $3$-parameter family of unique gapped systems.

In this process, what is the $2$-parameter family of (0+1)-dimensional invertible states that are pumped into the boundary? To clarify this, consider an effective model of the boundary. For the initial Hamiltonian Eq.(\ref{eq:model_with_boundary_0}), the boundary model is given by
\begin{eqnarray}\label{eq:boundary_model_0}
 H^{\rm bdy.}_{\rm in.}(\Vec{n}):=-\tau^{x}_{\frac{1}{2}}(\vec{n})\sigma^{z}_{1}-\tau^{z}_{\frac{1}{2}}(\Vec{n})\sigma^{x}_{1}\tau^{z}_{\frac{3}{2}}(\Vec{n})-\sigma^{z}_{1}\tau^{x}_{\frac{3}{2}}(\Vec{n}),
\end{eqnarray}
and for the final Hamiltonian Eq.(\ref{eq:model_with_boundary_2pi}), the boundary model is given by
\begin{eqnarray}\label{eq:boundary_model_1}
 H^{\rm bdy.}_{\rm fin.}(\Vec{n})=\tau^{x}_{\frac{1}{2}}(\vec{n})\sigma^{z}_{1}-\tau^{z}_{\frac{1}{2}}(\Vec{n})\sigma^{x}_{1}\tau^{z}_{\frac{3}{2}}(\Vec{n})-\sigma^{z}_{1}\tau^{x}_{\frac{3}{2}}(\Vec{n}).
\end{eqnarray}
Then, it can be seen that the ratio of the discrete Berry phases we calculated above is the same as that of these quantum mechanical systems over $\rp2$. Let's compute the discrete Berry phase of Hamiltonians Eq.(\ref{eq:boundary_model_0}) and Eq.(\ref{eq:boundary_model_1}), and confirm this point.

The ground state $\ket{{\rm G.S.}^{\rm bdy.}_{\rm in.}(\vec{n})}$ of $H^{\rm bdy.}_{\rm in.}(\Vec{n})$ is given by
\begin{eqnarray}
\ket{{\rm G.S.}^{\rm bdy.}_{\rm in.}(\vec{n})}&:=&\frac{1}{2}(\ket{\uparrow(\vec{n})+\uparrow(\vec{n})}+\ket{\uparrow(\vec{n})-\downarrow(\vec{n})}+\ket{\downarrow(\vec{n})-\uparrow(\vec{n})}+\ket{\downarrow(\vec{n})+\downarrow(\vec{n})}),\\
&=&V_{\tau}(\vec{n})_{\frac{1}{2}}V_{\tau}(\vec{n})_{\frac{3}{2}}\frac{1+\tau^{x}_{\frac{1}{2}}\sigma^{z}_{1}}{\sqrt{2}}\frac{1+\sigma^{z}_{1}\tau^{x}_{\frac{3}{2}}}{\sqrt{2}}\ket{\uparrow+\uparrow}.
\end{eqnarray}
Here, $\ket{\uparrow} = \frac{1}{\sqrt{2}}(1,1,0)^{\rm T}$ and $\ket{+}=\frac{1}{\sqrt{2}}(1,1)^{\rm T}$. 
The Berry connection is given by
\begin{eqnarray}
A^{\rm bdy.}_{\rm in.}(\vec{n})&:=&\bra{\mathrm{GS.}^{\rm bdy.}_{\rm in.}(\vec{n})}d\ket{\mathrm{GS.}^{\rm bdy.}_{\rm in.}(\vec{n})},\\
&=&\bra{\uparrow+\uparrow}
\frac{1+\tau^{x}_{\frac{1}{2}}\sigma^{z}_{1}}{\sqrt{2}}\frac{1+\sigma^{z}_{1}\tau^{x}_{\frac{3}{2}}}{\sqrt{2}}
(h_{\frac{1}{2}}(\vec{n})+h_{\frac{3}{2}}(\vec{n}))
\frac{1+\tau^{x}_{\frac{1}{2}}\sigma^{z}_{1}}{\sqrt{2}}\frac{1+\sigma^{z}_{1}\tau^{x}_{\frac{3}{2}}}{\sqrt{2}}\ket{\uparrow+\uparrow}.
\end{eqnarray}
Here, recall that $h_{j}(\vec{n})=V_{\tau}(\vec{n})_{j}^{\dagger}dV_{\tau}(\vec{n})_{j}$. We can check that
\begin{eqnarray}
V_{\tau}(\vec{n})_{j}^{\dagger}dV_{\tau}(\vec{n})_{j}=\begin{pmatrix}
0&&\\
&&-e^{i\phi}\\
&e^{i\phi}&
\end{pmatrix}d\theta
+\begin{pmatrix}
0&&\\
&i\sin^{2}(\theta)&ie^{-i\phi}\sin(\theta)\cos(\theta)\\
&ie^{i\phi}\sin(\theta)\cos(\theta)&-i\sin^{2}(\theta)
\end{pmatrix}d\phi,
\end{eqnarray}
and the Berry connection is
\begin{eqnarray}
A^{\rm bdy.}_{\rm in.}(\vec{n})&=&\bra{\uparrow+\uparrow}\frac{1+\tau^{x}_{\frac{1}{2}}\sigma^{z}_{1}}{\sqrt{2}}
h_{\frac{1}{2}}(\vec{n})
\frac{1+\tau^{x}_{\frac{1}{2}}\sigma^{z}_{1}}{\sqrt{2}}\ket{\uparrow+\uparrow}
+
\bra{\uparrow+\uparrow}\frac{1+\sigma^{z}_{1}\tau^{x}_{\frac{3}{2}}}{\sqrt{2}}
h_{\frac{3}{2}}(\vec{n})
\frac{1+\sigma^{z}_{1}\tau^{x}_{\frac{1}{2}}}{\sqrt{2}}\ket{\uparrow+\uparrow},\\
&=&\frac{\bra{\uparrow}h_{\frac{1}{2}}(\vec{n})\ket{\uparrow}}{2}+\frac{\bra{\downarrow}h_{\frac{1}{2}}(\vec{n})\ket{\downarrow}}{2}
+\frac{\bra{\uparrow}h_{\frac{3}{2}}(\vec{n})\ket{\uparrow}}{2}+\frac{\bra{\downarrow}h_{\frac{3}{2}}(\vec{n})\ket{\downarrow}}{2},\\
&=&\frac{i}{2}\sin^{2}(\theta)d\phi.
\end{eqnarray}
Thus, the Berry curvature is
\begin{eqnarray}
F^{\rm bdy.}_{\rm in.}(\vec{n}):=dA^{\rm bdy.}_{\rm in.}(\vec{n})
=\frac{i}{2}\sin(2\theta)d\theta d\phi.
\end{eqnarray}
Finally, let's compute the overlap $\braket{\mathrm{GS}^{\rm bdy.}_{\rm in.}(\vec{n})|\mathrm{GS}^{\rm bdy.}_{\rm in.}(-\vec{n})}$. Since, $V_{\tau}(\vec{n})_{j}^{\dagger}V_{\tau}(-\vec{n})_{j}=\tau^{x}_{j}-\ket{u_-^\perp(\vec{z}_0)}_j\bra{u_-^\perp(\vec{z}_0)}_j$,
\begin{eqnarray}
\braket{\mathrm{GS}^{\rm bdy.}_{\rm in.}(\vec{n})|\mathrm{GS}^{\rm bdy.}_{\rm in.}(-\vec{n})}&=&
\bra{\uparrow+\uparrow}
\frac{1+\tau^{x}_{\frac{1}{2}}\sigma^{z}_{1}}{\sqrt{2}}\frac{1+\sigma^{z}_{1}\tau^{x}_{\frac{3}{2}}}{\sqrt{2}}
\tau^{x}_{\frac{1}{2}}\tau^{x}_{\frac{3}{2}}\frac{1+\tau^{x}_{\frac{1}{2}}\sigma^{z}_{1}}{\sqrt{2}}\frac{1+\sigma^{z}_{1}\tau^{x}_{\frac{3}{2}}}{\sqrt{2}}\ket{\uparrow+\uparrow},\\
&=&\bra{\uparrow+\uparrow}
(1+\tau^{x}_{\frac{1}{2}}\sigma^{z}_{1})(1+\sigma^{z}_{1}\tau^{x}_{\frac{3}{2}})
\ket{\downarrow+\downarrow},\\
&=&\bra{\uparrow+\uparrow}
\tau^{x}_{\frac{1}{2}}(\sigma^{z}_{1})^2\tau^{x}_{\frac{3}{2}}
\ket{\downarrow+\downarrow},\\
&=&1.
\end{eqnarray}
Therefore, the discrete Berry phase along a nontrivial path $\gamma$ is
\begin{eqnarray}\label{eq:holonomy}
n^{\rm bdy.}_{\rm in.}(\gamma):=\exp(\int_{\gamma}A^{\rm bdy.}_{\rm in.}-\frac{1}{2}\int_{\Sigma}dA^{\rm bdy.}_{\rm in.})\braket{{\rm G.S.}_{\rm in.}^{\rm bdy.}(\gamma(2\pi))|{\rm G.S.}_{\rm in.}^{\rm bdy.}(\gamma(0))}=1,
\end{eqnarray}

Similarly, the ground state $\ket{\mathrm{GS.}^{\rm bdy.}_{\rm fin.}(\vec{n})}$ of $H^{\rm bdy.}_{\rm fin.}(\vec{n})$ is given by
\begin{eqnarray}
\ket{{\rm G.S.}^{\rm bdy.}_{\rm in.}(\vec{n})}&:=&
V_{\tau}(\vec{n})_{\frac{1}{2}}V_{\tau}(\vec{n})_{\frac{3}{2}}\frac{1-\tau^{x}_{\frac{1}{2}}\sigma^{z}_{1}}{\sqrt{2}}\frac{1+\sigma^{z}_{1}\tau^{x}_{\frac{3}{2}}}{\sqrt{2}}\ket{\uparrow+\uparrow}.
\end{eqnarray}
By the similar calculation, we can easily check that the Berry connection and curvature of $H^{\rm bdy.}_{\rm fin.}(\Vec{n})$ is the same as that of $H^{\rm bdy.}_{\rm in.}(\Vec{n})$:
\begin{eqnarray}
A^{\rm bdy.}_{\rm fin.}(\Vec{n}):=\bra{\mathrm{GS.}^{\rm bdy.}_{\rm fin.}(\vec{n})}d\ket{\mathrm{GS.}^{\rm bdy.}_{\rm fin.}(\vec{n})}=\frac{i}{2}\sin^{2}(\theta)d\phi,
\end{eqnarray}
and 
\begin{eqnarray}
F^{\rm bdy.}_{\rm fin.}(\Vec{n}):=dA^{\rm bdy.}_{\rm fin.}(\Vec{n})=\frac{i}{2}\sin(2\theta)d\theta d\phi.
\end{eqnarray}
On the other hand, the overlap $\braket{\mathrm{GS}^{\rm bdy.}_{\rm fin.}(-\vec{n})|\mathrm{GS}^{\rm bdy.}_{\rm fin.}(\vec{n})}$
is given by
\begin{eqnarray}
\braket{\mathrm{GS}^{\rm bdy.}_{\rm fin.}(-\vec{n})|\mathrm{GS}^{\rm bdy.}_{\rm fin.}(\vec{n})}&=&
\bra{\uparrow+\uparrow}
\frac{1-\tau^{x}_{\frac{1}{2}}\sigma^{z}_{1}}{\sqrt{2}}\frac{1+\sigma^{z}_{1}\tau^{x}_{\frac{3}{2}}}{\sqrt{2}}
\tau^{x}_{\frac{1}{2}}\tau^{x}_{\frac{3}{2}}\frac{1-\tau^{x}_{\frac{1}{2}}\sigma^{z}_{1}}{\sqrt{2}}\frac{1+\sigma^{z}_{1}\tau^{x}_{\frac{3}{2}}}{\sqrt{2}}\ket{\uparrow+\uparrow},\\
&=&\bra{\uparrow+\uparrow}
(1-\tau^{x}_{\frac{1}{2}}\sigma^{z}_{1})(1+\sigma^{z}_{1}\tau^{x}_{\frac{3}{2}})
\ket{\downarrow+\downarrow},\\
&=&-\bra{\uparrow+\uparrow}
\tau^{x}_{\frac{1}{2}}(\sigma^{z}_{1})^2\tau^{x}_{\frac{3}{2}}
\ket{\downarrow+\downarrow},\\
&=&-1.
\end{eqnarray}
Therefore, the discrete Berry phase is
\begin{eqnarray}\label{eq:holonomy fin}
n^{\rm bdy.}_{\rm fin.}(\gamma):=\exp(\int_{\gamma}A^{\rm bdy.}_{\rm fin.}-\frac{1}{2}\int_{\Sigma}dA^{\rm bdy.}_{\rm fin.})\braket{{\rm G.S.}_{\rm fin.}^{\rm bdy.}(\gamma(2\pi))|{\rm G.S.}_{\rm fin.}^{\rm bdy.}(\gamma(0))}=-1,
\end{eqnarray}
Thus the ratio of the discrete Berry phase is 
\begin{eqnarray}
\frac{n^{\rm bdy.}_{\rm in.}(\gamma)}{n^{\rm bdy.}_{\rm fin.}(\gamma)}=-1.
\end{eqnarray}

\subsubsection{Physical Interpretation II : Boundary Condition Obstacle }\label{sec:obstacle}

In Sec.\ref{sec:chern number pump}, we considered the system with boundary and discussed the flow of the discrete Berry phase, when comparing $t=0$ and $t=2\pi$. This breaking of the $2\pi$-periodicity is due to the fact that the boundary terms were not $2\pi$-periodic. In fact, the boundary term $-\tau^{x}_{\frac{1}{2}}(\vec{n})\sigma^{z}_{1}(t)$ which was added to the Hamiltonian is not $2\pi$-periodic. We can consider a boundary term like $\tau^{z}_{\frac{1}{2}}(\vec{n})$ that preserves $2\pi$-periodicity, but this time the boundary term is not global on $\rp{2}$. 

\begin{figure}[H]
 \begin{center}
  \includegraphics[width=90mm]{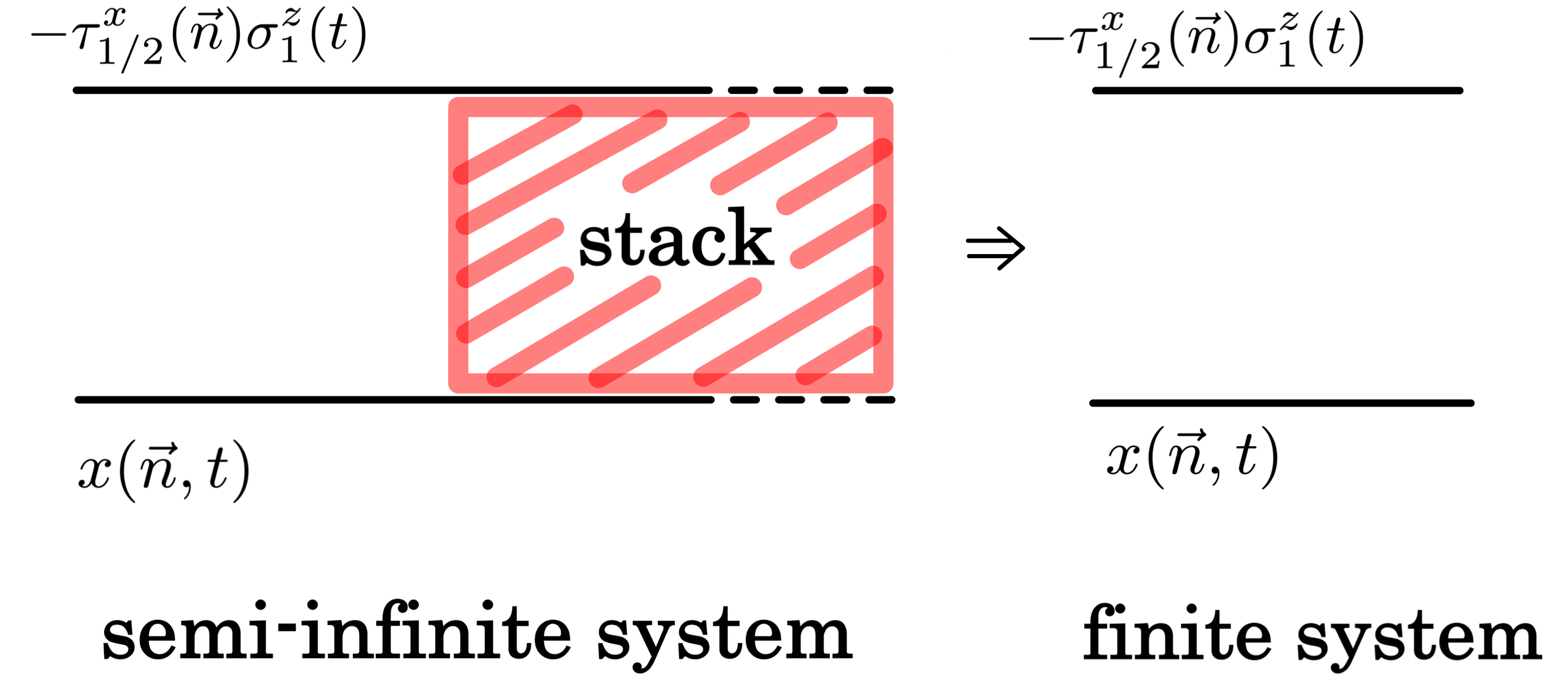}
 \end{center}
  \caption[]{By stacking two semi-infinite systems with different boundary conditions, we can trivialize the bulk of the system as a family. This results in the system having no more than a finite number of degrees of freedom. In particular, the discrete Berry phase is well-defined for any $t$. From its construction, the ratio of the discrete Berry phase at $t=0$ and $t=2\pi$ is $-1$. Therefore, there exists a singular point where the gap is closed.
}
\label{fig:boundary term obstacle}
\end{figure}

It is a natural question to ask whether there exists a term that is parameterized by $\rp{2}\times S^1$ globally and makes the system a unique gapped at all points in $\rp{2}\times S^1$. Let us suppose that there exists such a term, which we denote by $x(\vec{n},t)$. Then, the flow of the discrete Berry phase is trivial under this boundary condition. This implies that by stacking two semi-infinite chains with boundary condition $-\tau^{x}_{1/2}(\vec{n})\sigma^{z}_{1}(t)$ and $x(\vec{n},t)$, we obtain a nontrivial family of $(0+1)$-dimensional systems parametrized by  $\rp{2}\times \left[0,2\pi\right]$ whose ratio of the discrete Berry phase at $t=0$ and $t=2\pi$ is $-1$(Fig.\ref{fig:boundary term obstacle}). If there existed such a family, it would be inconsistent with the quantization of the discrete Berry phase. Therefore, there are no such boundary terms.

In general, when the higher pump is nontrivial, it is expected to give rise to a nontrivial flow of the discrete Berry phase or Berry curvature. Accepting this conjecture, it follows that there is no boundary term that is parameterized over the whole of $X$ and makes the system unique gap at all points $x\in X$, if its higher pump is nontrivial.

\subsection{$\mathrm{L}(3,1)\times S^1$ model (or $\zmod{3}$ charge pump model)}

\subsubsection{Definition of a Model}\label{sec:def of Z/3 cluster model}\label{sec:Z/3 cluster model}

Let's consider another model with nontrivial higher pump. As with the model in Sec.\ref{sec:model with rp2 times s1}, we will refer to integer sites as $\sigma$-sites and the others as $\tau$-sites. At each site, there is a $3$-dimensional Hilbert space. At $\tau$-site, we take the following orthonormal basis:
\begin{eqnarray}\label{eq:DDW basis 1}
\ket{u_{0}}=\frac{1}{\sqrt{3}}\begin{pmatrix}
1\\
\omega\\
\omega^2
\end{pmatrix},
\ket{u_{1}}=\frac{1}{\sqrt{3}}\begin{pmatrix}
1\\
1\\
1
\end{pmatrix},
\ket{u_{2}}=\frac{1}{\sqrt{3}}\begin{pmatrix}
1\\
\omega^2\\
\omega
\end{pmatrix}.
\end{eqnarray}
Here, $\omega = e^{\frac{2\pi i}{3}}$. 
At $\sigma$-site, we take the following orthonormal basis:
\begin{eqnarray}\label{eq:DDW basis 2}
\ket{\tilde{\sigma}_{0}}=\begin{pmatrix}
1\\
0\\
0
\end{pmatrix},
\ket{\tilde{\sigma}_{1}}=\begin{pmatrix}
0\\
1\\
0
\end{pmatrix},
\ket{\tilde{\sigma}_{2}}=\begin{pmatrix}
0\\
0\\
1
\end{pmatrix}.
\end{eqnarray}
We call basis Eq.(\ref{eq:DDW basis 1}) and Eq.(\ref{eq:DDW basis 2}) the decorated domain wall basis. On the other hand, we define
\begin{eqnarray}\label{eq:DDW basis 3}
\ket{\tilde{\tau}_{0}}=\begin{pmatrix}
1\\
0\\
0
\end{pmatrix},
\ket{\tilde{\tau}_{1}}=\begin{pmatrix}
0\\
1\\
0
\end{pmatrix},
\ket{\tilde{\tau}_{2}}=\begin{pmatrix}
0\\
0\\
1
\end{pmatrix},
\end{eqnarray}
and call Eq.(\ref{eq:DDW basis 2}) and Eq.(\ref{eq:DDW basis 3}) $z$-basis. The tilde on $\tau$ and $\sigma$ is a symbol to distinguish it from the $\zmod{2}$ model. In the following sections, the same calculations as for the $\zmod{2}$ model will be performed in parallel for the $\zmod{3}$ model. In that case, we always attach tildes to quantities related to the $\zmod{3}$ model.

We define the $\Zmod{3}$ spin operator acting on the local Hilbert space on $\tau$-sites by
\begin{eqnarray}
\tilde{\tau}^{x}:=\begin{pmatrix}
&&\omega\\
\omega&&\\
&\omega&
\end{pmatrix},
\tilde{\tau}^{z}:=\begin{pmatrix}
1&&\\
&\omega&\\
&&\omega^2
\end{pmatrix},
\end{eqnarray}
and the $\Zmod{3}$ spin operator acting on the local Hilbert space on $\sigma$-sites by
\begin{eqnarray}
\tilde{\sigma}^{x}:=\begin{pmatrix}
&&1\\
1&&\\
&1&
\end{pmatrix},
\tilde{\sigma}^{z}:=\begin{pmatrix}
1&&\\
&\omega&\\
&&\omega^2
\end{pmatrix}.
\end{eqnarray}
Remark that these matrices are not self-adjoint, and satisfy the following commutation relation:
\begin{eqnarray}
\tilde{\tau}^{z}\tilde{\tau}^{x}=\omega\tilde{\tau}^{x}\tilde{\tau}^{z}, \tilde{\sigma}^{z}\tilde{\sigma}^{x}=\omega\tilde{\sigma}^{x}\tilde{\sigma}^{z}.
\end{eqnarray}
We note that $\ket{u_{i}}$ is the basis for diagonalizing $\tilde{\tau}^{x}$, and it is cyclically shifted by $\tilde{\tau}^{z}$, and $\ket{\tilde{\sigma}_{i}}$ is the basis for diagonalizing $\tilde{\sigma}^{z}$, and it is cyclically shifted by $\tilde{\sigma}^{x}$:
\begin{eqnarray}
\tilde{\tau}^{z}\ket{u_{i}}=\ket{u_{i-1}},\tilde{\tau}^{x}\ket{u_{i}}&=&\omega^{i}\ket{u_{i}},\\
\tilde{\sigma}^{z}\ket{\tilde{\sigma}_{i}}=\omega^{i}\ket{\tilde{\sigma}_{i}},
\tilde{\sigma}^{x}\ket{\tilde{\sigma}_{i}}&=&\ket{\tilde{\sigma}_{i+1}}.
\end{eqnarray}
Here, the subscript of $u$ and $\tilde{\sigma}$ is defined modulo $3$. 

Now, we define the following Hamiltonian\cite{GM14}:
\begin{eqnarray}\label{eq:Z3 cluster}
H=\sum_{j}-\tilde{\tau}^{z\dagger}_{j-\frac{1}{2}}\tilde{\sigma}^{x}_{j}\tilde{\tau}^{z}_{j+\frac{1}{2}}-\tilde{\sigma}^{z}_{j}\tilde{\tau}^{x}_{j+\frac{1}{2}}\tilde{\sigma}^{z\dagger}_{j+1}-\tilde{\tau}^{z}_{j-\frac{1}{2}}\tilde{\sigma}^{x\dagger}_{j}\tilde{\tau}^{z\dagger}_{j+\frac{1}{2}}-\tilde{\sigma}^{z\dagger}_{j}\tilde{\tau}^{x\dagger}_{j+\frac{1}{2}}\tilde{\sigma}^{z}_{j+1}.
\end{eqnarray}
Remark that each term is commuted with the other, and the cube of each term is equal to $1$. We regard the second and fourth terms as the configuration terms and the first and third terms as fluctuation terms. 

\begin{figure}[H]
 \begin{center}
  \includegraphics[width=70mm]{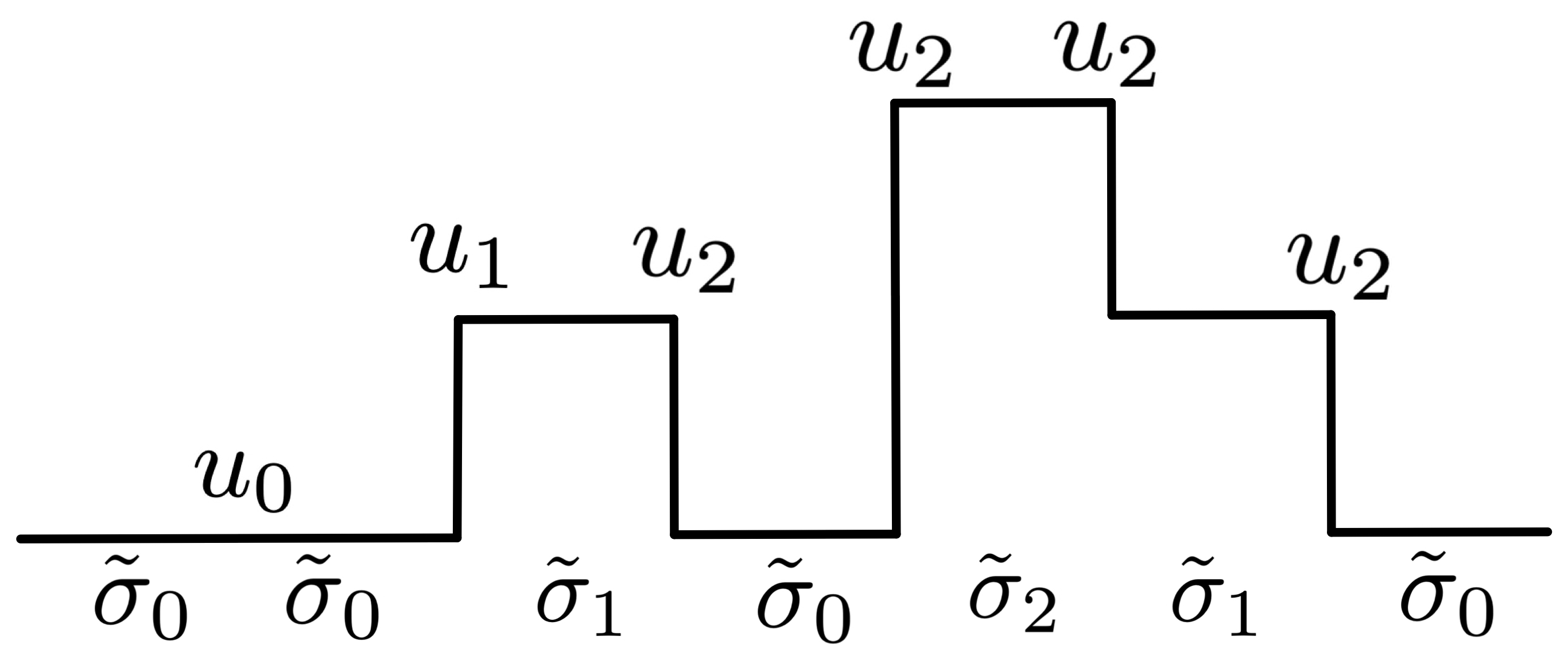}
 \end{center}
  \caption[]{
  An example of the decorated domain wall configuration.
}
\label{fig:DDWconfig}
\end{figure}

In order to write down the ground state of $H$, we introduce decorated domain wall state with respect to $\ket{u_{i}}$ and $\ket{\tilde{\sigma}_{i}}$. A typical decorated domain wall state is
\begin{eqnarray}\label{eq:Z/3 typical DDW}
\ket{\cdots u_{0}\tilde{\sigma}_{0}u_{1}\tilde{\sigma}_{1}u_{2}\tilde{\sigma}_{0}u_{2}\tilde{\sigma}_{2}u_{2}\tilde{\sigma}_{1}u_{2}\tilde{\sigma}_{0}\cdots},
\end{eqnarray}
i.e., put $u_{i_k}$ where the difference between $j_k-j_{k-1}\equiv i_k$ modulo $3$ (see Fig.\ref{fig:DDWconfig}). The place where $j_k-j_{k-1}\neq0$ is called the domain wall of $\tilde{\sigma}$-spin. Since we "decorate" $u_{i}$ on the domain wall, the state in Eq.(\ref{eq:Z/3 typical DDW}) is called a decorated domain wall state. This is a natural generalization of the decorated domain wall introduced in Sec.\ref{sec:model with rp2 times s1}. We will denote the set of decorated domain wall states as ${\rm DDW}_{3}$. The ground state of the Hamiltonian Eq.(\ref{eq:Z3 cluster}) is given by
\begin{eqnarray}\label{eq:GS of the Z/3 cluster}
\ket{\mathrm{G.S.}}:=\prod_{j\in\mathbb{Z}}\frac{\tilde{f}_{j}}{\sqrt{3}}\ket{\mathrm{Ref}},
\end{eqnarray}
where 
\begin{eqnarray}
\tilde{f}_{j}:=1+\tilde{\tau}^{z\dagger}_{j-\frac{1}{2}}\tilde{\sigma}^{x}_{j}\tilde{\tau}^{z}_{j+\frac{1}{2}}+\tilde{\tau}^{z}_{j-\frac{1}{2}}\tilde{\sigma}^{x\dagger}_{j}\tilde{\tau}^{z\dagger}_{j+\frac{1}{2}}
\end{eqnarray} 
and $\ket{\mathrm{Ref}}$ is a decorated domain wall state. Note that $\tilde f_j/3$s are orthogonal projections satisfying $(\tilde{f}_j/3)^\dag = \tilde f_j, (\tilde f_j/3)^2=\tilde f_j$, and $\tilde f_i \tilde f_j = \tilde f_j \tilde f_i$.
Note that $\ket{\mathrm{Ref}}$ is not unique but the ground state Eq.(\ref{eq:GS of the Z/3 cluster}) is independent of this choice. 
In other words, the ground state is a superposition of all decorated domain wall configurations with the same weights:
\begin{eqnarray}\label{eq:ground state of Z/3 cluster}
\ket{\mathrm{G.S.}}\propto\sum_{\{i_{k},j_{l}\}\in\mathrm{DDW}_{3}}\ket{u_{i_1}\tilde{\sigma}_{j_1}\cdots u_{i_L}\tilde{\sigma}_{j_L}}.
\end{eqnarray}   

Based on this model, let's construct a model parametrized by $\mathrm{L}(3,1)\times S^1$. First, we give $\mathrm{L}(3,1)$ dependence to $\tau$-sites. To this end, we define a unitary matrix
\begin{eqnarray}
\tilde{V}_{\tau}(\vec{z})&:=&\frac{1}{3}\begin{pmatrix}
1+z_1+z_1^{\ast}+z_{2}- z_{2}^{\ast}&\omega^{2}+z_1+\omega z_1^{\ast}+z_{2}-\omega z_{2}^{\ast}&\omega+z_1+\omega^{2} z_1^{\ast}+z_{2}-\omega^{2} z_{2}^{\ast}\\
\omega+z_{1}+\omega^{2}z_{1}^{\ast}+\omega^{2} z_{2}- z_{2}^{\ast}&1+z_{1}+z_{1}^{\ast}+\omega^2 z_{2}-\omega z_{2}^{\ast}&\omega^{2}+z_{1}+\omega z_{1}^{\ast}+\omega^{2} z_{2}-\omega^{2}z_{2}^{\ast}\\
\omega^{2}+z_{1}+\omega z_{1}^{\ast}+\omega z_{2}- z_{2}^{\ast}&\omega+z_{1}+\omega^{2} z_{1}^{\ast}+\omega z_{2}-\omega z_{2}^{\ast}&1+z_{1}+z_{1}^{\ast}+\omega z_{2}-\omega^{2} z_{2}^{\ast}
\end{pmatrix},\label{eq:Vtauz}
\end{eqnarray}
and by using this unitary matrix\footnote{We make a comment on the origin of this matrix in Appendix\ref{sec:comment on uni}.}, we define
\begin{eqnarray}
\tilde{\tau}^{z}(\vec{z})&:=&\tilde{V}_{\tau}(\vec{z})\tilde{\tau}^{z}\tilde{V}_{\tau}(\vec{z})^{\dagger},\\
\tilde{\tau}^{x}(\vec{z})&:=&\tilde{V}_{\tau}(\vec{z})\tilde{\tau}^{x}\tilde{V}_{\tau}(\vec{z})^{\dagger},
\end{eqnarray}
and 
\begin{eqnarray}
\ket{\tilde{\tau}_{i}(\vec{z})}&:=&\tilde{V}_{\tau}(\vec{z})\ket{\tilde{\tau}_{i}}.
\end{eqnarray}
Note that they meet the following relations:
\begin{eqnarray}\label{eq:prop of V tau}
\tilde{V}_{\tau}(\omega\vec{z})_{i,j}=\omega \tilde{V}_{\tau}(\vec{z})_{i,j+1}=(\tilde{V}_{\tau}(\vec{z})\tilde{\tau}^{x})_{i,j},
\end{eqnarray}
\begin{eqnarray}
\ket{\tilde{\tau}_{i}(\omega\vec{z})}=\tilde{\tau}^{x}_{i}(\vec{z})\ket{\tilde{\tau}_{i}(\vec{z})}, 
\end{eqnarray}
\begin{align}
    \tilde \tau^x(\omega \vec{z}) = \tilde \tau^x(\vec{z}), 
    \tilde \tau^z(\omega \vec{z}) = \omega^2 \tilde \tau^z(\vec{z}).
    \label{eq:tau_z_omega}
\end{align}
Next, we give $S^{1}$ dependence to $\sigma$-sites. We define a unitary matrix  
\begin{eqnarray}
\tilde{V}_{\sigma}(t):=\frac{1}{3}\begin{pmatrix}
1+\exp(i\frac{t}{3})+\exp(i\frac{2t}{3})&1+\omega\exp(i\frac{t}{3})+\omega^{2}\exp(i\frac{2t}{3})&1+\omega^{2}\exp(i\frac{t}{3})+\omega\exp(i\frac{2t}{3})\\
1+\omega^{2}\exp(i\frac{t}{3})+\omega\exp(i\frac{2t}{3})&1+\exp(i\frac{t}{3})+\exp(i\frac{2t}{3})&1+\omega\exp(i\frac{t}{3})+\omega^{2}\exp(i\frac{2t}{3})\\
1+\omega\exp(i\frac{t}{3})+\omega^{2}\exp(i\frac{2t}{3})&1+\omega^{2}\exp(i\frac{t}{3})+\omega\exp(i\frac{2t}{3})&1+\exp(i\frac{t}{3})+\exp(i\frac{2t}{3})
\end{pmatrix},\label{eq:Vsigmat}
\end{eqnarray}
and by using this matrix\footnote{We make a comment on the origin of this matrix in Appendix\ref{sec:comment on uni}.}, we define
\begin{eqnarray}
\tilde{\sigma}^{z}(t)&:=&\tilde{V}_{\sigma}(t)\tilde{\sigma}^{z}\tilde{V}_{\sigma}(t)^{\dagger},\\
\tilde{\sigma}^{x}(t)&:=&\tilde{V}_{\sigma}(t)\tilde{\sigma}^{x}\tilde{V}_{\sigma}(t)^{\dagger}(=\tilde \sigma^x),
\end{eqnarray}
and 
\begin{eqnarray}
\ket{\tilde{\sigma}_{i}(t)}=\tilde{V}_{\sigma}(t)\ket{\tilde{\sigma}_i}.
\end{eqnarray}
Note that they meet the following relations:
\begin{eqnarray}
    \tilde{V}_{\sigma}(t+2\pi)_{i,j}=\tilde{V}_{\sigma}(t)_{i,j+1},
\end{eqnarray}
\begin{eqnarray}
    \ket{\tilde{\sigma}_{i}(t+2\pi)}=\tilde{\sigma}^{x}_{i}(t)\ket{\tilde{\sigma}_{i}(t)},
\end{eqnarray}
\begin{align}
    \tilde \sigma^z(t+2\pi)=\omega^2 \tilde \sigma^z(t).\label{eq:sigma_t_2pi}
\end{align}
We define a model for $\vec{z} \in S^3$ and $t \in [0,2\pi]$ as
\begin{eqnarray}\label{eq:Z/3 cluster model with parameter}
H(\vec{z},t)&=&-\sum_{j\in\mathbb{Z}}\tilde{\tau}^{z\dagger}_{j-\frac{1}{2}}(\vec{z})\tilde{\sigma}^{x}_{j}(t)\tilde{\tau}^{z}_{j+\frac{1}{2}}(\vec{z})-\sum_{j\in\mathbb{Z}}\tilde{\sigma}^{z}_{j}(t)\tilde{\tau}^{x}_{j+\frac{1}{2}}(\vec{z})\tilde{\sigma}^{z\dagger}_{j+1}(t)\nonumber\\
&-&\sum_{j\in\mathbb{Z}}\tilde{\tau}^{z}_{j-\frac{1}{2}}(\vec{z})\tilde{\sigma}^{x\dagger}_{j}(t)\tilde{\tau}^{z\dagger}_{j+\frac{1}{2}}(\vec{z})-\sum_{j\in\mathbb{Z}}\tilde{\sigma}^{z\dagger}_{j}(t)\tilde{\tau}^{x\dagger}_{j+\frac{1}{2}}(\vec{z})\tilde{\sigma}^{z}_{j+1}(t).
\end{eqnarray}
Eqs. (\ref{eq:tau_z_omega}) and (\ref{eq:sigma_t_2pi}) guarantee that the Hamiltonian (\ref{eq:Z/3 cluster model with parameter}) is a model over $\mathrm{L}(3,1) \times S^1$.
The ground state of this model is the superposition of decorated domain wall configuration with the $\ket{u(\vec{z})}$ and $\ket{\tilde{\sigma}(\vec{z})}$ basis:
\begin{eqnarray}\label{eq:lens space}
\ket{\mathrm{G.S.}(\vec{z},t)}&\propto&
\sum_{\{i_k,j_l\}\in\mathrm{DDW}_3}\ket{u_{i_1}(\vec{z}),\tilde{\sigma}_{j_{1}}(t),...,u_{i_L}(\vec{z}),\tilde{\sigma}_{j_{L}}(t)},
\end{eqnarray}
or explicitly, 
\begin{eqnarray}\label{eq:GS of the Z/3 cluster with param}
\ket{\mathrm{G.S.}(\vec{z},t)}:=\prod_{j\in\mathbb{Z}}\frac{\tilde{f}_{j}(\vec{z},t)}{\sqrt{3}}\ket{\mathrm{Ref}(\vec{z},t)}.
\end{eqnarray}
Here 
\begin{eqnarray}
\tilde{f}_{j}(\vec{z},t):=1+\tilde{\tau}^{z\dagger}_{j-\frac{1}{2}}(\vec{z})\tilde{\sigma}^{x}_{j}(t)\tilde{\tau}^{z}_{j+\frac{1}{2}}(\vec{z})+\tilde{\tau}^{z}_{j-\frac{1}{2}}(\vec{z})\tilde{\sigma}^{x\dagger}_{j}(t)\tilde{\tau}^{z\dagger}_{j+\frac{1}{2}}(\vec{z}),
\end{eqnarray} 
and $\ket{\mathrm{Ref}(\vec{z},t)}$ is a simultaneous eigenstate of $\tilde{\sigma}^{z}_{j}(t)\tilde{\tau}^{x}_{j+\frac{1}{2}}(\vec{z})\tilde{\sigma}^{z\dagger}_{j+1}(t)$ and $\tilde{\sigma}^{z\dagger}_{j}(t)\tilde{\tau}^{x\dagger}_{j+\frac{1}{2}}(\vec{z})\tilde{\sigma}^{z}_{j+1}(t)$ with eigenvalue $1$.

Let's check explicitly that the ground state Eq.(\ref{eq:lens space}) is parametrized by $\mathrm{L}(3,1)\times S^1$. Let $\Delta\tilde{\sigma}_{i}(t)=\tilde{\sigma}_{j+1}(t)-\tilde{\sigma}_{j}(t)$. At $\omega\vec{z}=(\omega z_1,\omega z_2)$, 
\begin{eqnarray}
\ket{\mathrm{G.S.}(\omega\vec{z},t)}&=&\sum_{\{i_k,j_l\}\in\mathrm{DDW}_3}\prod_{j}e^{i\frac{\pi}{3}\Delta\tilde{\sigma}_{j}}\ket{u_{i_1}(\vec{z}),\tilde{\sigma}_{j_{1}}(t),...,u_{i_L}(\vec{z}),\tilde{\sigma}_{j_{L}}(t)},\\
&=&\sum_{\{i_k,j_l\}\in\mathrm{DDW}_3}e^{i\frac{\pi}{3}\sum_{j}\Delta\tilde{\sigma}_{j}(t)}\ket{u_{i_1}(\vec{z}),\tilde{\sigma}_{j_{1}}(t),...,u_{i_L}(\vec{z}),\tilde{\sigma}_{j_{L}}(t)},\\
&=&\ket{\mathrm{G.S.}(\vec{z},t)}.
\end{eqnarray}
Here, we used
\begin{eqnarray}
\sum_{j}\Delta\tilde{\sigma}_{j}(t)=0,
\end{eqnarray}
under the periodic boundary condition. Also, at $t+2\pi$,
\begin{eqnarray}
    \ket{\mathrm{G.S.}(\vec{z},t+2\pi)}&=&\sum_{\{i_k,j_l\}\in\mathrm{DDW}_3}\prod_{j}\tilde{\sigma}_j(t)\ket{u_{i_1}(\vec{z}),\tilde{\sigma}_{j_{1}}(t),...,u_{i_L}(\vec{z}),\tilde{\sigma}_{j_{L}}(t)},\\
    &=&\sum_{\{i_k,j_l\}\in\mathrm{DDW}_3}\ket{u_{i_1}(\vec{z}),\tilde{\sigma}_{j_{1}}(t),...,u_{i_L}(\vec{z}),\tilde{\sigma}_{j_{L}}(t)},\\
    &=&\ket{\mathrm{G.S.}(\vec{z},t)}.
\end{eqnarray}
Therefore, $\ket{\mathrm{G.S.}(\vec{z},t)}$ is a state over $\mathrm{L}(3,1)\times S^1$. In the following, we verify the nontriviality of this model\footnote{Since $\cohoZ{3}{\mathrm{L}(3,1)\times S^1}\simeq\zmod{3}$, it can be nontrivial as a family of invertible states.} as a family of invertible states over $\mathrm{L}(3,1)\times S^{1}$. 

\subsubsection{Physical Interpretation I: Discrete Berry Phase Pumping}\label{sec:Z/3 chernnumber pump}
As in the case of $\rp{2}\times S^1$ model, we can see that the quantum mechanical system parametrized by $\mathrm{L}(3,1)$ is pumped to the edge. In fact, by deforming the system along the $S^1$ direction, we can see the flow of the effective discrete Berry phase, as we will see below. 

Let's cut the system between site $0$ and site $\frac{1}{2}$, and create a boundary such that $\tau$-site appears at the edge:
\begin{eqnarray}
H(\vec{z},t)&=&\sum_{j=1}^{\infty}-\tilde{\tau}^{z\dagger}_{j-\frac{1}{2}}(\vec{z})\tilde{\sigma}^{x}_{j}(t)\tilde{\tau}^{z}_{j+\frac{1}{2}}(\vec{z})-\tilde{\sigma}^{z}_{j}(t)\tilde{\tau}^{x}_{j+\frac{1}{2}}(\vec{z})\tilde{\sigma}^{z\dagger}_{j+1}(t)\nonumber\\
&-&\tilde{\tau}^{z}_{j-\frac{1}{2}}(\vec{z})\tilde{\sigma}^{x\dagger}_{j}(t)\tilde{\tau}^{z\dagger}_{j+\frac{1}{2}}(\vec{z})-\tilde{\sigma}^{z\dagger}_{j}(t)\tilde{\tau}^{x\dagger}_{j+\frac{1}{2}}(\vec{z})\tilde{\sigma}^{z}_{j+1}(t).
\end{eqnarray}
To remove the ground state degeneracy, we need to add a boundary term. 
We choose $-\tilde{\tau}^{x}_{\frac{1}{2}}(\vec{z})\tilde{\sigma}^{z}_{1}(t)$ as a boundary term, and  consider the following initial($t=0$) and final($t=2\pi$) Hamiltonians\footnote{Remark that this term is not $2\pi$-periodic.}: 
\begin{eqnarray}
H_{\rm in.}(\vec{z})&:=&-\tilde{\tau}^{x}_{\frac{1}{2}}(\vec{z})\tilde{\sigma}^{z}_{1}-\sum_{j=1}^{\infty}\tilde{\tau}^{z\dagger}_{j-\frac{1}{2}}(\vec{z})\tilde{\sigma}^{x}_{j}\tilde{\tau}^{z}_{j+\frac{1}{2}}(\vec{z})-\sum_{j=1}^{\infty}\tilde{\sigma}^{z}_{j}\tilde{\tau}^{x}_{j+\frac{1}{2}}(\vec{z})\tilde{\sigma}^{z\dagger}_{j+1}+h.c.,\\
H_{\rm fin.}(\vec{z})&:=&-\omega^2\tilde{\tau}^{x}_{\frac{1}{2}}(\vec{z})\tilde{\sigma}^{z}_{1}-\sum_{j=1}^{\infty}\tilde{\tau}^{z\dagger}_{j-\frac{1}{2}}(\vec{z})\tilde{\sigma}^{x}_{j}\tilde{\tau}^{z}_{j+\frac{1}{2}}(\vec{z})-\sum_{j=1}^{\infty}\tilde{\sigma}^{z}_{j}\tilde{\tau}^{x}_{j+\frac{1}{2}}(\vec{z})\tilde{\sigma}^{z\dagger}_{j+1}+h.c.
\end{eqnarray}
Here, we used $\tilde{\sigma}^{z}_{j}(2\pi)=\omega^2\tilde{\sigma}^{z}_{j}$. Let's compute the discrete Berry phase of these Hamiltonians.

\begin{figure}[H]
 \begin{center}
  \includegraphics[width=35mm]{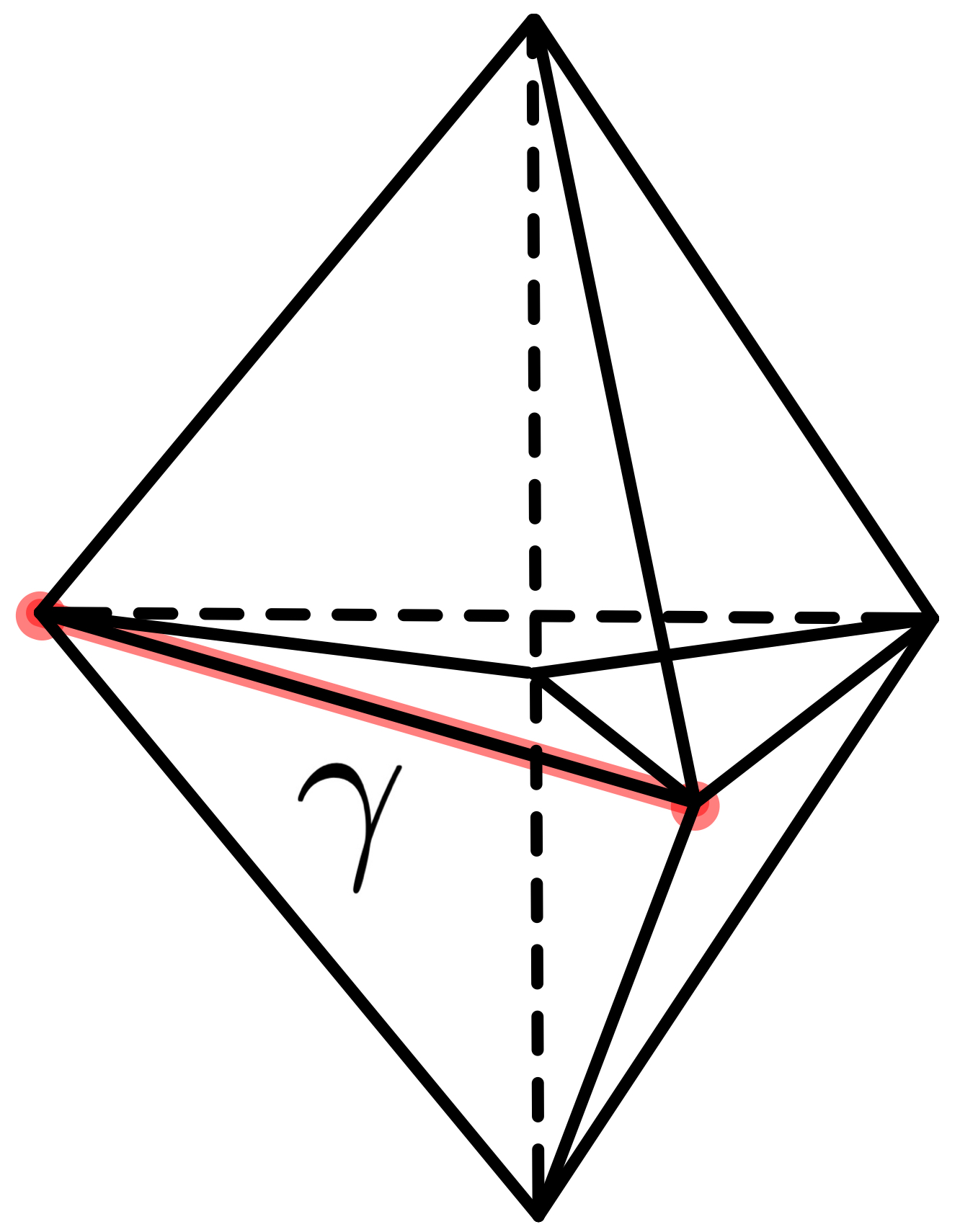}
 \end{center}
  \caption[]{
  A nontrivial path in $\mathrm{L}(3,1)$. As is well known, the lens space can be constructed from a $3$-dimensional ball. The surface of a $3$-dimensional ball is a 
$2$-dimensional sphere, and divide this sphere into the northern and southern hemispheres. Then, the southern hemisphere is rotated by $2\pi/3$ with respect to the northern hemisphere and glued together. Consider a path $\gamma$ starting from a point on the equator and arriving at a point rotated along the equator by $2\pi/3$. The fundamental group of $\mathrm{L}(3,1)$ is $\zmod{3}$ and a representative path of the generator of $\zmod{3}$ is $\gamma$.
}
\label{fig:lens space path}
\end{figure}

First, the ground states of $H_{\rm in.}(\vec{z})$ and $H_{\rm fin.}(\vec{z})$ are
\begin{eqnarray}
\ket{\mathrm{G.S.}_{\rm in.}(\vec{z})}&:=&\prod_{j=1}^{\infty}\tilde{V}_{\tau}(\vec{z})_{j-\frac{1}{2}}\prod_{j=1}^{\infty}\frac{\tilde{f}_{j}}{\sqrt{3}}\ket{\mathrm{Ref}_{\rm in.}},\\
&=&\prod_{j=1}^{\infty}\tilde{V}_{\tau}(\vec{z})_{j-\frac{1}{2}}\ket{\mathrm{G.S.}_{\rm in.}(\vec{z}=(1,0))},\\
\ket{\mathrm{G.S.}_{\rm fin.}(\vec{z})}&:=&\prod_{j=1}^{\infty}\tilde{V}_{\tau}(\vec{z})_{j-\frac{1}{2}}\prod_{j=1}^{\infty}\frac{\tilde{f}_{j}}{\sqrt{3}}\ket{\mathrm{Ref}_{\rm fin.}},\\
&=&\prod_{j=1}^{\infty}\tilde{V}_{\tau}(\vec{z})_{j-\frac{1}{2}}\ket{\mathrm{G.S.}_{\rm fin.}(\vec{z}=(1,0))},
\end{eqnarray}
where $\tilde{V}_{\tau}(\vec{z})_{j-\frac{1}{2}}$ is a unitary operator $\tilde{V}_{\tau}(\vec{z})$ acting on a $\tau$-site $j-\frac{1}{2}$, and $\ket{\mathrm{Ref}_{\rm in.}(\vec{z})}$ is a simultaneous eigenstate of first and third term of $H_{\rm in.}(\vec{z}=(1,0),t=0)$ with eigenvalue $1$, i.e., 
\begin{eqnarray}
\tilde{\tau}^{x}_{\frac{1}{2}}\tilde{\sigma}^{z}_{1}\ket{\mathrm{Ref}_{\rm in.}}&=&\ket{\mathrm{Ref}_{\rm in.}},\\
\tilde{\sigma}^{z}_{j}\tilde{\tau}^{x}_{j+\frac{1}{2}}\tilde{\sigma}^{z\dagger}_{j+1}\ket{\mathrm{Ref}_{\rm in.}}&=&\ket{\mathrm{Ref}_{\rm in.}}, \label{eq:l31_bdy_ddw_2}
\end{eqnarray}
for all $j \in \N$. Note that the eigenspace with eigenvalue 1 is the same for both $\tilde{\tau}^{x}_{\frac{1}{2}}\tilde{\sigma}^{z}_{1}$ and its real part $(\tilde{\tau}^{x}_{\frac{1}{2}}\tilde{\sigma}^{z}_{1}+h.c.)/2$. The same is true for (\ref{eq:l31_bdy_ddw_2}). Similarly, $\ket{\mathrm{Ref}_{\rm fin.}(\vec{z})}$ is a simultaneous eigenstate of first and third term of $H_{\rm fin.}(\vec{z}=(1,0),t=0)$ with eigenvalue $1$, i.e., 
\begin{eqnarray}
\tilde{\tau}^{x}_{\frac{1}{2}}\tilde{\sigma}^{z}_{1}\ket{\mathrm{Ref}_{\rm fin.}}&=&\ket{\mathrm{Ref}_{\rm fin.}},\\
\omega^2\tilde{\sigma}^{z}_{j}\tilde{\tau}^{x}_{j+\frac{1}{2}}\tilde{\sigma}^{z\dagger}_{j+1}\ket{\mathrm{Ref}_{\rm fin.}}&=&\ket{\mathrm{Ref}_{\rm fin.}},
\end{eqnarray}
for all $j \in \N$. We define 
\begin{eqnarray}
\tilde{h}_{j-\frac{1}{2}}(\vec{z}):=\tilde{V}_{\tau}(\vec{z})_{j-\frac{1}{2}}^{\dagger}d\tilde{V}_{\tau}(\vec{z})_{j-\frac{1}{2}}.
\end{eqnarray}
Then the Berry connections of $H_{\rm in.}(\vec{z})$ and $H_{\rm fin.}(\vec{z})$ is given by
\begin{eqnarray}
\tilde{A}_{\rm in.}(\vec{z})&:=&\bra{\mathrm{G.S.}_{\rm in.}(\vec{z})}d\ket{\mathrm{G.S.}_{\rm in.}(\vec{z})},\\
&=&\sum_{j=1}^{\infty}\bra{\mathrm{G.S.}_{\rm in.}(\vec{z})}\tilde{h}_{j-\frac{1}{2}}(\vec{z})\ket{\mathrm{G.S.}_{\rm in.}(\vec{z})},\\
&=&\frac{1}{9}\sum_{j=1}^{\infty} \bra{\mathrm{Ref}_{\rm in.}} \tilde{f}_{j}\tilde{f}_{j-1} \tilde{h}_{j-\frac{1}{2}}(\vec{z}) \tilde{f}_{j-1}\tilde{f}_{j}\ket{\mathrm{Ref}_{\rm in.}},\label{eq:Z/3 Berry connection at t=0}\\
\tilde{A}_{\rm fin.}(\vec{z})&:=&\bra{\mathrm{G.S.}_{\rm fin.}(\vec{z})}d\ket{\mathrm{G.S.}_{\rm fin.}(\vec{z})},\\
&=&\sum_{j=1}^{\infty}\bra{\mathrm{G.S.}_{\rm fin.}(\vec{z})}\tilde{h}_{j-\frac{1}{2}}(\vec{z})\ket{\mathrm{G.S.}_{\rm fin.}(\vec{z})},\\
&=&\frac{1}{9}\sum_{j=1}^{\infty} \bra{\mathrm{Ref}_{\rm fin.}} \tilde{f}_{j}\tilde{f}_{j-1} \tilde{h}_{j-\frac{1}{2}}(\vec{z}) \tilde{f}_{j-1}\tilde{f}_{j}\ket{\mathrm{Ref}_{\rm fin.}},\label{eq:Z/3 Berry connection at t=2pi}
\end{eqnarray}
where $\tilde{f}_{0}:=\sqrt{3}$. By using these quantities, the discrete Berry phases are given by
\begin{eqnarray}
\tilde{n}_{\rm in.}(\tilde{\gamma})&:=&\exp(\int_{\tilde{\gamma}}\tilde{A}_{\rm in.}-\frac{1}{3}\int_{\tilde{\Sigma}}d\tilde{A}_{\rm in.})\braket{\mathrm{G.S.}_{\rm in.}(\vec{z}=\tilde{\gamma}_{0})|\mathrm{G.S.}_{\rm in.}(\vec{z}=\tilde{\gamma}_{1})},\label{eq:Z/3 holonomy 1}\\
\tilde{n}_{\rm fin.}(\tilde{\gamma})&:=&\exp(\int_{\tilde{\gamma}}\tilde{A}_{\rm fin.}-\frac{1}{3}\int_{\tilde{\Sigma}}d\tilde{A}_{\rm fin.})\braket{\mathrm{G.S.}_{\rm fin.}(\vec{z}=\tilde{\gamma}_{0})|\mathrm{G.S.}_{\rm fin.}(\vec{z}=\tilde{\gamma}_{1})}.\label{eq:Z/3 holonomy 2}
\end{eqnarray}
Here $\tilde{\gamma}$ is a nontrivial path in $\mathrm{L}(3,1)$ as in the Fig.\ref{fig:lens space path}, and $\tilde{\gamma}_{0}=\tilde{\gamma}(0)$, $\tilde{\gamma}_{1}=\tilde{\gamma}(2\pi)$ and $\partial\tilde{\Sigma}=3\cdot\tilde{\gamma}$. As in the case of Sec.\ref{sec:chern number pump}, being a semi-infinite system, the values of each discrete Berry phase Eq.(\ref{eq:Z/3 holonomy 1}) and Eq.(\ref{eq:Z/3 holonomy 2}) do not converge in general. However, in the pump model, the bulk states coincide at $t=0$ and $2\pi$, so we can choose reference states at $t=0$ and $2\pi$ with the same bulk configuration. Then, only the edge contribution remains in the ratio of the holonomy
\begin{eqnarray}\label{eq:Z/3 ratio}
\tilde{r}:=\frac{\tilde{n}_{\rm in.}(\tilde{\gamma})}{\tilde{n}_{\rm fin.}(\tilde{\gamma})}=\exp(\int_{\tilde{\gamma}}(\tilde{A}_{\rm in.}-\tilde{A}_{\rm fin.})-\frac{1}{3}\int_{\tilde{\Sigma}}(d\tilde{A}_{\rm in.}-d\tilde{A}_{\rm fin.}))\frac{\braket{\mathrm{G.S.}_{\rm in.}(\vec{z}=\tilde{\gamma}_{1})|\mathrm{G.S.}_{\rm in.}(\vec{z}=\tilde{\gamma}_{0})}}{\braket{\mathrm{G.S.}_{\rm fin.}(\vec{z}=\tilde{\gamma}_{1})|\mathrm{G.S.}_{\rm fin.}(\vec{z}=\tilde{\gamma}_{0})}}.
\end{eqnarray}
We choose $\ket{\mathrm{Ref}_{\rm in.}}$ and $\ket{\mathrm{Ref}_{\rm fin.}}$ as
\begin{eqnarray}
\ket{\mathrm{Ref}_{\rm in.}}&=&\ket{u_{0}\tilde{\sigma}_{0}u_{0}\tilde{\sigma}_{0}\cdots},\\
\ket{\mathrm{Ref}_{\rm fin.}}&=&\ket{u_{1}\tilde{\sigma}_{0}u_{0}\tilde{\sigma}_{0}\cdots}.
\end{eqnarray}
Then, by using Eqs.(\ref{eq:Z/3 Berry connection at t=0}) and Eq.(\ref{eq:Z/3 Berry connection at t=2pi}), 
\begin{eqnarray}
\tilde{A}_{\rm in.}(\vec{z})-\tilde{A}_{\rm fin.}(\vec{z})&=&\frac{1}{3}\bra{u_{0}}(1+\tilde{\tau}^{z}_{\frac{1}{2}}+\tilde{\tau}^{z\dagger}_{\frac{1}{2}})\tilde{h}_{\frac{1}{2}}(\vec{z})(1+\tilde{\tau}^{z}_{\frac{1}{2}}+\tilde{\tau}^{z\dagger}_{\frac{1}{2}})\ket{u_{0}}\\
&-&\frac{1}{3}\bra{u_{1}}(1+\tilde{\tau}^{z}_{\frac{1}{2}}+\tilde{\tau}^{z\dagger}_{\frac{1}{2}})\tilde{h}_{\frac{1}{2}}(\vec{z})(1+\tilde{\tau}^{z}_{\frac{1}{2}}+\tilde{\tau}^{z\dagger}_{\frac{1}{2}})\ket{u_{1}},\\
&=&0.
\end{eqnarray}
Here, first term is a contribution from $\tilde{A}_{\rm in.}(\vec{z})$ and second term is a contribution from $\tilde{A}_{\rm fin.}(\vec{z})$. On the other hand, 
\begin{eqnarray}
\frac{\braket{\mathrm{G.S.}_{\rm in.}(\vec{z}=\tilde{\gamma}_{1})|\mathrm{G.S.}_{\rm in.}(\vec{z}=\tilde{\gamma}_{0})}}{\braket{\mathrm{G.S.}_{\rm fin.}(\vec{z}=\tilde{\gamma}_{1})|\mathrm{G.S.}_{\rm fin.}(\vec{z}=\tilde{\gamma}_{0})}}
&=&\frac{\braket{\mathrm{G.S.}_{\rm in.}(\vec{z}=\omega\tilde{\gamma}_{0})|\mathrm{G.S.}_{\rm in.}(\vec{z}=\tilde{\gamma}_{0})}}{\braket{\mathrm{G.S.}_{\rm fin.}(\vec{z}=\omega\tilde{\gamma}_{0})|\mathrm{G.S.}_{\rm fin.}(\vec{z}=\tilde{\gamma}_{0})}},\\
&=&\frac{\bra{\mathrm{G.S.}_{\rm in.}(\vec{z}=\tilde{\gamma}_{0})}\tilde{\tau}^{x\dagger}_\frac{1}{2}(\vec{z}=\gamma_0)\ket{\mathrm{G.S.}_{\rm in.}(\vec{z}=\tilde{\gamma}_{0})}}{\bra{\mathrm{G.S.}_{\rm fin.}(\vec{z}=\tilde{\gamma}_{0})}\tilde{\tau}^{x\dagger}_{\frac{1}{2}}(\vec{z}=\gamma_0)\ket{\mathrm{G.S.}_{\rm fin.}(\vec{z}=\tilde{\gamma}_{0})}},\\
&=&\omega.
\end{eqnarray}
Therefore, the ratio Eq.(\ref{eq:Z/3 ratio}) is 
\begin{eqnarray}
\tilde{r}=\omega.
\end{eqnarray}

In this process, what is the $3$-parameter family of (0+1)-dimensional invertible states that are pumped into the boundary? To clarify this, consider an effective model of the boundary. For the initial Hamiltonian Eq.(\ref{eq:model_with_boundary_0}), the boundary model is given by
\begin{eqnarray}\label{eq:Z/3 boundary_model_0}
 H^{\rm bdy.}_{\rm in.}(\Vec{z}):=-\tilde{\tau}^{x}_{\frac{1}{2}}(\vec{z})\tilde{\sigma}^{z\dagger}_{1}-\tilde{\tau}^{z\dagger}_{\frac{1}{2}}(\Vec{z})\tilde{\sigma}^{x}_{1}\tilde{\tau}^{z}_{\frac{3}{2}}(\Vec{z})-\tilde{\sigma}^{z}_{1}\tilde{\tau}^{x}_{\frac{3}{2}}(\Vec{z})+h.c.,
\end{eqnarray}
and for the final Hamiltonian Eq.(\ref{eq:model_with_boundary_2pi}), the boundary model is given by
\begin{eqnarray}\label{eq:Z/3 boundary_model_1}
 H^{\rm bdy.}_{\rm fin.}(\Vec{z}):=-\omega^2\tilde{\tau}^{x}_{\frac{1}{2}}(\vec{z})\tilde{\sigma}^{z\dagger}_{1}-\tilde{\tau}^{z\dagger}_{\frac{1}{2}}(\Vec{z})\tilde{\sigma}^{x}_{1}\tilde{\tau}^{z}_{\frac{3}{2}}(\Vec{z})-\tilde{\sigma}^{z}_{1}\tilde{\tau}^{x}_{\frac{3}{2}}(\Vec{z})+h.c.,
\end{eqnarray}
Then, it can be seen that the ratio of the discrete Berry phases we calculated above is the same as that of these quantum mechanical systems. Let's check this point. The ground state of the initial Hamiltonian $H^{\rm bdy.}_{\rm in.}(\vec{z})$ is given by
\begin{eqnarray}
\ket{\mathrm{G.S.}^{\rm bdy.}_{\rm in.}(\vec{z})}&=&\frac{1+\tilde{\tau}^{x}_{\frac{1}{2}}(\vec{z})\tilde{\sigma}^{z\dagger}_{1}+\tilde{\tau}^{x\dagger}_{\frac{1}{2}}(\vec{z})\tilde{\sigma}^{z}_{1}}{\sqrt{3}}\frac{1+\tilde{\sigma}^{z}_{1}\tilde{\tau}^{x}_{\frac{3}{2}}(\Vec{z})+\tilde{\sigma}^{z\dagger}_{1}\tilde{\tau}^{x\dagger}_{\frac{3}{2}}(\Vec{z})}{\sqrt{3}}\ket{\tilde{\tau}_{0}(\vec{z})v_{0}\tilde{\tau}_{0}(\vec{z})},\\
&=&\tilde{V}_{\tau}(\vec{z})_{\frac{1}{2}}\tilde{V}_{\tau}(\vec{z})_{\frac{3}{2}}
\frac{1+\tilde{\tau}^{x}_{\frac{1}{2}}\tilde{\sigma}^{z\dagger}_{1}+\tilde{\tau}^{x\dagger}_{\frac{1}{2}}\tilde{\sigma}^{z}_{1}}{\sqrt{3}}\frac{1+\tilde{\sigma}^{z}_{1}\tilde{\tau}^{x}_{\frac{3}{2}}+\tilde{\sigma}^{z\dagger}_{1}\tilde{\tau}^{x\dagger}_{\frac{3}{2}}}{\sqrt{3}}\ket{\tilde{\tau}_{0}v_{0}\tilde{\tau}_{0}},
\label{eq:l31_bdy_gs_gauge}
\end{eqnarray}
where $\ket{v_{0}}=(1,1,1)^{\rm T}/\sqrt{3}$ is a eigenstate of $\tilde{\sigma}^{x}$ with eigenvalue $1$. 
Introducing the spherical coordinates\footnote{Although the Hopf coordinates $z_1=e^{i\alpha}\cos(\beta),z_2=e^{i\alpha'}\sin(\beta),\alpha,\alpha'\in\left[0,2\pi\right),\beta\in\left[0,\pi/2\right)$ is useful to see for the $\zmod{3}$ action in lens space $\mathrm{L}(3,1)$, we use spherical coordinates since only a path on the equator is used here.}
\begin{eqnarray}
z_1&=&\sin(\chi)\sin(\theta)e^{i\phi},\\
z_2&=&\cos(\chi)+i\sin(\chi)\cos(\theta), 
\end{eqnarray}
a loop $\gamma$ generating the first homotopy group $\pi_1(\mathrm{L}(3,1))$ is given by 
\begin{align}
    \gamma= \{(\chi=\frac{\pi}{2},\theta=\frac{\pi}{2},\phi) | \phi \in [0,\frac{2\pi}{3}]\}. 
\end{align}
With the gauge (\ref{eq:l31_bdy_gs_gauge}), it is straightforward to show that the Berry connection is trivial:
\begin{eqnarray}
A^{\rm bdy.}_{\rm in.}(\vec{z})=0. 
\end{eqnarray}
In addition, since $\tilde{V}_{\tau}(\omega\vec{z})^{\dagger}\tilde{V}_{\tau}(\vec{z})=\tilde{\tau}^{x\dagger}$, 
\begin{align}
\braket{\mathrm{G.S.}^{\rm bdy.}_{\rm in.}(\omega\vec{z})|\mathrm{G.S.}^{\rm bdy.}_{\rm in.}(\vec{z})}\nonumber
&=\bra{\tilde{\tau}_{0}v_{0}\tilde{\tau}_{0}}\frac{1+\tilde{\tau}^{x}_{\frac{1}{2}}\tilde{\sigma}^{z\dagger}_{1}+\tilde{\tau}^{x\dagger}_{\frac{1}{2}}\tilde{\sigma}^{z}_{1}}{\sqrt{3}}\frac{1+\tilde{\sigma}^{z}_{1}\tilde{\tau}^{x}_{\frac{3}{2}}+\tilde{\sigma}^{z\dagger}_{1}\tilde{\tau}^{x\dagger}_{\frac{3}{2}}}{\sqrt{3}}\nonumber\\
&\times \tilde{\tau}^{x\dagger}_{\frac{1}{2}}\tilde{\tau}^{x\dagger}_{\frac{3}{2}}
\frac{1+\tilde{\tau}^{x}_{\frac{1}{2}}\tilde{\sigma}^{z\dagger}_{1}+\tilde{\tau}^{x\dagger}_{\frac{1}{2}}\tilde{\sigma}^{z}_{1}}{\sqrt{3}}\frac{1+\tilde{\sigma}^{z}_{1}\tilde{\tau}^{x}_{\frac{3}{2}}+\tilde{\sigma}^{z\dagger}_{1}\tilde{\tau}^{x\dagger}_{\frac{3}{2}}}{\sqrt{3}}\ket{\tilde{\tau}_{0}v_{0}\tilde{\tau}_{0}}=1.
\end{align}
Thus, the discrete Berry phase of the initial Hamiltonian $H^{\rm bdy.}_{\rm in.}(\vec{z})$ is $1$:
\begin{eqnarray}
\tilde{n}^{\rm bdy.}_{\rm in.}(\tilde{\gamma}):=\exp(\int_{\tilde{\gamma}}\tilde{A}^{\rm bdy.}_{\rm in.}-\frac{1}{3}\int_{\tilde{\Sigma}}d\tilde{A}^{\rm bdy.}_{\rm in.})\braket{\mathrm{G.S.}^{\rm bdy.}_{\rm in.}(\vec{z}=\tilde{\gamma}_{0})|\mathrm{G.S.}^{\rm bdy.}_{\rm in.}(\vec{z}=\tilde{\gamma}_{1})}=1.
\end{eqnarray}

Similarly, the ground state of the final Hamiltonian $H^{\rm bdy.}_{\rm fin.}(\vec{z})$ is given by
\begin{eqnarray}
\ket{\mathrm{G.S.}^{\rm bdy.}_{\rm in.}(\vec{z})}&=&\frac{1+\omega^2\tilde{\tau}^{x}_{\frac{1}{2}}(\vec{z})\tilde{\sigma}^{z\dagger}_{1}+\omega\tilde{\tau}^{x\dagger}_{\frac{1}{2}}(\vec{z})\tilde{\sigma}^{z}_{1}}{\sqrt{3}}\frac{1+\tilde{\sigma}^{z}_{1}\tilde{\tau}^{x}_{\frac{3}{2}}(\Vec{z})+\tilde{\sigma}^{z\dagger}_{1}\tilde{\tau}^{x\dagger}_{\frac{3}{2}}(\Vec{z})}{\sqrt{3}}\ket{\tilde{\tau}_{0}(\vec{z})v_{0}\tilde{\tau}_{0}(\vec{z})},\\
&=&\tilde{V}_{\tau}(\vec{z})_{\frac{1}{2}}\tilde{V}_{\tau}(\vec{z})_{\frac{3}{2}}
\frac{1+\omega^2\tilde{\tau}^{x}_{\frac{1}{2}}\tilde{\sigma}^{z\dagger}_{1}+\omega\tilde{\tau}^{x\dagger}_{\frac{1}{2}}\tilde{\sigma}^{z}_{1}}{\sqrt{3}}\frac{1+\tilde{\sigma}^{z}_{1}\tilde{\tau}^{x}_{\frac{3}{2}}+\tilde{\sigma}^{z\dagger}_{1}\tilde{\tau}^{x\dagger}_{\frac{3}{2}}}{\sqrt{3}}\ket{\tilde{\tau}_{0}v_{0}\tilde{\tau}_{0}},
\end{eqnarray}
and we can easily check that the Berry connection of $A^{\rm bdy.}_{\rm fin.}(\vec{z})$ is also trivial:
\begin{eqnarray}
A^{\rm bdy.}_{\rm fin.}(\vec{z}) = 0.
\end{eqnarray}
On the other hand, 
\begin{align}
\braket{\mathrm{G.S.}^{\rm bdy.}_{\rm fin.}(\omega\vec{z})|\mathrm{G.S.}^{\rm bdy.}_{\rm fin.}(\vec{z})}\nonumber
&=\bra{\tilde{\tau}_{0}v_{0}\tilde{\tau}_{0}}\frac{1+\omega^2 \tilde{\tau}^{x}_{\frac{1}{2}}\tilde{\sigma}^{z\dagger}_{1}+\omega\tilde{\tau}^{x\dagger}_{\frac{1}{2}}\tilde{\sigma}^{z}_{1}}{\sqrt{3}}\frac{1+\tilde{\sigma}^{z}_{1}\tilde{\tau}^{x}_{\frac{3}{2}}+\tilde{\sigma}^{z\dagger}_{1}\tilde{\tau}^{x\dagger}_{\frac{3}{2}}}{\sqrt{3}}\nonumber\\
&\times \tilde{\tau}^{x\dagger}_{\frac{1}{2}}\tilde{\tau}^{x\dagger}_{\frac{3}{2}}
\frac{1+\omega^2 \tilde{\tau}^{x}_{\frac{1}{2}}\tilde{\sigma}^{z\dagger}_{1}+\omega \tilde{\tau}^{x\dagger}_{\frac{1}{2}}\tilde{\sigma}^{z}_{1}}{\sqrt{3}}\frac{1+\tilde{\sigma}^{z}_{1}\tilde{\tau}^{x}_{\frac{3}{2}}+\tilde{\sigma}^{z\dagger}_{1}\tilde{\tau}^{x\dagger}_{\frac{3}{2}}}{\sqrt{3}}\ket{\tilde{\tau}_{0}v_{0}\tilde{\tau}_{0}}=\omega^2.
\end{align}
Thus, the discrete Berry phase of the initial Hamiltonian $H^{\rm bdy.}_{\rm in.}(\vec{z})$ is $\omega^2$.
\begin{eqnarray}
\tilde{n}^{\rm bdy.}_{\rm fin.}(\tilde{\gamma}):=\exp(\int_{\tilde{\gamma}}\tilde{A}^{\rm bdy.}_{\rm fin.}-\frac{1}{3}\int_{\tilde{\Sigma}}d\tilde{A}^{\rm bdy.}_{\rm fin.})\braket{\mathrm{G.S.}^{\rm bdy.}_{\rm fin.}(\vec{z}=\tilde{\gamma}_{0})|\mathrm{G.S.}^{\rm bdy.}_{\rm fin.}(\vec{z}=\tilde{\gamma}_{1})}=\omega^2.
\end{eqnarray}

Therefore, the ratio of the discrete Berry phases of these quantum mechanical systems is 
\begin{eqnarray}
\frac{\tilde{n}^{\rm bdy.}_{\rm in.}(\tilde{\gamma})}{\tilde{n}^{\rm bdy.}_{\rm fin.}(\tilde{\gamma})}=\omega.
\end{eqnarray}

\subsubsection{Physical Interpretation II : Boundary Condition Obstacle}\label{sec:Z/3 obstacle}

In Sec.\ref{sec:Z/3 chernnumber pump}, we considered the system with boundary and discussed the flow of the discrete Berry phase, when comparing $t=0$ and $t=2\pi$. This breaking of the $2\pi$-periodicity is due to the fact that the boundary terms were not $2\pi$-periodic. In fact, the boundary term $-\tilde{\tau}^{x}_{1/2}(\vec{n})\tilde{\sigma}^{z}_{1}(t)$ which was added to the Hamiltonian is not $2\pi$-periodic. We can consider a boundary term like $\tilde{\tau}^{z}(\vec{n})$ that preserves $2\pi$-periodicity, but this time the boundary term is not global on $\mathrm{L}(3,1)$. Similarly to the discussion in Sec.\ref{sec:obstacle}, we can show that there is no boundary term that is parameterized over the whole of $\mathrm{L}(3,1)\times S^1$ and makes the system unique gap at all points $\mathrm{L}(3,1)\times S^1$.

Let us suppose that there exists such a term, which we denote by $\tilde{x}(\vec{n},t)$. Then, the flow of the discrete Berry phase is trivial under this boundary condition. This implies that by stacking two semi-infinite chains with boundary condition $-\tilde{\tau}^{x}_{1/2}(\vec{n})\tilde{\sigma}^{z}_{1}(t)$ and $\tilde{x}(\vec{n},t)$, we obtain a nontrivial family of $(0+1)$-dimensional systems parametrized by  $\mathrm{L}(3,1)\times \left[0,2\pi\right]$ whose ratio of the discrete Berry phase at $t=0$ and $t=2\pi$ is $\omega$. If there existed such a family, it would be inconsistent with the quantization of the discrete Berry phase. Therefore, there are no such boundary terms.

\section{Quick Review of the Smooth Deligne Cohomology}\label{sec:smooth Deligne}

In this section, first, we review the smooth Deligne cohomology\cite{Brylinski} and its integration theory. In Sec.\ref{sec:def of smooth}, we introduce the higher analogue of the discrete Berry phase based on the integration theory.  The smooth Deligne cohomology is isomorphic to the differential cohomology group\cite{Brylinski,BB}. In fact, the integration map for the smooth Deligne cohomology
gives such explicit isomorphism\cite{Terashima00,CJM04}. This isomorphism is an analogy of the de Rham isomorphism. In Sec.\ref{sec:ordinarypump}, as an application example, we reformulate the invariant of a fermion parity pump proposed in \cite{OSS22} as an integration of a smooth Deligne cohomology class.

\subsection{Definition and Integration}\label{sec:def of smooth}

Let $X$ be a smooth manifold. The smooth Deligne complex of $X$ is the complex of sheaves  
$$
\D (p): \underline{\C}^* \os{d \log}{\to} \ua^{1} \os{d}{\to} \dots \os{d}{\to} \ua^{p-1},
$$
where $\underline{\C}^*$ is the sheaf of $\C^*$-valued smooth functions on $X$ and 
$\ua^k$ is the sheaf of smooth $k$-forms on $X$\footnote{$\C^*:=\C\backslash\{0\}$.}.  The smooth Deligne cohomology is the 
hypercohomology $H^n (X; \D (p))$ of the smooth Deligne complex $\D (p)$. Fixing a good open covering $\oc=\{ U_\a \}_{\a \in I}$ of $X$, a smooth Deligne cohomology class in $H^{p} (X; \D (p))$ is represented by a cocycle $c=(w_{\a_0 \dots \a_{p-1}},\th^1_{\a_0 \dots \a_{p-2}}, \dots , 
\th^{p-1}_{\a_0})$ where $w_{\a_0 \dots \a_{p-1}}$ is a   smooth function on the intersection $U_{\a_0} \cap \dots \cap  U_{\a_{p-1}}$ of open sets in values with nonzero complex numbers, and $\th^k_{\a_0 \dots \a_{p-k-1}}$ is a  smooth $k$-form on the intersection $U_{\a_0} \cap \dots \cap  U_{\a_{p-k-1}}$. The cocycle condition on $c$ is equivalent to the condition $\delta(\theta^{k})_{\alpha_{0},...,\alpha_{p-k}}=(-1)^{p-k}d\theta^{k-1}_{\alpha_{0},...,\alpha_{p-k}}$ for $k=2,...,p-1$,  $\delta(\theta^{1})_{\alpha_{0},...,\alpha_{p-1}}=(-1)^{p-1}d\log(w_{\alpha_{0},...,\alpha_{p-1}})$ and $\delta(w)_{\alpha_{0},...,\alpha_{p}}=1$. Here $\delta$ is the \v{C}eck derivative. This condition on $c$ shows that the differential forms $d \th^{p-1}_\a$ and $d \th^{p-1}_{\a'}$ are equal on the intersection $U_\a \cap U_{\a'}$. So, we have a global closed $p$-form $\eta$ given by $\eta |_{U_\a}=d \th^{p-1}_\a$ which is called {\it higher curvature form} of $c$. A smooth Deligne cocycle $c$ is {\it flat} if the higher curvature form of $c$ is zero. An important example of a flat cocycle which is used in this paper is $c=(w_{\a_0 \dots \a_{p-1}}, 0, \dots , 0)$ such that $w_{\a_0 \dots \a_{p-1}}$ is a constant function on connected components of $U_{\a_0 \dots \a_{p-1}}$ with $(\delta w)_{\a_0 \dots  \a_{p}}=1$.

A main tool in this paper is an integration theory 
for the smooth Deligne cohomology developed in\cite{Gawedzki88,GT00,GT01,CJM04,GT09}. See
also related papers \cite{Alvarez85, Kapustin00,BM96,Gajer97,BKS10,DL05,HLZ03,HS05} and books \cite{BB,Brylinski}. For a smooth Deligne cohomology class $c$ in $H^p (X; \D (p))$ 
and a $(p-1)$-dimensional closed oriented submanifold $Y$ in $X$, we construct a paring $\hol_Y(c)$ with values in $\C^*$, called 
{\it higher holonomy} of $c$ along $Y$, as follows: 

First, fixing a good open covering $\oc=\{ U_\a \}_{\a \in I}$, 
we choose a representative 
cocycle 
\begin{eqnarray}
(w_{\a_0 \dots \a_{p-1}}, \th^1_{\a_0 \dots \a_{p-2}}, \dots , 
\th^{p-1}_{\a_0})    
\end{eqnarray}
of $c$. Second, we choose a triangulation $K$ 
of $Y$ which is sufficiently fine such that there exists 
a map $\phi :K \to I$ satisfying $\si \subset U_{\phi_\si}$ for each $\si \in K$.  
 Such map is called index map. Third, we fix an index map 
$\phi :K \to I$. 

Then, we define the higher holonomy as  
\begin{eqnarray}
\hol_Y(c):=\exp \left( \sum_{i=0}^{p-2} \sum_{\si \in F(i)} \int_{\si^{p-i-1}} 
\th^{p-i-1}_{\phi_{\si^{p-1}} \phi_{\si^{p-2}} \dots \phi_{\si^{p-i-1}}} \right) \times \prod_{\si \in F(p-1)} w_{\phi_{\si^{p-1}} \phi_{\si^{p-2}} \dots \phi_{\si^{0}}} (\si^0),
\end{eqnarray}
where $F(i)$ is the set of flags of simplices
\begin{eqnarray}
F(i):=\{ \si =(\si^{p-i-1},\dots ,\si^{p-1})|\ \dim \si^k=k,\ \si^{p-i-1} \subset \dots \subset \si^{p-2} \subset \si^{p-1} \}.
\end{eqnarray}
This definition is independent of all choices. Remark that if we have a representative constant cocycle $(w_{\a_0 \dots \a_{p-1}}, 0, \dots , 
0)$, then the higher holonomy is simplified as follows:
\begin{eqnarray}
\hol_Y(c)=\prod_{\si \in F(p)} w_{\phi_{\si^{p-1}} \phi_{\si^{p-2}} \dots \phi_{\si^{0}}} (\si^0).
\end{eqnarray}

In general, $\mathrm{Hol}_{Y}(c)$ takes value in $\mathrm{U}(1)$, and not quantized. However, if there is $k\in\mathbb{N}$ such that $k[Y]=0\in\mathrm{H}_{n-1}(X;\mathbb{Z})$, we can extract the information of the torsion part of $\cohoZ{n}{X}$, by considering the following quantity:
\begin{eqnarray}
n_{\rm top.}(Y):=\mathrm{Hol}_{Y}(c)\exp\left(\frac{1}{k}\int_{\Sigma}\eta\right)\in\zmod{k},
\end{eqnarray}
where $\Sigma$ is a bounding manifold of $k\cdot Y$, i.e., $\partial \Sigma=k\cdot Y$. This can be considered as a generalization of an expression given in \cite{Freed86} on Chern-Simons forms which live on total spaces of principal bundles. This value is only depend on ${w_{\alpha_{0},...,\alpha_{p-1}}}$ and $\left[Y\right]\in\mathrm{H}_{n-1}(X;\mathbb{Z})$, and not depend on the choice of connections $\th^1_{\a_0 \dots \a_{p-2}}, \dots , 
\th^{p-1}_{\a_0}$. We call $n_{\rm top.}(Y)$ as the discrete higher Berry phase of $c$ along $Y$.

In the case of $p=2$, $n_{\rm top.}(Y)$ is the discrete Berry phase in the usual sense. In Sec.\ref{sec:ordinarypump}, we will verify that a fermion parity pump invariant proposed in \cite{OSS22} can be written as the discrete Berry phase in the usual sense($p=2$), and in Sec.\ref{sec:defofpumpinv}, a higher pump invariant can be written as the discrete higher Berry phase. See Appendix.\ref{sec:complex line bundle} for a clarification of terminology and basic facts about complex line bundles.

\subsection{Example: Ordinary Pump Invariant}\label{sec:ordinarypump}

Let us reformulate the invariants of the fermion parity pump proposed in \cite{OSS22} as an integration of a smooth Deligne cohomology class. In the case of usual pumps, we consider a family of invertible states parametrized by $S^1$. Therefore, take an open covering $\{U_{\alpha}\}_{\alpha\in I}$ of $S^{1}$ and consider a family of $\zmod{2}$-graded injective $2n\times 2n$ MPS matrices $\{A^{i}_{\alpha},u_{\alpha}\}$\cite{BWHV17,PTSC21} on each patch $U_{\alpha}$. For simplicity, consider the case where $\{A^{i}_{\alpha},u_{\alpha}\}$ has the Wall invariant $(+)$. Then, by using the fundamental theorem for $\zmod{2}$-graded injective MPS\cite{OSS22}, there exists unique $U(1)$ phases  $e^{i\phi_{\alpha\beta}},e^{i\varphi_{\alpha\beta}}\in \mathrm{U}(1)$ and unique projective unitary matrix $V_{\alpha\beta}\in\mathrm{PU}(n)$ on each intersections $U_{\alpha\beta}:=U_{\alpha}\cap U_{\beta}$ so that
\begin{eqnarray}
A^{i}_{\alpha}&=&e^{i\varphi_{\alpha\beta}}V_{\alpha\beta}A^{i}_{\beta}V_{\alpha\beta}^{\dagger},\\
u_{\alpha}&=&e^{i\phi_{\alpha\beta}}V_{\alpha\beta}u_{\beta}V_{\alpha\beta}^{\dagger}.\label{eq:wallmat0}
\end{eqnarray}
We claim that the quantity $(w_{\alpha\beta},\theta_{\alpha})=(e^{i\phi_{\alpha\beta}},\frac{1}{2}d\log \tr{u_{\alpha}^{2}})$ is a $2$-cocycle of the smooth Deligne cohomology, and the pump invariant is given as its (ordinary) discrete Berry phase. 

Let's check cocycle conditions $(\delta w)_{\alpha\beta\gamma}=1$ and $(\delta \theta)_{\alpha\beta}=d\log w_{\alpha\beta}$. On $U_{\alpha}\cap U_{\beta}\cap U_{\gamma}$, 
\begin{eqnarray}
A^{i}_{\alpha}&=&e^{i\varphi_{\alpha\beta}}V_{\alpha\beta}A^{i}_{\beta}V_{\alpha\beta}^{\dagger}=e^{i\varphi_{\alpha\beta}}e^{i\varphi_{\beta\gamma}}V_{\alpha\beta}V_{\beta\gamma}A^{i}_{\gamma}V_{\beta\gamma}^{\dagger}V_{\alpha\beta}^{\dagger},
\end{eqnarray}
and
\begin{eqnarray}
A^{i}_{\alpha}&=&e^{i\varphi_{\alpha\gamma}}V_{\alpha\gamma}A^{i}_{\gamma}V_{\alpha\gamma}^{\dagger}.
\end{eqnarray}
Since $\{A^{i}_{\alpha}\}$ are $\zmod{2}$-graded injective with the Wall invariant $(+)$, there is some $\mathrm{U}(1)$ phase $e^{i\omega_{\alpha\beta\gamma}}$ so that 
\begin{eqnarray}\label{eq:cocycle}
V_{\alpha\beta}V_{\beta\gamma}=e^{i\omega_{\alpha\beta\gamma}}V_{\alpha\gamma}.
\end{eqnarray} 
Similarly, on $U_{\alpha}\cap U_{\beta}\cap U_{\gamma}$, 
\begin{eqnarray}\label{eq:wallmat1}
u_{\alpha}=e^{i\phi_{\alpha\beta}}e^{i\phi_{\beta\gamma}}V_{\alpha\beta}V_{\beta\gamma}u_{\gamma}V_{\beta\gamma}^{\dagger}V_{\alpha\beta}^{\dagger},
\end{eqnarray}
and 
\begin{eqnarray}\label{eq:wallmat2}
u_{\alpha}=e^{i\phi_{\alpha\gamma}}V_{\alpha\gamma}u_{\gamma}V_{\alpha\gamma}^{\dagger}.
\end{eqnarray}
Comparing Eq.(\ref{eq:wallmat1}) and Eq.(\ref{eq:wallmat2}) and using Eq.(\ref{eq:cocycle}), we obtain 
\begin{eqnarray}
e^{i\phi_{\alpha\beta}}e^{i\phi_{\beta\gamma}}=e^{i\phi_{\alpha\gamma}}\Leftrightarrow(\delta w)_{\alpha\beta\gamma}=1.
\end{eqnarray}
Next, taking the square of both sides of Eq.(\ref{eq:wallmat0}), we obtain
\begin{eqnarray}
u_{\alpha}^{2}&=&e^{2i\phi_{\alpha\beta}}V_{\alpha\beta}u_{\beta}^{2}V_{\alpha\beta}^{\dagger},
\end{eqnarray}
and taking $\log\mathrm{tr}$ of both sides of this equation, 
\begin{eqnarray}
\log\tr{u_{\alpha}^{2}}&=&\log e^{2i\phi_{\alpha\beta}}+ \log\tr{u_{\beta}^{2}} \mathrm{\;\;(mod.} 2\pi i\Z),\\
\Leftrightarrow \log e^{i\phi_{\alpha\beta}} &=& \frac{1}{2}\log\tr{u_{\alpha}^{2}}- \frac{1}{2}\log\tr{u_{\beta}^{2}} \mathrm{\;\;(mod.} \pi i\Z).
\end{eqnarray}
Therefore,
\begin{eqnarray}
d\log e^{i\phi_{\alpha\beta}} = \frac{1}{2}d\log\tr{u_{\alpha}^{2}}- \frac{1}{2}d\log\tr{u_{\beta}^{2}}\Leftrightarrow(\delta \theta)_{\alpha\beta}=d\log w_{\alpha\beta}.
\end{eqnarray}
Thus, $(e^{i\phi_{\alpha\beta}},\frac{1}{2}d\log \tr{u_{\alpha}^{2}})$ is a cocycle of degree $2$ in the sense of the smooth Deligne cohomology.

\begin{figure}[H]
 \begin{center}
  \includegraphics[width=50mm]{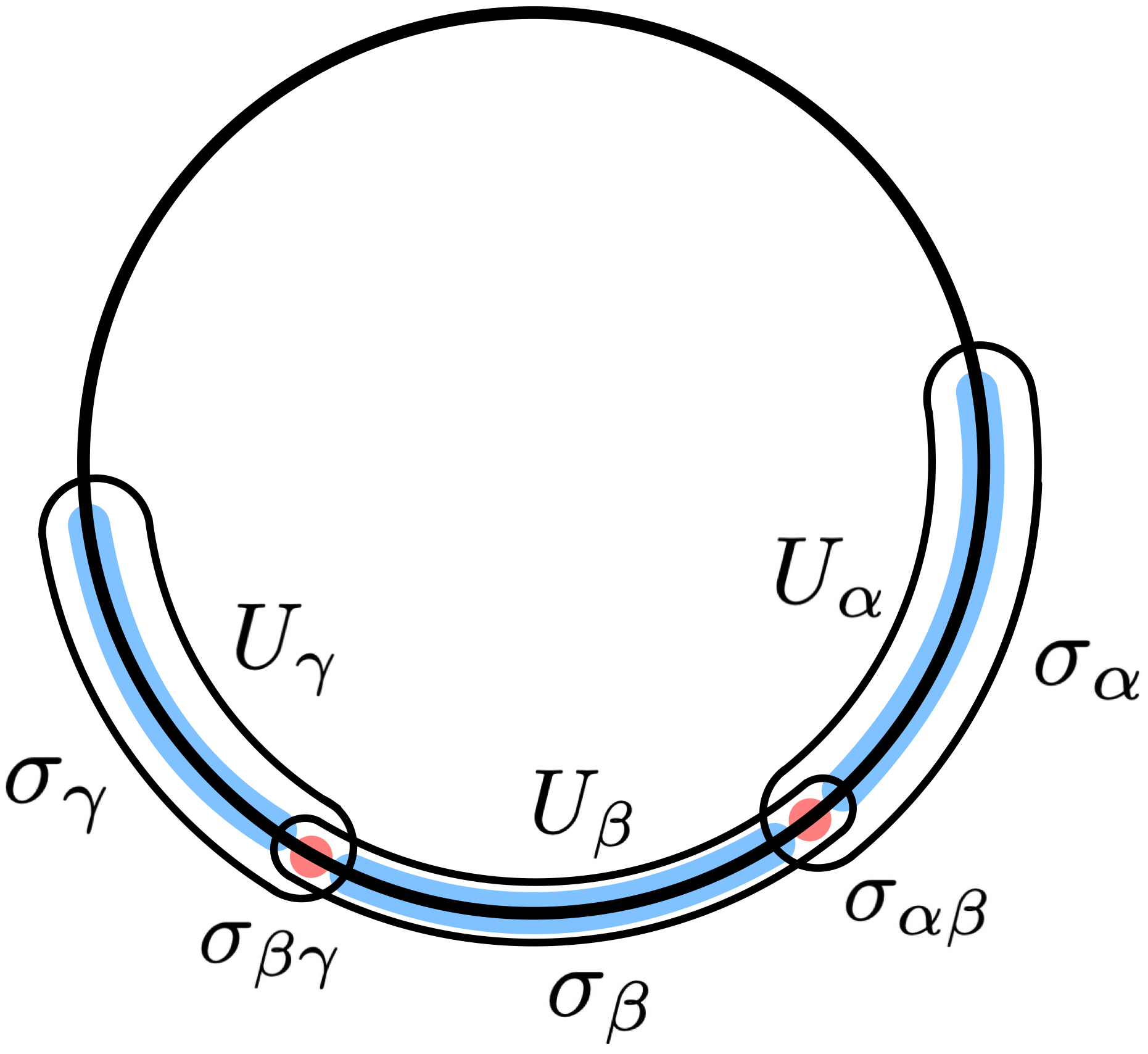}
 \end{center}
  \caption[]{A part of a triangulation of $S^{1}$. The elongated circles represent open coverings, the red dots represent $0$-simplices, and the blue edges represent $1$-simplices.
}
\label{fig:triangulationofS1}
\end{figure}

To write the discrete Berry phase explicitly, we take a triangulation as shown in the Fig.\ref{fig:triangulationofS1} and take an index map defined by $\phi(\sigma_{\alpha})=\alpha$ for all $\alpha\in I$ and  $\phi(\sigma_{\alpha\beta})=\beta$ for all $\alpha,\beta\in I$ such that $U_{\alpha}\cap U_{\beta}\neq\emptyset$. In this case, the discrete Berry phase of $(w_{\alpha\beta},\theta_{\alpha})=(e^{i\phi_{\alpha\beta}},\frac{1}{2}d\log \tr{u_{\alpha}^{2}})$ is given by 
\begin{eqnarray}\label{eq:holos1}
n_{\rm top.}(S^1)=\exp{\left(\frac{1}{2}\sum_{\alpha}\int_{\sigma_{\alpha}}d\log\tr{u^{2}_{\alpha}}\right)}\times\prod_{\sigma_{\alpha\beta}}e^{i\phi_{\alpha\beta}},
\end{eqnarray}
and it takes value in $\pi i \Z/2\pi i \Z\simeq\zmod{2}$. Note that the correction by the $2$-form curvature has disappeared, since the cocycle $(w_{\alpha\beta},\theta_{\alpha})$ is always flat. We will check that the analytic and algebraic invariants proposed in \cite{OSS22} are realized as a special case of Eq.(\ref{eq:holos1}) by taking suitable gauges. First, we take the gauge with $w_{\alpha\beta}=1$ for any $\alpha$ and $\beta$. From the cocycle condition, we obtain 
\begin{eqnarray}
\frac{1}{2}d\log\tr{u_{\alpha}^{2}}- \frac{1}{2}d\log\tr{u_{\beta}^{2}}=0\Leftrightarrow(\delta \theta)_{\alpha\beta}=0,
\end{eqnarray}
and this implies that  $\frac{1}{2}\log\tr{u_{\alpha}^{2}}$ is global $1$-form over $S^1$. 
Therefore, the discrete Berry phase Eq.(\ref{eq:holos1}) is recast into
\begin{eqnarray}\label{eq:holoanalytic}
n_{\rm top.}(S^{1})=\exp{\left(\frac{1}{2}\int_{S^{1}}d\log\tr{u^{2}_{\alpha}}\right)}\in\zmod{2},
\end{eqnarray}
and this is nothing but the analytic invariant of fermion parity pump. Next, we take the gauge with $d\log\tr{u_{\alpha}^{2}}=0$. From the cocycle condition, we obtain $d\log e^{i\phi_{\alpha\beta}} = 0$, and this implies that $\log e^{i\phi_{\alpha\beta}}=c_{\alpha\beta}+\pi i n_{\alpha\beta}$ for some $c_{\alpha\beta}\in\R$ and $n_{\alpha\beta}\in\{0,1\}$. Remark that $\sum_{\sigma_{\alpha\beta}}c_{\alpha\beta}=0 \mathrm{\;\;(mod.} \pi i\Z)$. In fact, $2c_{\alpha\beta}=\log\tr{u_{\alpha}^{2}}- \log\tr{u_{\beta}^{2}}\mathrm{\;\;(mod.}2\pi i\Z)$ at $\sigma_{\alpha\beta}$, and  $\int_{\sigma_{\beta}}d\log\tr{u^{2}_{\beta}}=\log\tr{u_{\beta}^{2}}\left|\right._{\sigma_{\alpha\beta}}-\log\tr{u_{\beta}^{2}}\left|\right._{\sigma_{\beta\gamma}}=0$. Thus
\begin{eqnarray}
\sum_{\sigma_{\alpha\beta}}c_{\alpha\beta}&=&\frac{1}{2}\sum_{\sigma_{\alpha\beta}}(\log\tr{u_{\alpha}^{2}}- \log\tr{u_{\beta}^{2}})\left|\right._{\sigma_{\alpha\beta}}\mathrm{\;\;(mod.}\pi i\Z),\\
&=&\sum_{\sigma_{\alpha\beta}}(-\frac{1}{2}\log\tr{u^{2}_{\beta}}\left|\right._{\sigma_{\alpha\beta}}+\frac{1}{2}\log\tr{u^{2}_{\beta}}\left|\right._{\sigma_{\beta\gamma}})\mathrm{\;\;(mod.}\pi i\Z),\\
&=&0\mathrm{\;\;(mod.}\pi i\Z).
\end{eqnarray}
Therefore, the holonomy Eq.(\ref{eq:holos1}) is recast into
\begin{eqnarray}\label{eq:holoalgebraic}
n_{\rm top.}(S^{1})=\prod_{\sigma_{\alpha\beta}}w_{\alpha\beta}\in\zmod{2},
\end{eqnarray}
and this is nothing but the algebraic invariant of fermion parity pump.

\section{Matrix Product State Representation and Higher Pump Invariant}\label{sec:mps}

In this section, we define the higher pump invariant as an integration of the smooth Deligne cohomology. To this end, we utilize an injective MPS bundle over the parameter space. By using this bundle, we can construct a smooth Deligne cocycle, and define the higher pump invariant as an integration of the cocycle. In Sec.\ref{sec:defofpumpinv}, we explain the construction of the higher pump invariant. In Sec.\ref{sec:computation of n Z/2} and Sec.\ref{sec:computation of n Z/3}, we compute the higher pump invariant for the models introduced in Sec.\ref{sec:model} and Sec.\ref{sec:Z/3 cluster model} respectively. 

\subsection{Definition of the Higher Pump Invariant}\label{sec:defofpumpinv}

Fix an $n$-dimensional manifold $X$ as a parameter space and its good open covering $\{U_{\alpha}\}_{\alpha\in I}$. Let $\{A^{i}_{\alpha}(x)\}_{i}$ be an $n\times n$ injective matrices \cite{OSS22} on $U_{\alpha}$ and assume that $\{A^{i}_{\alpha}(x)\}_{i}$ is in the right canonical form \cite{P-GVWC07}, i.e.,
\begin{eqnarray}
    \sum_{i}A^{i}_{\alpha}(x)A^{i\dagger}_{\alpha}(x)=1_n,
\end{eqnarray}
for any $\alpha$ and $x$.We call this an injective MPS bundle over $X$. On $U_{\alpha\beta}:=U_{\alpha}\cap U_{\beta}$, $\{A^{i}_{\alpha}(x)\}_{i}$ and $\{A^{i}_{\beta}(x)\}_{i}$ give the same MPS. Thus, by using the fundamental theorem for bosonic injective MPS\cite{P-GVWC07}, we obtain a $\mathrm{PU}(n)$-valued function $\{g_{\alpha\beta}(x)\}$ and $\mathrm{U}(1)$-valued function $\{e^{i\theta_{\alpha\beta}}\}$ on $U_{\alpha\beta}$ such that
\begin{eqnarray}\label{eq:transition}
A^{i}_{\alpha}(x)=e^{i\theta_{\alpha\beta}}g_{\alpha\beta}(x)A^{i}_{\beta}(x)g_{\alpha\beta}(x)^{\dagger}.
\end{eqnarray}
Remark that $\{g_{\alpha\beta}(x)\}$ and $\{e^{i\theta_{\alpha\beta}}\}$ are unique. By using the uniqueness of $\{g_{\alpha\beta}(x)\}$, we can easily check that $\{g_{\alpha\beta}(x)\}$ satisfy the cocycle condition 
\begin{eqnarray}\label{eq:cocycle condition}
(\delta g)_{\alpha\beta\gamma}=1.
\end{eqnarray}
We call $\{g_{\alpha\beta}(x)\}$ a transition function.\footnote{We may regard $g_{\alpha\beta}(x)$ as a $\mathrm{U}(n)$-valued function, but in that case, the definition Eq.(\ref{eq:transition}) remains redundant in the $\mathrm{U}(1)$ phase of $g_{\alpha\beta}(x)$. Thus, it is natural to regard $g_{\alpha\beta}(x)$ as a function that takes value in  $\mathrm{PU}(n)=\mathrm{U}(n)/\mathrm{U}(1)$.}

Next, we take a $\mathrm{U}(n)$-lift $\{\hat{g}_{\alpha\beta}(x)\}$ of the transition function, i.e., $\{\hat{g}_{\alpha\beta}(x)\}$ is a $\mathrm{U}(n)$-valued continuous function such that $\pi(\hat{g}_{\alpha\beta})=g_{\alpha\beta}$. Here, $\pi:\mathrm{U}(n)\to\mathrm{PU}(n)$ is a projection. Since $\{g_{\alpha\beta}\}$ satisfy the cocycle condition Eq.(\ref{eq:cocycle condition}), $(\delta \hat{g})_{\alpha\beta\gamma}$ takes value in $\mathrm{U}(1)$. We define 
\begin{eqnarray}\label{eq:DD}
c_{\alpha\beta\gamma}:=\hat{g}_{\alpha\beta}\hat{g}_{\beta\gamma}\hat{g}_{\gamma\alpha}.
\end{eqnarray} 
By definition, $c_{\alpha\beta\gamma}$ is a cocycle, i.e., $(\delta c)_{\alpha\beta\gamma\delta}=1$. Thus $c_{\alpha\beta\gamma}$ defines a cohomology class  $\left[c_{\alpha\beta\gamma}\right]\in\coho{2}{X}{\underline{\mathrm{U}(1)}}\simeq \cohoZ{3}{X}$. $\left[c_{\alpha\beta\gamma}\right]$ is called the Dixmier-Douady class\cite{DD63}. Since $\left[c_{\alpha\beta\gamma}\right]$ takes values in $\cohoZ{3}{X}$, we may already consider this as a higher pump invariant. However, to extract this quantity numerically by integration, we need to introduce a $1$-from and $2$-form connection $\theta^{1}_{\alpha\beta}$ and $\theta^{2}_{\alpha}$ as an element $c=(c_{\alpha\beta\gamma},\theta^{1}_{\alpha\beta},\theta^{2}_{\alpha})\in H^3 (X; \D (3))$ and $3$-form curvature $\eta\left.\right|_{U_\alpha}=d\theta^{2}_{\alpha}$. This procedure can be explained as an analogy with complex line bundles. A complex line bundle over $X$ is completely characterized by a transition function $\{g_{\alpha\beta}\}$ and it defines an element of $\left[g_{\alpha\beta}\right]\in\cohoZ{2}{X}$. Since isomorphism classes of complex line bundles are classified by $\cohoZ{2}{X}$, $\left[g_{\alpha\beta}\right]$ is a topological invariant. However, in order to numerically extract the Chern number and the discrete Berry phase, we need to take a connection $A_{\alpha}$ and a curvature $F\left.\right|_{U_\alpha}=dA_{\alpha}$ and integrate it. A complex line bundle with connection $(g_{\alpha\beta},A_{\alpha})$ determines an element of $H^2 (X; \D (2))$ and computation of the Chern number and the discrete Berry phase can be regarded as an integration of the smooth Deligne cohomology class as we saw in Sec.\ref{sec:smooth Deligne}.

The existence of $\theta^{1}_{\alpha\beta}$ and $\theta^{2}_{\alpha}$ is guaranteed from the generalized Mayer-Vietoris theorem\cite{Bott-Tu}, and the value of the integration does not depend on how to take $\theta^{1}_{\alpha\beta}$ and $\theta^{2}_{\alpha}$. In this case, for example, we can take $\theta^{1}_{\alpha\beta}$ and $\theta^{2}_{\alpha}$ as follows:
\begin{eqnarray}
\theta^{2}_{\alpha}=0,\theta^{1}_{\alpha\beta}=d\log\det(\hat{g}_{\alpha\beta}).
\end{eqnarray}
Since $c_{\alpha\beta\gamma}$ is defined by Eq.(\ref{eq:DD}), $(\theta^{2}_{\alpha},\theta^{1}_{\alpha\beta},c_{\alpha\beta\gamma})$ is $3$-cocycle. Consequently, our pump invariant $n_{\rm top.}$ is the discrete Berry phase of $(c_{\alpha\beta\gamma},\theta^{1}_{\alpha\beta},\theta^{2}_{\alpha})$ over a suitable $2$-cycle $Y\in\zmod{k}\subset\mathrm{H}_{2}(X;\mathbb{Z})_{\rm tor.}$:
\begin{eqnarray}
n_{\rm top.}(Y):=\mathrm{Hol}_{Y}(c_{\alpha\beta\gamma},\theta^{1}_{\alpha\beta},\theta^{2}_{\alpha})\exp(\frac{1}{k}\int_{\Sigma}\eta)\in\cohoZ{3}{X}_{\rm tor.},
\end{eqnarray}
where $\cohoZ{3}{X}_{\rm tor.}$ is the torsion part of $\cohoZ{3}{X}$ and $\Sigma$ is a bounding manifold of $k\cdot Y$, i.e., $\partial \Sigma=k\cdot Y$. Here, the choice of the homology class of $Y$ depends on which charge we measure. From the universal coefficient theorem, there is an isomorphism between $\mathrm{H}_{2}(X;\mathbb{Z})_{\rm tor.}$ and $\cohoZ{3}{X}_{\rm tor.}$. If $\cohoZ{3}{X}_{\rm tor.}$ has more than one component, we integrate over the $2$-cycle $Y$, which is determined as a pullback by the isomorphism of the generator of the component. For example, consider the case where the parameter space $X$ is given by the direct product of some manifold $M^n$ and $S^1$: $X=M^n\times S^1$. In this case, a model of quantum mechanics parametrized by $M^n$ on the boundary is pumped by the deformation of the system along the $S^1$ direction. If we are interested in whether the discrete Berry phase along a path $\gamma\subset M^n$ is pumped, we choose $Y=\gamma\times S^1\subset X$ and compute $n_{\rm top.}(Y)$.

\subsection{Computation of the Higher Pump Invariant:$\rp{2}\times S^1$ model}\label{sec:computation of n Z/2}

\subsubsection{MPS representation and Transition Function}
In this section, we compute our higher pump invariant for the model (\ref{eq:model with rp2}). 
The ground state of (\ref{eq:model with rp2}) is known to be the cluster state, of which an MPS representation is given by~\cite{P-GVWC07,Tasaki20}
\begin{align}
\ket{{\rm GS}(\vec{n},t)}
= \sum_{\{i_k\}, \{j_l\}} {\rm tr} (A_{\tau(\vec n)}^{i_1} A^{j_l}_{\sigma(t)} \cdots A_{\tau(\vec n)}^{i_L} A^{j_L}_{\sigma(t)}) \ket{\tau_{i_1}(\vec n) \sigma_{j_1}(t) \cdots \tau_{i_L}(\vec n) \sigma_{j_L}(t)}
\label{eq:rp2s1_sg_nt_mps}
\end{align}
with 
\begin{align}
    &A_{\tau(\vec n)}^{\uparrow}= \begin{pmatrix}
        1&1\\
        0&0\\
    \end{pmatrix},
    A_{\tau(\vec n)}^\downarrow= \begin{pmatrix}
        0&0\\
        1&-1\\
    \end{pmatrix},
    A_{\tau(\vec n)}^0= \begin{pmatrix}
        0&0\\
        0&0\\
    \end{pmatrix}, \\
    &A_{\sigma(t)}^\uparrow= \begin{pmatrix}
        1&1\\
        0&0\\
    \end{pmatrix},
    A_{\sigma(t)}^\downarrow= \begin{pmatrix}
        0&0\\
        1&-1\\
    \end{pmatrix}.
\end{align}
Here, $\ket{\tau_i(\vec n)}$ is basis diagonalizing $\tau^z(\vec n)$. 
Explicitly, 
\begin{align}
&\ket{\tau(\vec n)_\uparrow}
=\frac{1}{\sqrt{2}}(\ket{u_{+}(\vec{n})}+\ket{u_{-}(\vec{n})})=\frac{1}{\sqrt{2}}\begin{pmatrix}
1\\
n_{3}\\
n_{1}+in_{2}
\end{pmatrix}, \nonumber \\
&\ket{\tau_\downarrow(\vec{n})}
=\frac{1}{\sqrt{2}}(\ket{u_{+}(\vec{n})}-\ket{u_{-}(\vec{n})})=\frac{1}{\sqrt{2}}\begin{pmatrix}
1\\
-n_{3}\\
-n_{1}-in_{2}
\end{pmatrix},\nonumber \\
&\ket{\tau_0(\vec{n})}=\ket{u_{-}^{\perp}(\vec{n})}=\begin{pmatrix}
0\\
-n_{1}+in_{2}\\
n_{3}
\end{pmatrix}. 
\end{align}
In the expression (\ref{eq:rp2s1_sg_nt_mps}), the basis depends on the parameters $\vec n,t$, whereas the MPS matrices do not. 
We move to parameter-independent basis $\ket{\tau_i}= \ket{\tau_i(\vec n=\vec z_0)}$ and $\ket{\sigma_j} = \ket{\sigma_j(t=0)}$ to get 
\begin{align}
\ket{{\rm GS}(\vec{n},t)}
= \sum_{\{i_k\}, \{j_l\}} {\rm tr} (A_{\tau}^{i_1}(\vec n) A^{j_l}_{\sigma}(t) \cdots A_{\tau}^{i_L}(\vec n) A^{j_L}_{\sigma}(t)) \ket{\tau_{i_1}\sigma_{j_1}\cdots \tau_{i_L}\sigma_{j_L}}
\label{eq:mpsrepofGS}
\end{align}
with 
\begin{align}
    &A_\tau^i(\vec n) := \sum_k K_{\tau}(\vec n)_{i,k} A^k_{\tau(\vec n)}, \nonumber\\
    &A_\sigma^i(t) := \sum_k K_{\sigma}(t)_{i,k} A^k_{\sigma(t)}. 
\end{align}
Here, $K_{\tau}(\vec n)$ and $K_{\sigma}(t)$ are basis transformations defined by $\ket{\tau_k(\vec n)} = \sum_i \ket{\tau_i(\vec n=\vec z_0)} K_{\tau}(\vec n)_{i,k}$ and $\ket{\sigma_k(t)} =\sum_i \ket{\sigma_i(t=0)} K_{\sigma}(t)_{i,k}$, explicitly given by 
\begin{align}
    &K_{\tau}(\vec n) = \begin{pmatrix}
        \frac{1+n_3}{2}&\frac{1-n_3}{2}&-\frac{n_1-i n_2}{\sqrt{2}}\\
        \frac{1-n_3}{2}&\frac{1+n_3}{2}&\frac{n_1-i n_2}{\sqrt{2}}\\
        \frac{n_1+in_2}{\sqrt{2}}&-\frac{n_1+in_2}{\sqrt{2}}&n_3\\
    \end{pmatrix},  \\
    &K_{\sigma}(t)=\begin{pmatrix}
        \cos \frac{t}{4} & -i \sin \frac{t}{4}\\
        -i \sin \frac{t}{4} & \cos \frac{t}{4}\\
    \end{pmatrix}.
\end{align}
Considering $\tau$ and $\sigma$ spins as a unit site, we have a translational invariant MPS
\begin{eqnarray}
A^{i,j}_{\tau,\sigma}(\vec{n},t):=A^i_{\tau}(\vec{n})A^j_{\sigma}(t).\label{eq:rp2s1model_mps}
\end{eqnarray}

The ground state Eq.(\ref{eq:mpsrepofGS}) is parametrized by $\rp{2}\times S^{1}$ as a family of physical states, while the matrices Eq.(\ref{eq:rp2s1model_mps}) are not, i.e.,  
\begin{align}
    A^{i,j}_{\tau\sigma}(-\vec n,t) \neq A^{i,j}_{\tau\sigma}(\vec n,t),
    A^{i,j}_{\tau\sigma}(\vec n,t+2\pi) \neq A^{i,j}_{\tau\sigma}(\vec n,t).
\end{align}
The fundamental theorem for matrix product state \cite{P-GVWC07} implies the existence of unitary matrices $g_{\rp{2}}(\vec n,t),g_{S^1}(\vec n,t)$ and phases $e^{i\alpha},e^{i\beta}$ so that 
\begin{eqnarray}
A^{i,j}_{\tau\sigma}(\vec n,t)=e^{i\alpha}g_{\rp{2}}(\vec n,t) A^{i,j}_{\tau\sigma}(-\vec n,t)g_{\rp{2}}(\vec n,t)^{\dagger},
\end{eqnarray}
and 
\begin{eqnarray}
A^{i,j}_{\tau\sigma}(\vec n,t)=e^{i\beta}g_{S^1}(\vec n,t)A^{i,j}_{\tau\sigma}(\vec n,t+2\pi)g_{S^1}(\vec n,t)^{\dagger}.
\end{eqnarray}
It is easy to find that 
\begin{align}
    &e^{i\alpha}=1, g_{\rp{2}}(\vec n,t) = \sigma_x, \\
    &e^{i\beta}=i, g_{S^1}(\vec n,t) = \sigma_z.
\end{align}
It is important to note that the unitary matrix given by the fundamental theorem is unique up to a $\U{1}$ phase factor. 
Thus we should regard $\sigma^{x}$ and $\sigma^{z}$ as elements of $\PU{2}$, not of $\mathrm{U}(2)$. 
To indicate this explicitly, we write an element of $\PU{2}$ as $[\sigma^{x}]$ instead of $\sigma^{x}$, for example.
 \begin{figure}[H]
 \begin{center}
  \includegraphics[width=70mm]{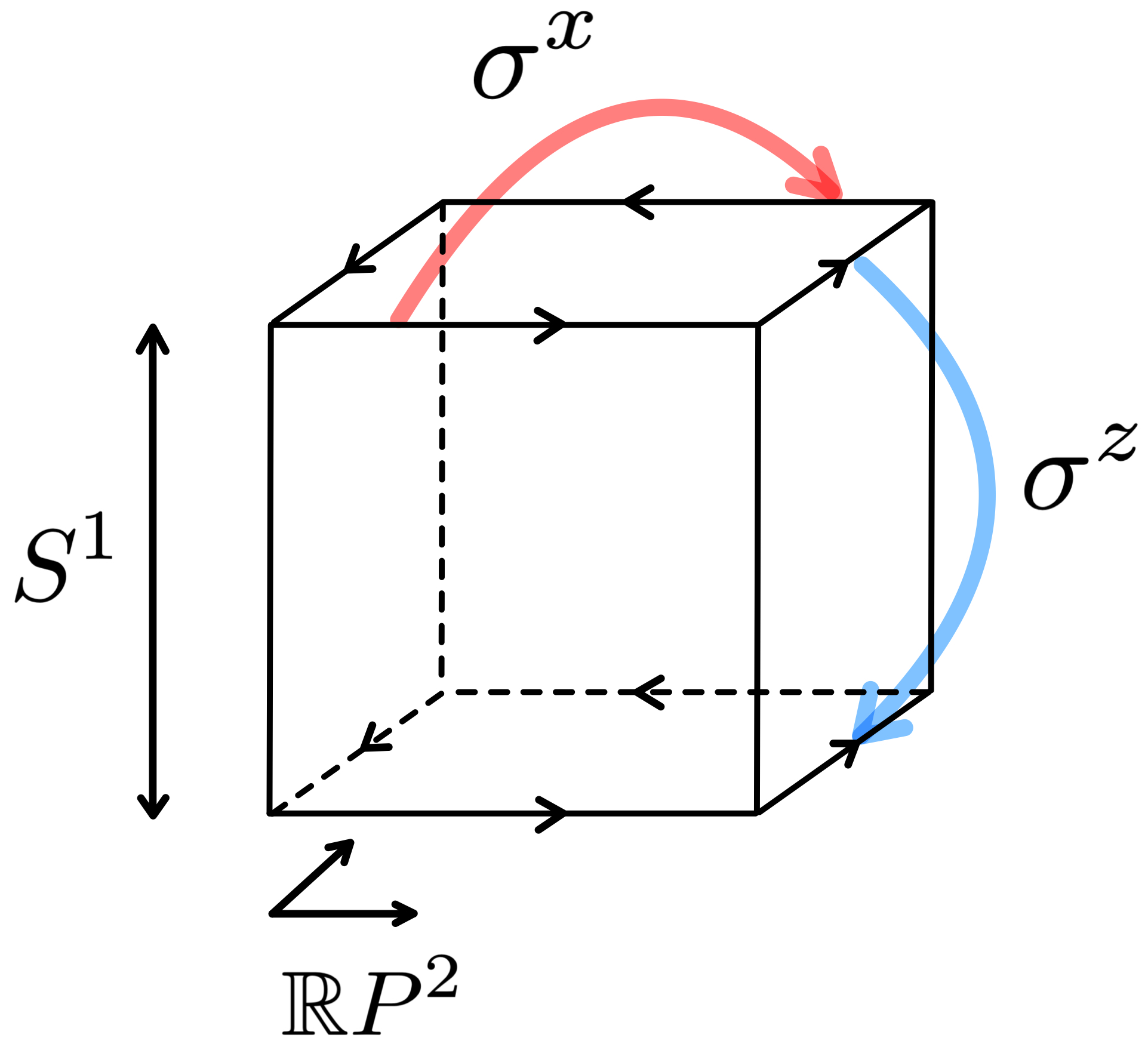}
 \end{center}
  \caption[]{Transition functions on $\rp{2}\times S^{1}$.
}
\label{fig:rp2timess1}
\end{figure}

\subsubsection{Calculation of the Higher Pump Invariant}\label{sec:integration of rp2 model}

As we showed in Sec.\ref{sec:chern number pump}, we found that a quantum mechanical system parametrized by $\rp{2}$ was pumped by adiabatic deformation along $S^1$, and it had a nontrivial discrete Berry phase along the nontrivial path $\gamma$ of $\rp{2}$. This implies that the higher pump invariant defined in the Sec.\ref{sec:defofpumpinv} along the surface $Y=\gamma\times S^1\subset\rp{2}\times S^1$ is nontrivial. Hence, let's compute the discrete higher Berry phase along this surface $Y$.

\begin{figure}[H]
 \begin{center}
  \includegraphics[width=60mm]{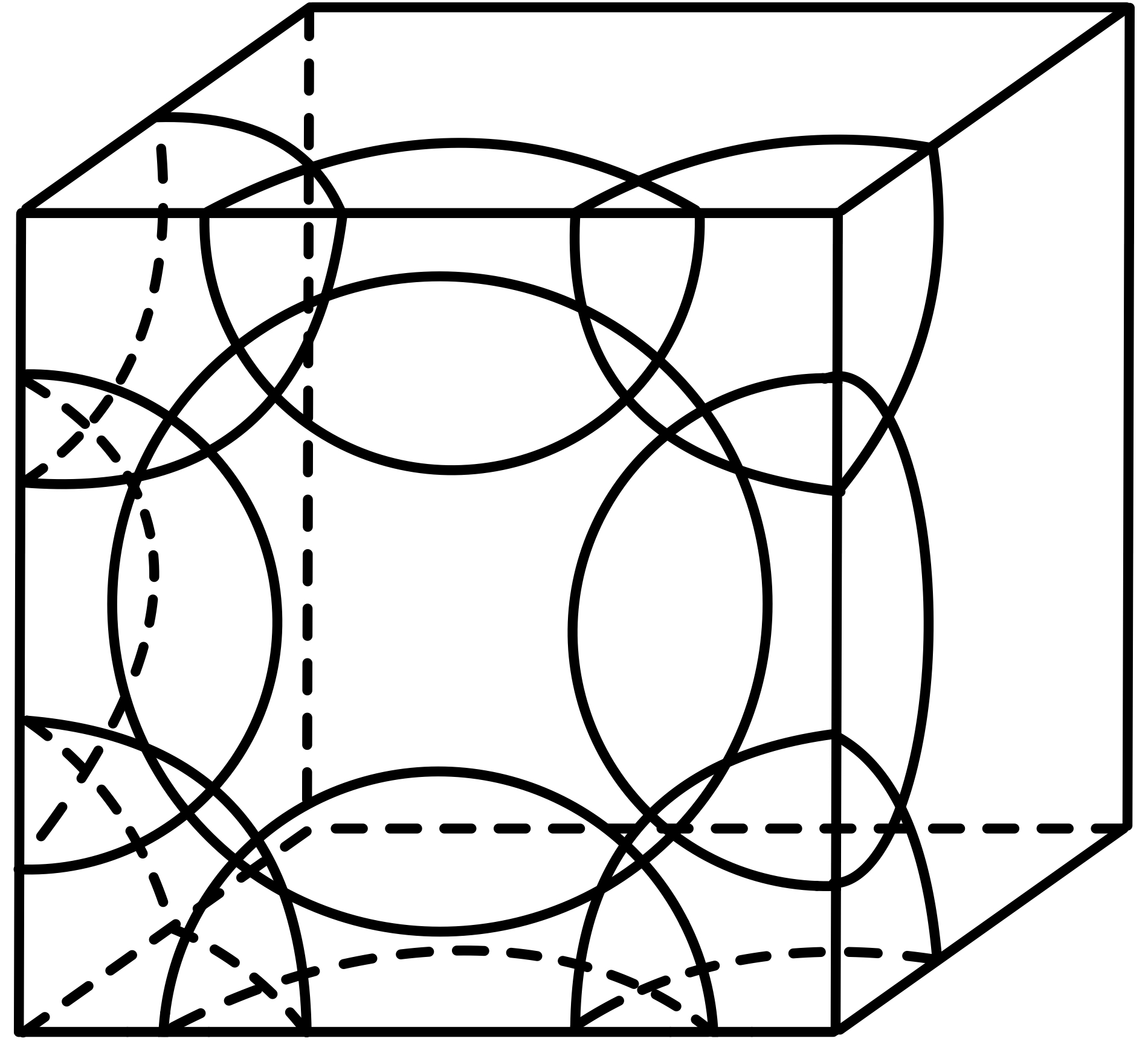}
 \end{center}
  \caption[]{A part of an open covering of $\rp{2}\times S^{1}$. These consist of one open ball centered at the center of the cube, one open ball centered at the vertex, three open balls centered at the midpoints of the edges, and three open balls centered at the midpoints of the faces.
}
\label{fig:opencover}
\end{figure}

To calculate this invariant, first, we need to take a good open  cover of $X=\rp{2}\times S^{1}$. We take the good open cover as in Fig.\ref{fig:opencover}. Note that the intersections of the common part of the balls are taken to be contained within the interior of the cube. In the following, we formally write this open covering as $\oc=\{U_{i}\}_{i \in I}$.

Next, we take a polyhedral decomposition $\tt$ of $\rp{2}\times S^{1}$ which is compatible with the open covering $\oc$. We take the polyhedral decomposition $\tt$ as in Fig.\ref{fig:triangulation}:
\begin{figure}[H]
 \begin{center}
  \includegraphics[width=60mm]{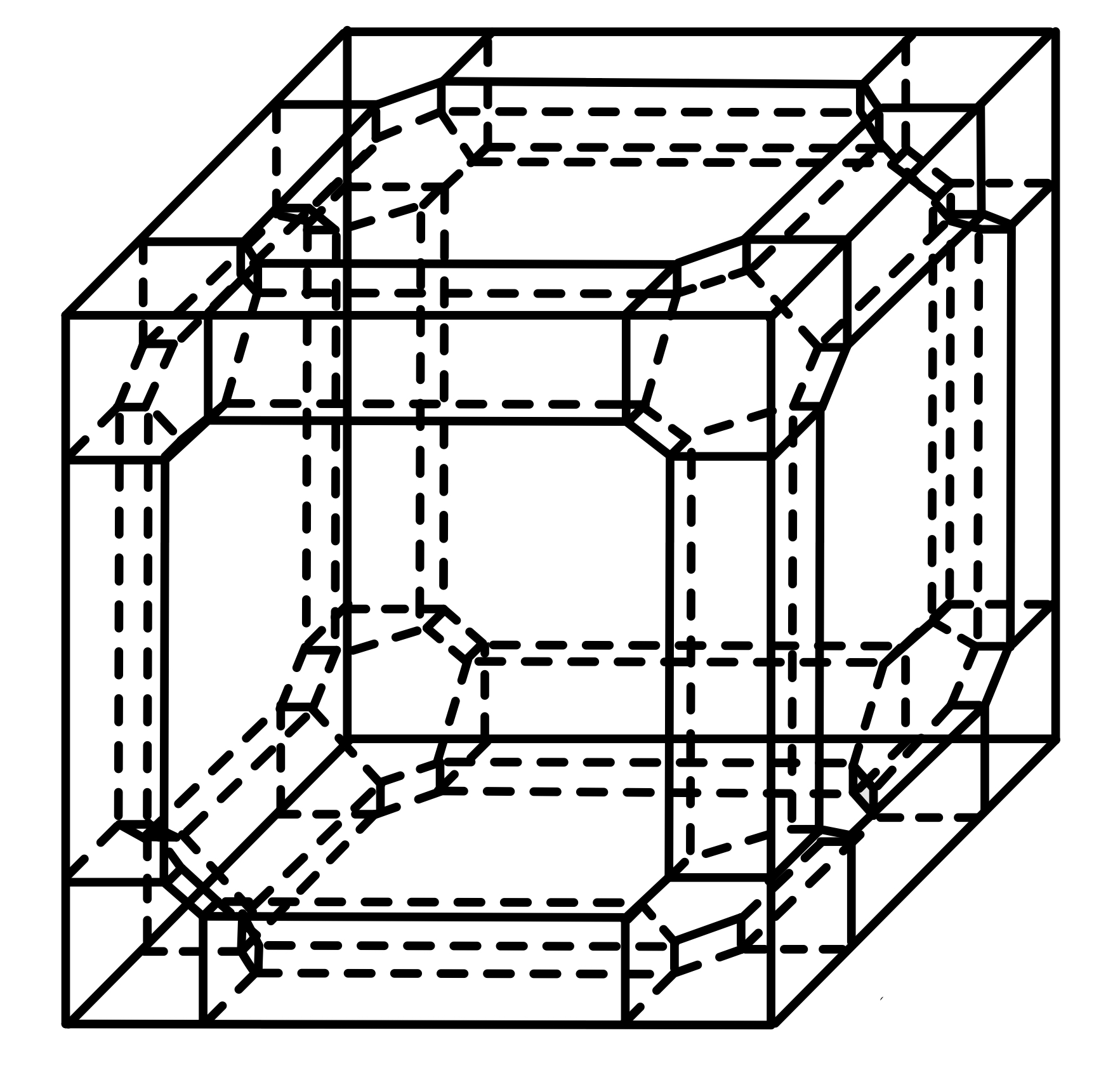}
 \end{center}
  \caption[]{A polyhedral decomposition of $\rp{2}\times S^{1}$. This is compatible with the open cover, that is, for any dimensional cells, there is a patch so that the cell is included within the patch. 
}
\label{fig:triangulation}
\end{figure}

\begin{figure}[H]
 \begin{center}
  \includegraphics[width=120mm]{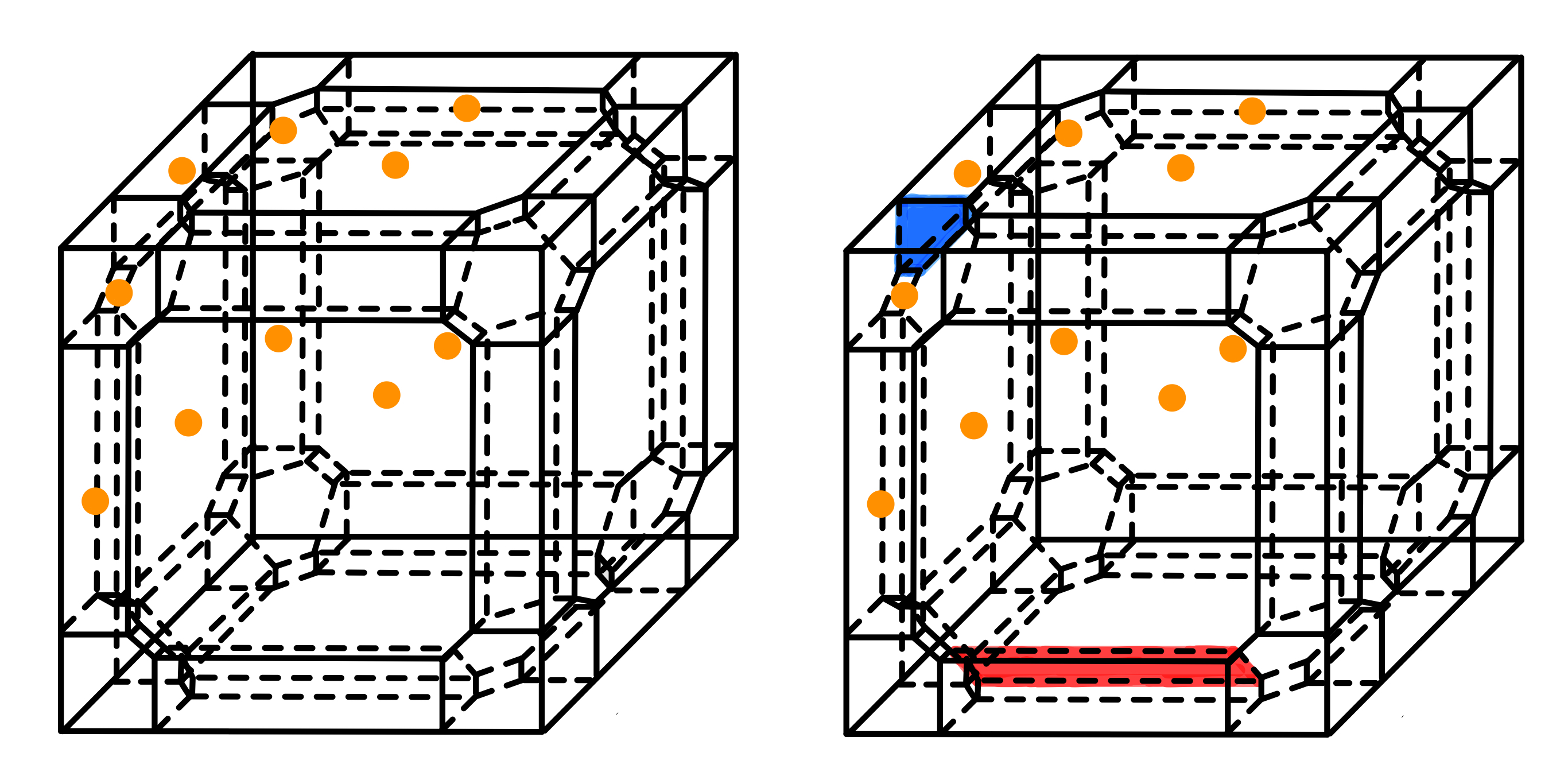}
 \end{center}
  \caption[]{(Left) A based polyhedral decomposition of $\rp{2}\times S^{1}$. The two in the middle belong to the back patch and the center patch of the cube, respectively. (Right) For example, the transition function on the blue surface is $\left[1_{2}\right]$ since the base points of the front-side and back-side patches are directly connected. On the other hand, the transition function on the red surface is $[\sigma^{x}\sigma^{z}]$ since the path connecting the base points of the front-side and back-side patches have to get through the bottom and side faces of the cube one at a time. 
}
\label{fig:basedtriangulation}
\end{figure}

The open set corresponding to a simplex $\tau$ in $\tt$ is written by $U_{i_\tau}$. To implement the injective MPS bundle on this cube, we assign transition functions on the faces of the polyhedral decomposition. We would like to perform this assignment systematically. To this end, we take base points of each patch, and define  transition functions on the intersection $U_{12}=U_{1}\cap U_{2}$ under the following rules:
\begin{enumerate}
    \item Take a path starting from the base point of the patch $U_{1}$ and passing through $U_{12}$ and terminating at the base point of $U_{2}$.
    \item If the path is through the side of the cube, we give $\left[\sigma^{x}\right]$, and if  the path is through the top or bottom, we give $\left[\sigma^{z}\right]$.
\end{enumerate}
Under these assignment rules, the configuration of the injective MPS bundle is determined by fixing the base points of each patch. We fix the base points as in Fig.\ref{fig:basedtriangulation}. In the following, we formally write the transition functions as $\{g_{\alpha\beta}:U_{\alpha}\cap U_{\beta}\to\PU{2}\}$.

\begin{figure}[H]
 \begin{center}
  \includegraphics[width=100mm]{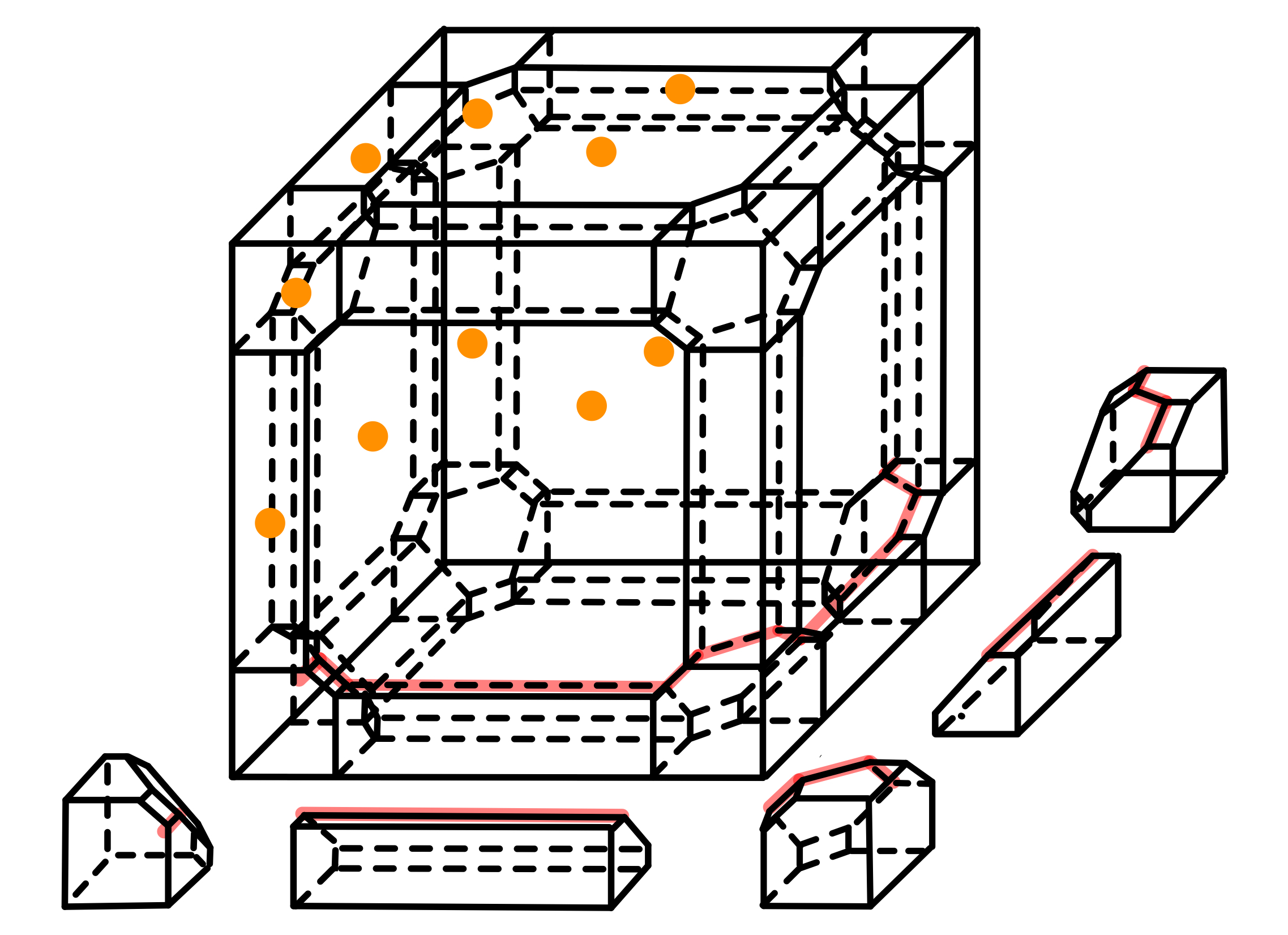}
 \end{center}
  \caption[]{The configuration of $c_{\alpha\beta\gamma}$. On the red lines, $c_{\alpha\beta\gamma}=-1$ on them and $c_{\alpha\beta\gamma}=1$ on the others. 
}
\label{fig:violationofcocycle}
\end{figure}

The Dixmier-Douady class can be viewed as a violation of the cocycle condition for lifting a transition function that takes values in $\PU{2}$ to $\U{2}$. We take the following lifts:
\begin{eqnarray}\label{eq:lifting rule}
\left[1_{2}\right]\mapsto1_{2},
\left[\sigma^{x}\right]\mapsto\sigma^{x},
\left[\sigma^{z}\right]\mapsto\sigma^{z},
\left[\sigma^{x}\sigma^{z}\right]\mapsto\sigma^{x}\sigma^{z},
\end{eqnarray}
In the following, we formally write the lifted transition functions as $\{\hat{g}_{\alpha\beta}:U_{\alpha}\cap U_{\beta}\to\U{2}\}$. Under these lifts, the violation of the cocycle condition can only occur if the surfaces with the transition functions $\left[\sigma^{x}\right]$,  $\left[\sigma^{z}\right]$, and $\left[\sigma^{x}\sigma^{z}\right]=\left[\sigma^{z}\sigma^{x}\right]$ intersect each other, and the violation occurs by $-1$ on them. For example, let the bottom patch of the front-side be $U_{\alpha}$, the center patch of the front-side be $U_{\beta}$, and the center patch of the cube $U_{\gamma}$. By the assignment rule of transition functions, we can easily check that $g_{\alpha\beta}=\left[\sigma^{z}\right]$, $g_{\beta\gamma}=\left[\sigma^{x}\right]$, and $g_{\alpha\gamma}=\left[\sigma^{x}\sigma^{z}\right]$. Thus, lifted transition functions are $\hat{g}_{\alpha\beta}=\sigma^{z}$, $\hat{g}_{\beta\gamma}=\sigma^{x}$, and $\hat{g}_{\alpha\gamma}=\sigma^{x}\sigma^{z}$. The Dixmier-Douady class $c_{\alpha\beta\gamma}$ on the line is given by
\begin{eqnarray}
c_{\alpha\beta\gamma}=\hat{g}_{\alpha\beta}\hat{g}_{\beta\gamma}\hat{g}_{\gamma\alpha}=-1.
\end{eqnarray}
Consequently, we can identify the edge with nontrivial $c_{\alpha\beta\gamma}$ over whole $\rp{2}\times S^1$, as in Fig.\ref{fig:violationofcocycle}. Note that $c_{\alpha\beta\gamma}$ defines a cohomology class $\left[c_{\alpha\beta\gamma}\right]\in\mathrm{H}^{2}(\rp{2}\times S^{1};\mathrm{U}(1))\simeq\cohoZ{3}{\rp{2}\times S^1}\simeq\Zmod{2}$ and a cohomology class $[(c_{\alpha\beta\gamma},0,0)]  \in H^3(\rp{2}\times S^{1};\D(3))$ which is flat. The group $\cohoU{2}{\rp{2}\times S^{1}}$ is a subgroup of $H^3(\rp{2}\times S^{1};\D(3))$ under the map $\left[c_{\alpha\beta\gamma}\right] \mapsto [(c_{\alpha\beta\gamma},0,0)]$.

\begin{figure}[H]
 \begin{center}
  \includegraphics[width=40mm]{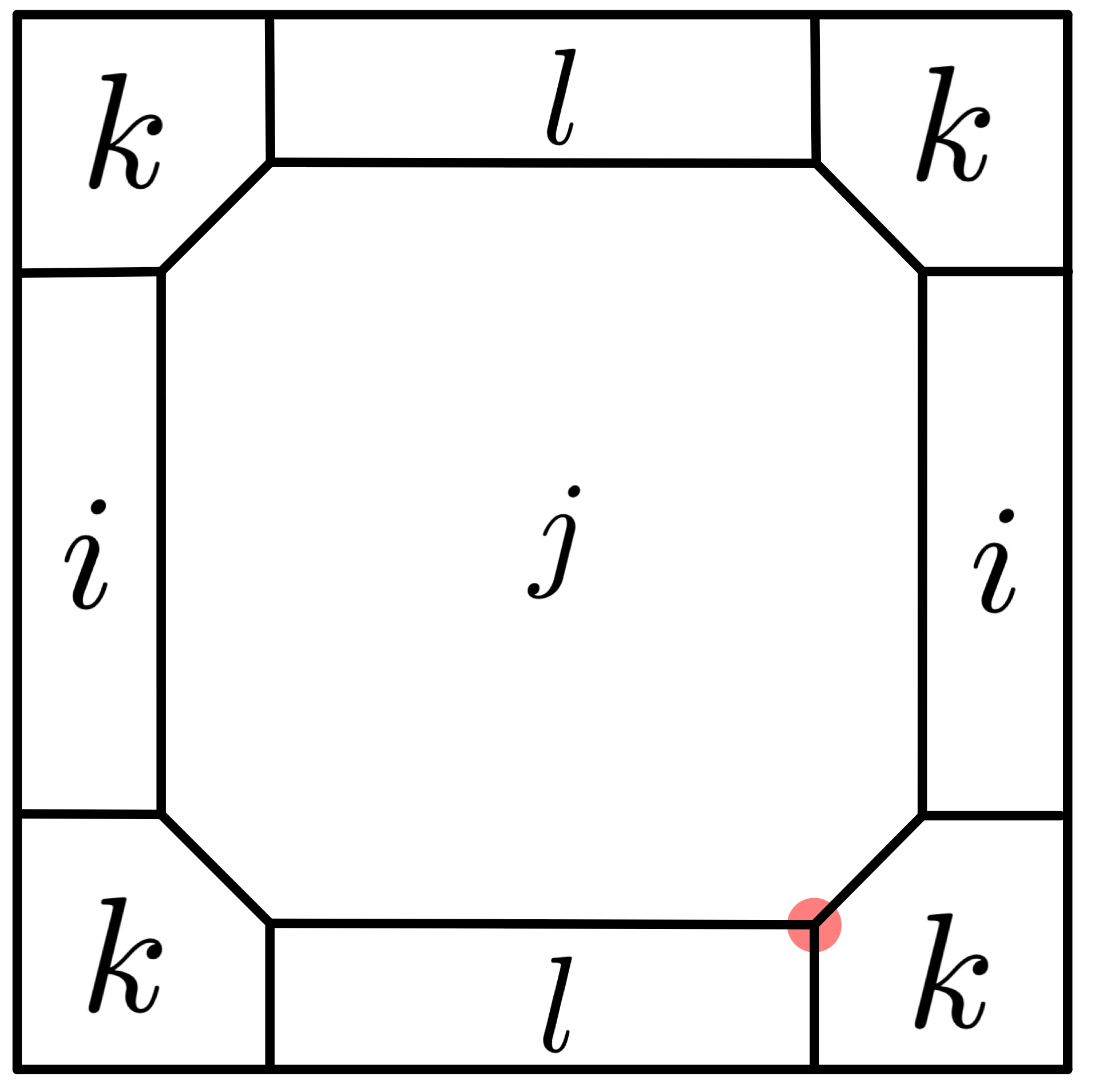}
 \end{center}
  \caption[]{The polyhedral decomposition of $Y$ induced from that of $\rp{2}\times S^1$, and a vertex on which the Dixmier-Douady class has a value $-1$.
}
\label{fig:nontrivial cocycle on Y}
\end{figure}

\begin{figure}[H]
 \begin{center}
  \includegraphics[width=40mm]{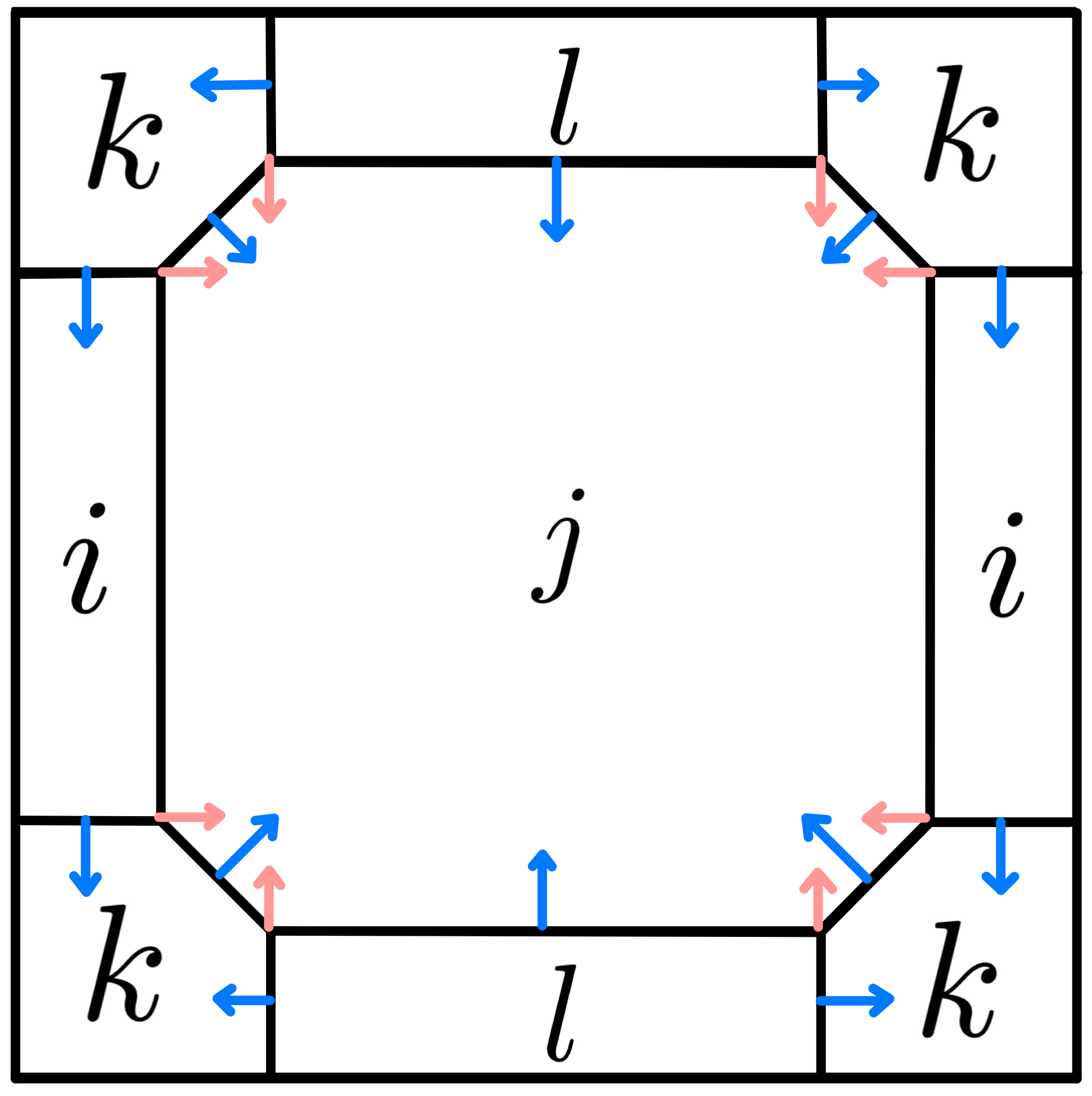}
 \end{center}
  \caption[]{An index map of the polyhedral decomposition of $Y$. Blue arrows are index maps for edges, and the direction of it indicates the patch to which the edge belongs. Similarly, orange arrows are index maps for vertices, and the direction of it indicates the patch to which the vertex belongs.
}
\label{fig:index map on Y}
\end{figure}

\begin{figure}[H]
 \begin{center}
  \includegraphics[width=40mm]{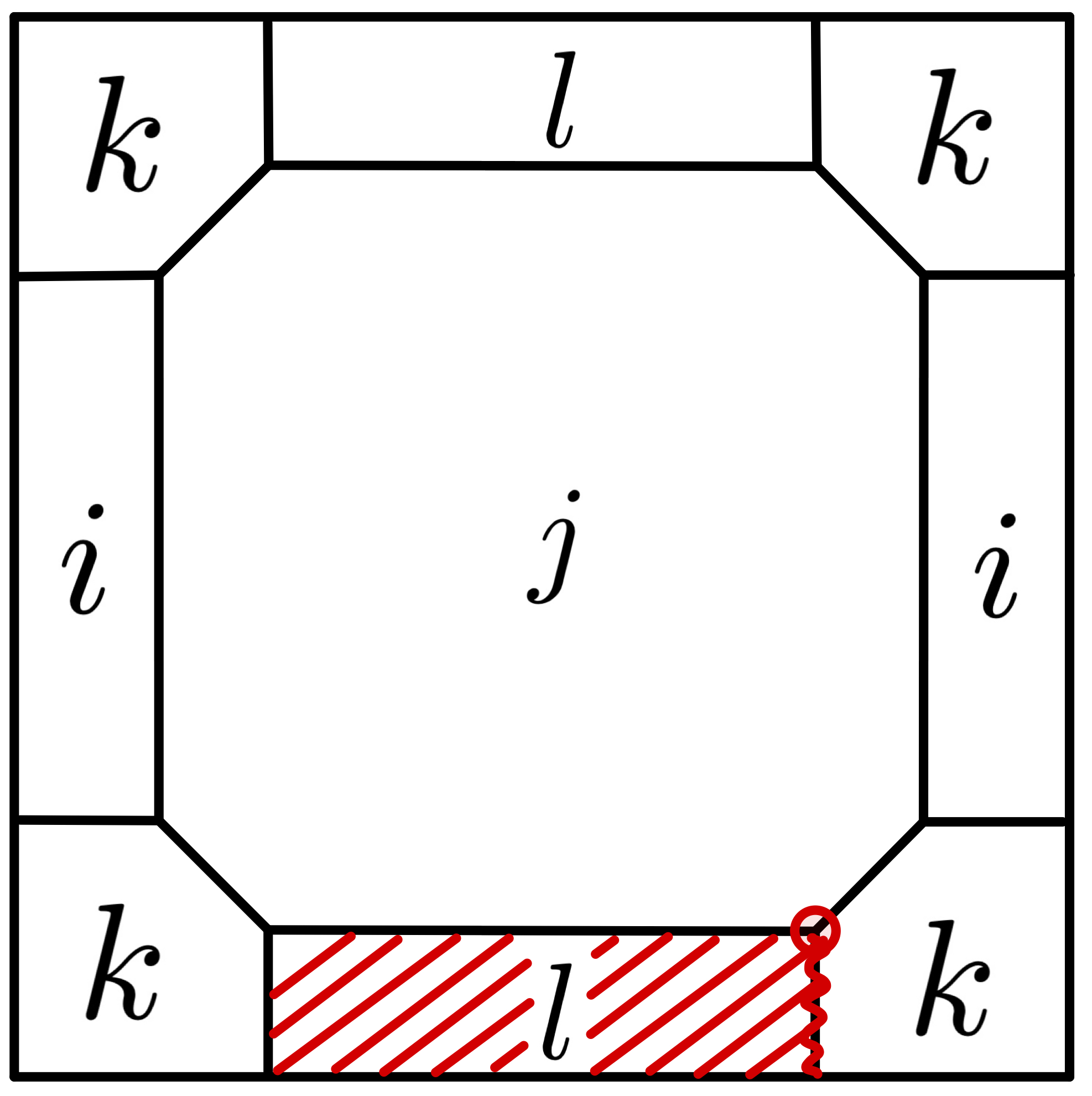}
 \end{center}
  \caption[]{A flag contributing to integration. The red circle represents $s^0$, the red wavy line represents $s^1$, and the red shaded face represents $s^2$.
}
\label{fig:nontrivial flag}
\end{figure}

Let's compute the higher holonomy of $c=(c_{\alpha\beta\gamma},0,0)$ along this surface $Y=\gamma\times S^1$. We take the diagonal line of the top square of the cube as a nontrivial path in $\rp{2}$ and take a polyhedral decomposition\footnote{Although what is shown in Fig.\ref{fig:nontrivial cocycle on Y} is a polyhedral decomposition, we theoretically consider a triangulation that subdivides it and assume that the index map takes the same value for all simplices in each polyhedron.} of $Y$ induced from that of $\rp{2}\times S^1$. We label the open sets of $Y$ with Roman letters ($i,j,k,...$) instead of Greek letters ($\alpha,\beta,\gamma,...$). We show the nontrivial cocycle on $Y$ in the Fig.\ref{fig:nontrivial cocycle on Y}.
It is necessary to take an index map $\phi$ to perform an integration. However, to compute the invariant on $Y$, it is sufficient to determine the index map on $Y$. We take an index map as in the Fig.\ref{fig:index map on Y}.
Then the higher holonomy is given by the following  formula:
\begin{eqnarray}
\hol_Y(c)=\prod_{\si=(\si^0 \subset \si^1 \subset \si^2) \in F(2)} c_{\phi_{\si^{2}} \phi_{\si^{1}} \phi_{\si^{0}}} (\si^0).
\end{eqnarray}
As seen in Fig.\ref{fig:nontrivial cocycle on Y}, the only intersection where the Dixmier-Douady class is nontrivial is $U_{ijk}$.
Moreover, a full flag $\si=(\si^0 \subset \si^1 \subset \si^2)$ satisfying $\{ l,k,j \}=\{\phi_{\si^{2}},  \phi_{\si^{1}}, \phi_{\si^{0}}\}$ is only one $(s^0 \subset s^1 \subset s^2)$. We show this flag in Fig.\ref{fig:nontrivial flag}. Therefore, we have 
\begin{align}
\hol_Y(c)&=\prod_{\si=(\si^0 \subset \si^1 \subset \si^2) \in F(2)} c_{\phi_{\si^{2}} \phi_{\si^{1}} \phi_{\si^{0}}} (\si^0), \\
&=c_{\phi_{s^{2}} \phi_{s^{1}} \phi_{s^{0}}} (s^0),\\
&=c_{lkj}, \\
&=-1.
\end{align}
Since the cocycle $c$ is flat, there is no correction by the $3$-form curvature. As a result, we have that the higher pump invariant is 
\begin{eqnarray}
n_{\rm top.}(Y)=\hol_Y(c)=-1\in\zmod{2}.
\end{eqnarray}
Therefore, the higher pump invariant is nontrivial. 

\subsection{Computation of the Higher Pump Invariant: $\mathrm{L}(3,1)\times S^1$ model}\label{sec:computation of n Z/3}

\subsubsection{MPS representation and Transition Function}
Let's compute an MPS representation of the ground state (\ref{eq:lens space}). 
We find that the following MPS representation 
\begin{align}\label{eq:l31s1_MPS}
    \ket{\{A^i_{u(\vec z)},A^j_{\tilde \sigma(t)}\}}
    =\sum_{\{i_k,j_l\}} \tr{A_{u(\vec z)}^{i_1}A_{\tilde \sigma(t)}^{j_1} \cdots} \ket{u_{i_1}(\vec z)\tilde \sigma_{j_1}(t)\cdots }
\end{align}
with 
\begin{align}
&A_{u(\vec z)}^0
=
\frac{1}{3}\begin{pmatrix}
1&1&1\\
\omega^2&1&\omega\\
\omega^2&\omega&1
\end{pmatrix},
A_{u(\vec z)}^1
=\frac{1}{3}\begin{pmatrix}
1&\omega&\omega^2\\
\omega^2&\omega&1\\
\omega^2&\omega^2&\omega^2
\end{pmatrix},
A_{u(\vec z)}^2
=\frac{1}{3}\begin{pmatrix}
1&\omega^2&\omega\\
\omega^2&\omega^2&\omega^2\\
\omega^2&1&\omega
\end{pmatrix}, \\
&A_{\tilde \sigma(t)}^0
=\frac{1}{\sqrt{3}}\begin{pmatrix}
1&0&0\\
1&0&0\\
1&0&0
\end{pmatrix},
A_{\tilde \sigma(t)}^1
=\frac{1}{\sqrt{3}}\begin{pmatrix}
0&\omega&0\\
0&1&0\\
0&\omega^2&0
\end{pmatrix},
A_{\tilde \sigma(t)}^2
=\frac{1}{\sqrt{3}}\begin{pmatrix}
0&0&\omega\\
0&0&\omega^2\\
0&0&1
\end{pmatrix}.
\end{align}
In fact, these matrices satisfy the desired properties: 
Firstly, 
\begin{align}
    A_{\tilde \sigma(t)}^i A_{u(\vec z)}^j A_{\tilde \sigma(t)}^k 
    = \left\{ \begin{array}{ll}
        0 & (i+j \neq k \mod 3),  \\
        p_{ik} A_{\tilde \sigma(t)}^i A_{\tilde \sigma(t)}^k & (i+j = k \mod 3), 
    \end{array}\right.
\end{align}
holds, where $p_{ik}$ are phase factors 
\begin{eqnarray}
p_{00}&=&1, p_{01}=1, p_{02}=1,\\
p_{10}&=&\omega^2, p_{11}=1, p_{12}=\omega,\\
p_{20}&=&\omega^2,p_{21}=\omega, p_{22}=1.  
\end{eqnarray}
Therefore, the MPS is a superposition of decorated domain wall states with the additional decoration of $\mathrm{U}(1)$ phases $p_{ik}$. 
We can also show that
\begin{eqnarray}
p_{i0}p_{0j}A^{i}_{\tilde \sigma(t)}A^{0}_{\tilde \sigma(t)}A^{j}_{\tilde \sigma(t)}=p_{i1}p_{1j}A^{i}_{\tilde \sigma(t)}A^{1}_{\tilde \sigma(t)}A^{j}_{\tilde \sigma(t)}=p_{i2}p_{2j}A^{i}_{\tilde \sigma(t)}A^{2}_{\tilde \sigma(t)}A^{j}_{\tilde \sigma(t)}.
\end{eqnarray}
Hence, all weights are equal to $\tr{A_{\tilde \sigma(t)}^0 A_{\tilde \sigma(t)}^0 \cdots A_{\tilde \sigma(t)}^0}=1$. 
Therefore, $\ket{\{A^i_{u(\vec z)},A^j_{\tilde \sigma(t)}\}}$ is an MPS representation of the ground state (\ref{eq:lens space}).

The basis of the MPS (\ref{eq:l31s1_MPS}) depends on $\vec z$ and $t$, but the matrices do not. 
We rewrite the MPS (\ref{eq:l31s1_MPS}) in the basis of $\ket{\tilde \tau_i}$ and $\ket{\tilde \sigma_j}$. 
To do so, using the basis transformation $\ket{u_k} = \sum_i \ket{\tilde \tau_i} T_{i,k}$ with 
\begin{align}
T=\frac{1}{\sqrt{3}}\begin{pmatrix}
1&1&1\\
\omega&1&\omega^2\\
\omega^2&1&\omega\\
\end{pmatrix}, 
\end{align}
and 
\begin{align}
    &\ket{u_k(\vec z)} 
    = \tilde{V}_{\tau}(\vec z) \ket{u_k} 
    = \sum_{i} \tilde{V}_{\tau}(\vec z) \ket{\tilde \tau_i} T_{i,k}
    = \sum_{i} \ket{\tilde \tau_i} [\tilde{V}_{\tau}(\vec z) T]_{i,k}, \\
    &\ket{\tilde \sigma_k(t)} = \tilde{V}_{\sigma}(t) \ket{\tilde \sigma_k} = \sum_j \ket{\tilde \sigma_j} \tilde{V}_{\sigma}(t)_{j,k}
\end{align}
with matrices $\tilde{V}_{\tau}(\vec z)$ and $\tilde{V}_{\sigma}(t)$ introduced before in Eq.(\ref{eq:Vtauz}) and Eq.(\ref{eq:Vsigmat}), respectively, we have 
\begin{align}
    \ket{\{A^i_{\tilde \tau}(\vec z),A^j_{\tilde \sigma}(t)\}}
    =\sum_{\{i_k,j_l\}} \tr{A_{\tilde \tau}^{i_1}(\vec z)A_{\tilde \sigma}^{j_1}(t) \cdots} \ket{\tilde \tau_{i_1}\tilde \sigma_{j_1}\cdots }
\end{align}
with 
\begin{align}
    A^i_{\tilde \tau}(\vec z) = \sum_k [\tilde{V}_{\tau}(\vec z)T]_{i,k} A^k_{u(\vec z)}, 
    A^j_{\tilde \sigma}(t) = \sum_k \tilde{V}_{\sigma}(t)_{j,k} A^k_{\tilde \sigma(t)}.
\end{align}
Regarding $\tau$ and $\sigma$ spins as a unit site, we get the translational invariant MPS 
\begin{align}
    A^{i,j}_{\tilde{\tau} \tilde{\sigma}}(\vec z,t) := A^i_{\tilde \tau}(\vec z)A^j_{\tilde \sigma}(t). 
\end{align}
With this MPS, it is straightforward to show that the transition functions are given as 
\begin{eqnarray}
A^{i,j}_{\tilde{\tau}\tilde{\sigma}}(\omega\vec{z},t)=\tilde{g}_{\mathrm{L}(3,1)}A^{i,j}_{\tau\sigma}(\vec{z},t)\tilde{g}_{\mathrm{L}(3,1)}^{\dagger}, 
A^{i,j}_{\tau\sigma}(\vec{z},t+2\pi)=\tilde{g}_{S^{1}}A^{i,j}_{\tau\sigma}(\vec{z},t)\tilde{g}_{S^{1}}^{\dagger},
\end{eqnarray}
where
\begin{eqnarray}
\tilde{g}_{\mathrm{L}(3,1)}:=\left[\begin{pmatrix}
\omega&&\\
&1&\\
&&\omega^{2}
\end{pmatrix}\right], 
\tilde{g}_{S^{1}}:=\left[\begin{pmatrix}
&1&\\
&&\omega^{2}\\
\omega&&
\end{pmatrix}\right].
\end{eqnarray}
Here, $\left[\bullet\right]$ implies that $\tilde{g}_{\mathrm{L}(3,1)}$ and $\tilde{g}_{S^{1}}$ are not an element of $\mathrm{U}(3)$, but also $\mathrm{PU}(3)$. 

\subsubsection{Calculation of the Higher Pump Invariant}
Finally, let's compute the discrete higher Berry phase defined in the Sec.\ref{sec:defofpumpinv}. As we showed in Sec.\ref{sec:Z/3 chernnumber pump}, we found that a quantum mechanical system parametrized by $\mathrm{L}(3,1)$ was pumped by adiabatic deformation along $S^1$ and it had a nontrivial discrete Berry phase along the nontrivial path $\gamma$ of $\mathrm{L}(3,1)$. This implies that the higher pump invariant defined in Sec.\ref{sec:defofpumpinv} along the surface $Y=\gamma\times S^1\subset\mathrm{L}(3,1)\times S^1$ is nontrivial. Let's compute the discrete higher Berry phase over $\gamma\times S^1\subset \mathrm{L}(3,1)\times S^{1}$.

\begin{figure}[H]
 \begin{center}
  \includegraphics[width=50mm]{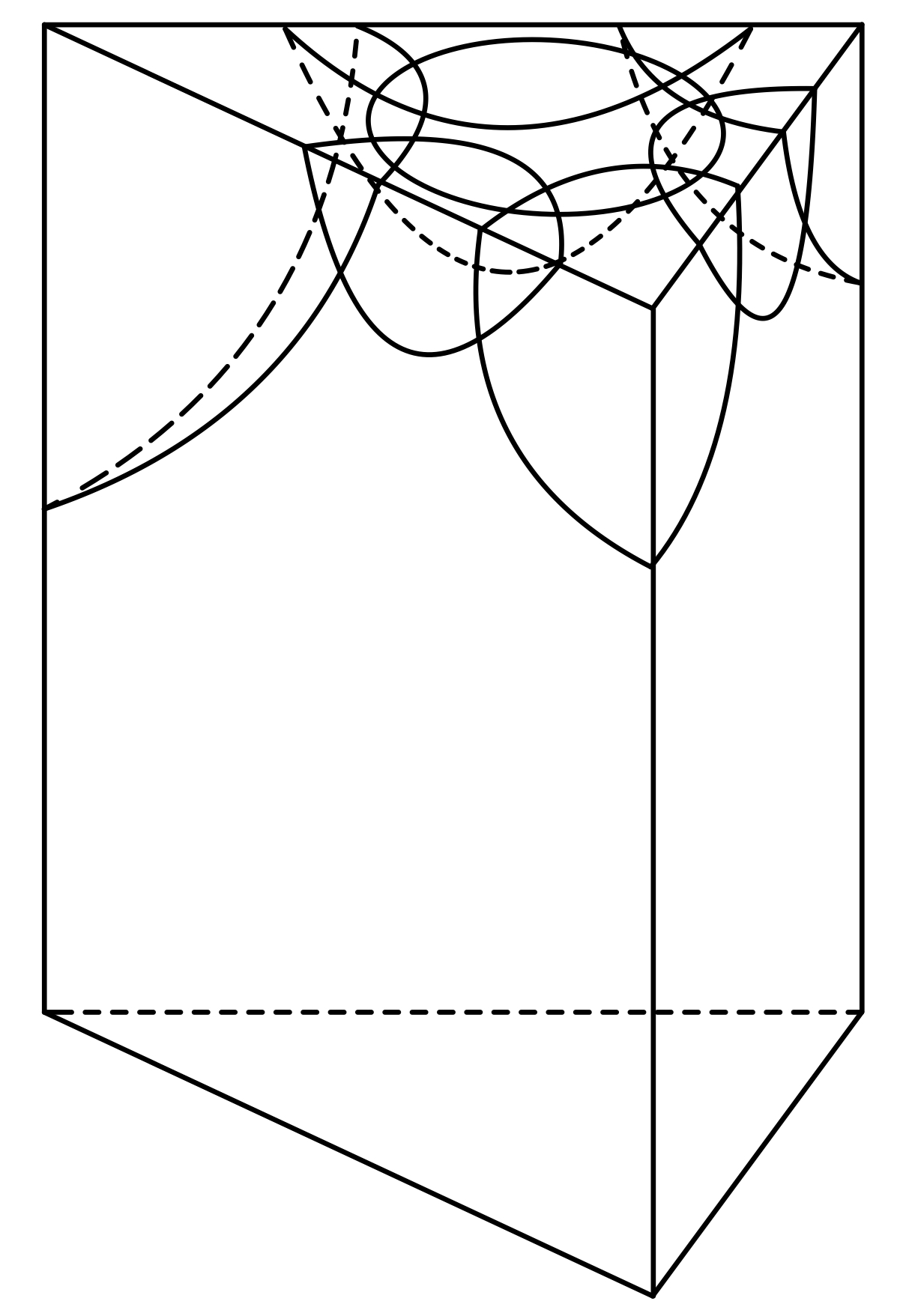}
 \end{center}
  \caption[]{
  A part of the open covering of $\mathrm{L}(3,1)\times S^1$. The bottom and top triangles represent the surface of the ball, and the vertical direction represents the $S^1$ direction. The open covers consist of one open ball centered at the center of the prism, one open ball centered at the vertex, two open balls centered at the midpoints of the edges, and two open balls centered at the midpoints of the faces.}
\label{fig:open cover of lens times s1}
\end{figure}

To integrate the discrete Berry phase, we need to take an open cover of $X=\mathrm{L}(3,1)\times S^{1}$. Since $\mathrm{L}(3,1)\times S^{1}$ is a $4$-dimensional manifold, we cannot draw a picture similar to the $\rp{2}\times S^1$ model. Thus, instead of drawing the entire $\mathrm{L}(3,1)\times S^1$, we draw the direct product of the surface of the ball\footnote{Strictly speaking, the surface of the ball is not a manifold. In fact, the neighborhood of the point on the equator is not homeomorphic to the Euclidian space, because there are three directions to move. However, for the purpose of performing the integration, it is sufficient to know the cocycle on the integration surface $\gamma\times S^1$, so in this paper, it is sufficient to examine its surface instead of the ball.} and $S^{1}$, and take an open cover as in the Fig.\ref{fig:open cover of lens times s1}.

Next, we take a polyhedral decomposition $\tt$ of $\mathrm{L}(3,1)\times S^{1}$ which is compatible with the open covering $\oc$. We take the polyhedral decomposition $\tt$ as in Fig.\ref{fig:triangulation of lens space times S1}:
\begin{figure}[H]
 \begin{center}
  \includegraphics[width=50mm]{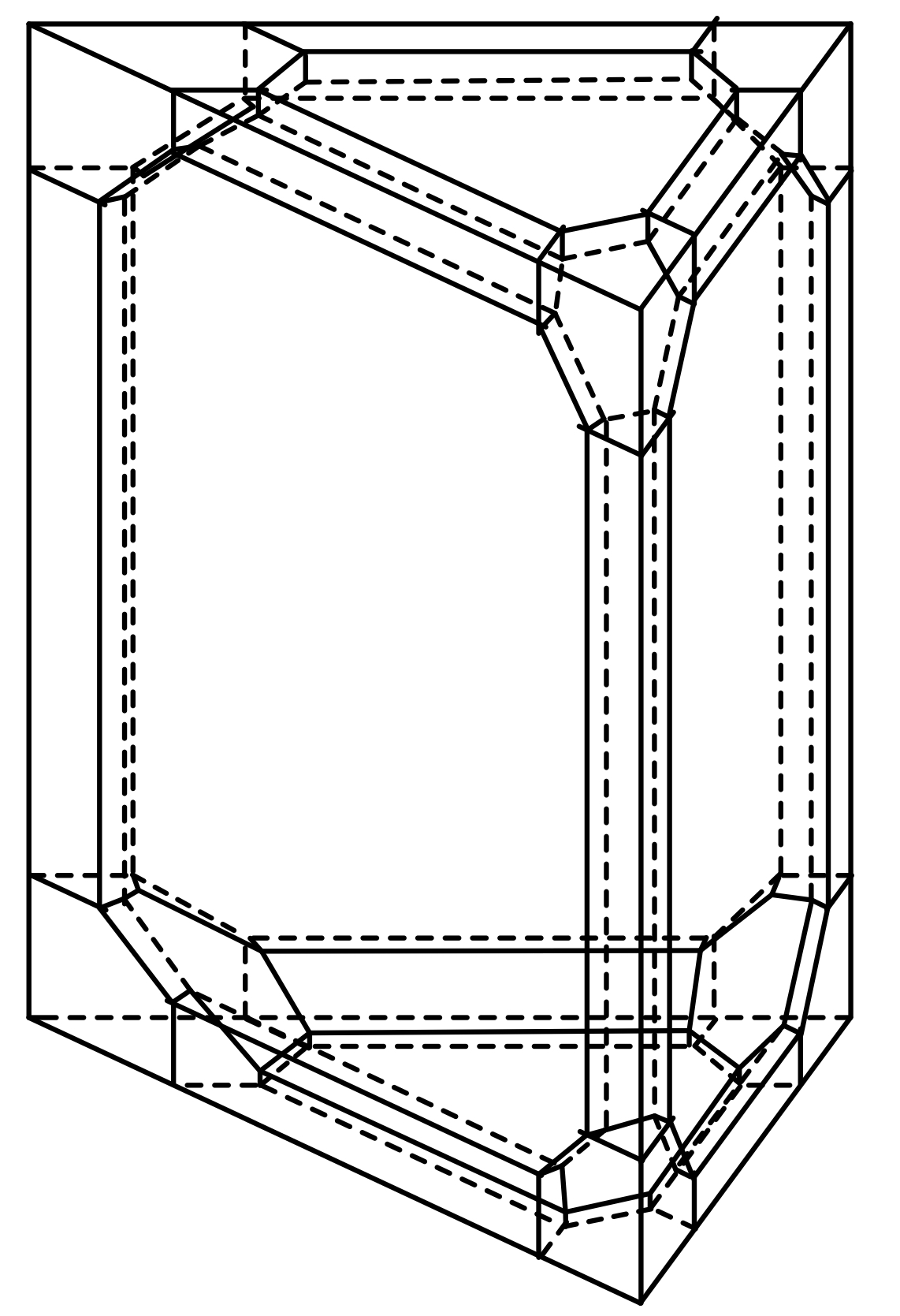}
 \end{center}
  \caption[]{
  A polyhedral decomposition of $\mathrm{L}(3,1)\times S^1$.
}
\label{fig:triangulation of lens space times S1}
\end{figure}

\begin{figure}[H]
 \begin{center}
  \includegraphics[width=50mm]{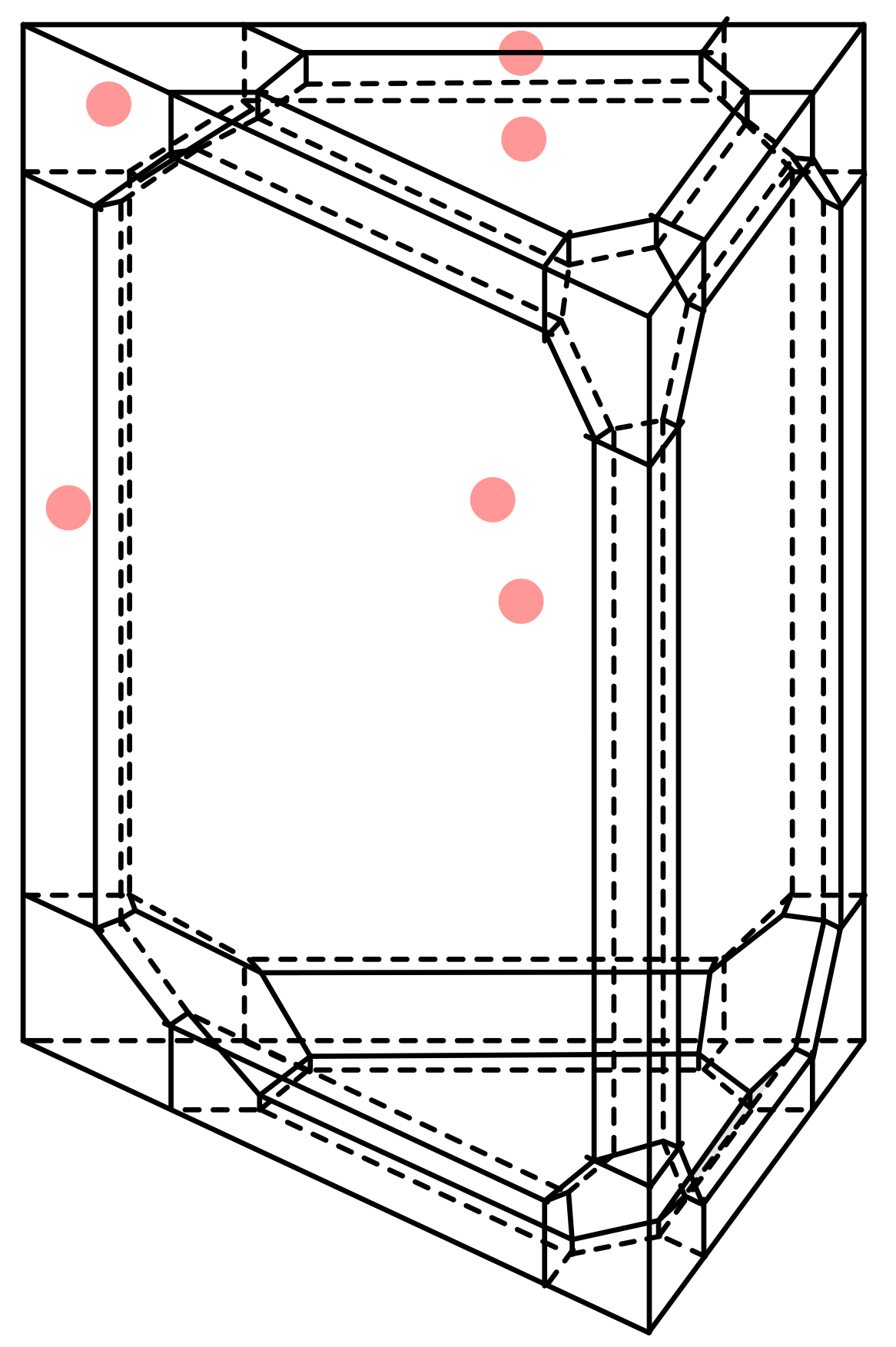}
 \end{center}
  \caption[]{A base point of the polyhedral decomposition of $Y$. The two in the middle belong to the back patch and the center patch of the prism, respectively. 
}
\label{fig:basedtriangulation of lens space}
\end{figure}

The open set corresponding to a simplex $\tau$ in $\tt$ is written by $U_{i_\tau}$. To implement the injective MPS bundle on this triangular prism, we assign transition functions on the faces of the polyhedral decomposition. We would like to perform this assignment systematically. To do this, we take base points of each patch, and define  transition functions on the intersection $U_{12}$ of patches $U_{1}$ and $U_{2}$ under the following rules:
\begin{enumerate}
    \item Take a path starting from the base point of the patch $U_{1}$ and passing through $U_{12}$ and terminating at the base point of $U_{2}$.
    \item If the path is through the side of the triangular prism, we give $\tilde{g}_{\mathrm{L}(3,1)}$, and if  the path is through the top or bottom, we give $\tilde{g}_{S^1}$.
\end{enumerate}
Under these assignment rules, the configuration of the transition function of the injective MPS bundle is determined by fixing the base points of each patch. We fix the base points as in Fig.\ref{fig:basedtriangulation of lens space}. In the following, we formally write the transition functions as $\{g_{\alpha\beta}:U_{\alpha}\cap U_{\beta}\to\PU{3}\}$.

\begin{figure}[H]
 \begin{center}
  \includegraphics[width=90mm]{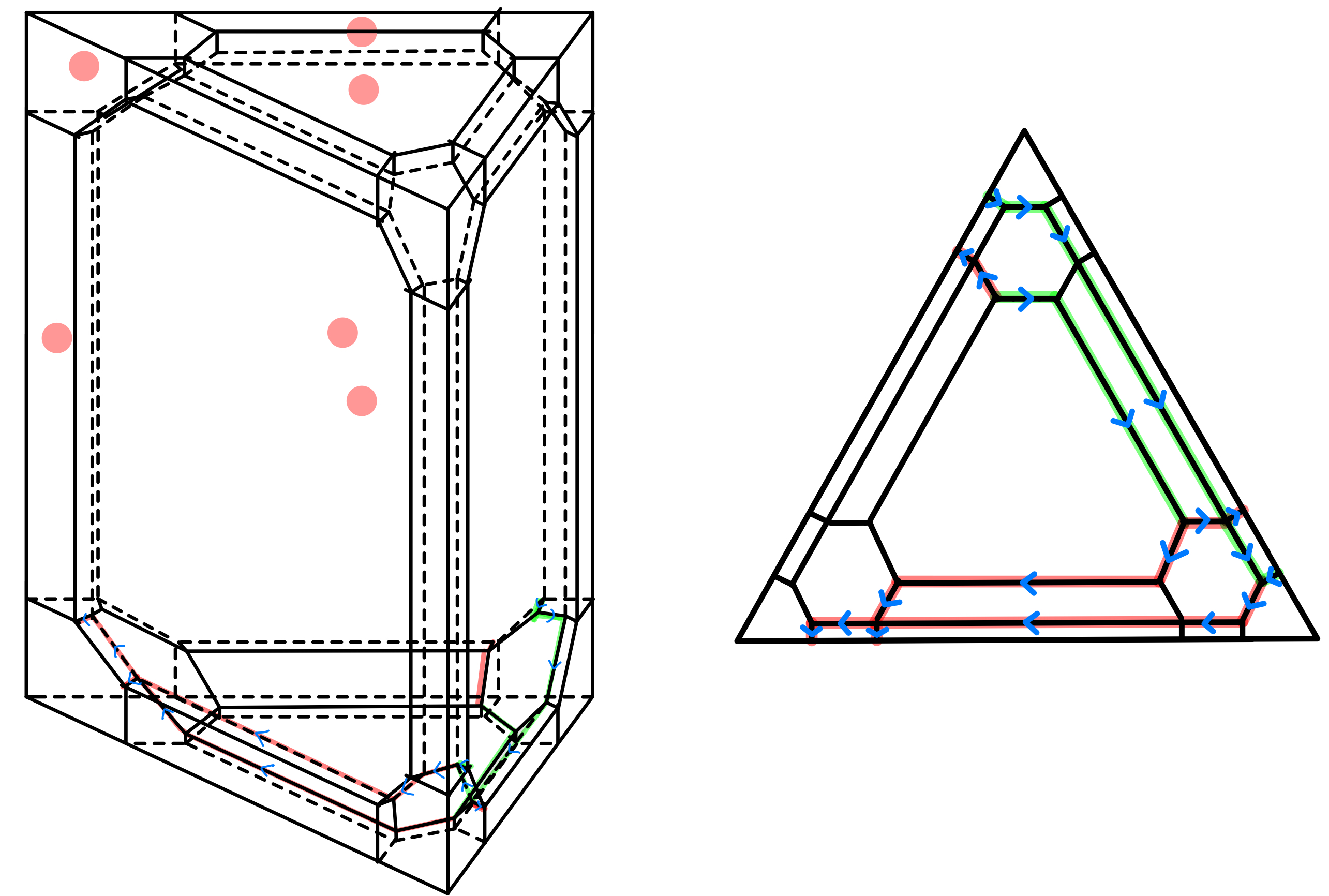}
 \end{center}
  \caption[]{The configuration of $c_{\alpha\beta\gamma}$. (Left)  $c_{\alpha\beta\gamma}=\omega^2$ on red lines,  $c_{\alpha\beta\gamma}=\omega$ on green lines, and $c_{\alpha\beta\gamma}=1$ on the others. (Right) Configuration of $c_{\alpha\beta\gamma}$ projected on the bottom.
}
\label{fig:DD class lens}
\end{figure}

We take a lift of the transition functions as follows:
\begin{eqnarray}
\left[1_3\right]\mapsto1_3, \tilde{g}_{\mathrm{L}(3,1)}\mapsto
\begin{pmatrix}
\omega&&\\
&1&\\
&&\omega^{2}
\end{pmatrix}, 
\tilde{g}_{S^1}\mapsto
\begin{pmatrix}
&1&\\
&&\omega^{2}\\
\omega&&
\end{pmatrix},
\tilde{g}_{\mathrm{L}(3,1)}\tilde{g}_{S^1}\mapsto
\begin{pmatrix}
\omega&&\\
&1&\\
&&\omega^{2}
\end{pmatrix} 
\begin{pmatrix}
&1&\\
&&\omega^{2}\\
\omega&&
\end{pmatrix},
\end{eqnarray}
Under these lifts, the violation of the cocycle condition can only occur if the surfaces with the transition functions $\tilde{g}_{\mathrm{L}(3,1)}$,  $\tilde{g}_{S^1}$, and $\tilde{g}_{\mathrm{L}(3,1)}\tilde{g}_{S^1}=\tilde{g}_{S^1}\tilde{g}_{\mathrm{L}(3,1)}$ intersect each other, and the violation occurs by $\omega$ or $\omega^2$ on them. Thus, we can easily identify the edge with nontrivial $c_{\alpha\beta\gamma}$ as in Fig.\ref{fig:DD class lens}. Note that $c_{\alpha\beta\gamma}$ defines a cohomology class $\left[c_{\alpha\beta\gamma}\right]\in\mathrm{H}^{2}(\mathrm{L}(3,1)\times S^{1};\mathrm{U}(1))\simeq\cohoZ{3}{\mathrm{L}(3,1)\times S^1}\simeq\Zmod{3}$ and a cohomology class $[(c_{\alpha\beta\gamma},0,0)]  \in \mathrm{H}^3(\mathrm{L}(3,1)\times S^{1};\D(3))$ which is flat. The group $\cohoU{2}{\mathrm{L}(3,1)\times S^{1}}$ is a subgroup of $\mathrm{H}^3(\mathrm{L}(3,1)\times S^{1};\D(3))$ under the map $\left[c_{\alpha\beta\gamma}\right] \mapsto [(c_{\alpha\beta\gamma},0,0)]$.

\begin{figure}[H]
 \begin{center}
  \includegraphics[width=40mm]{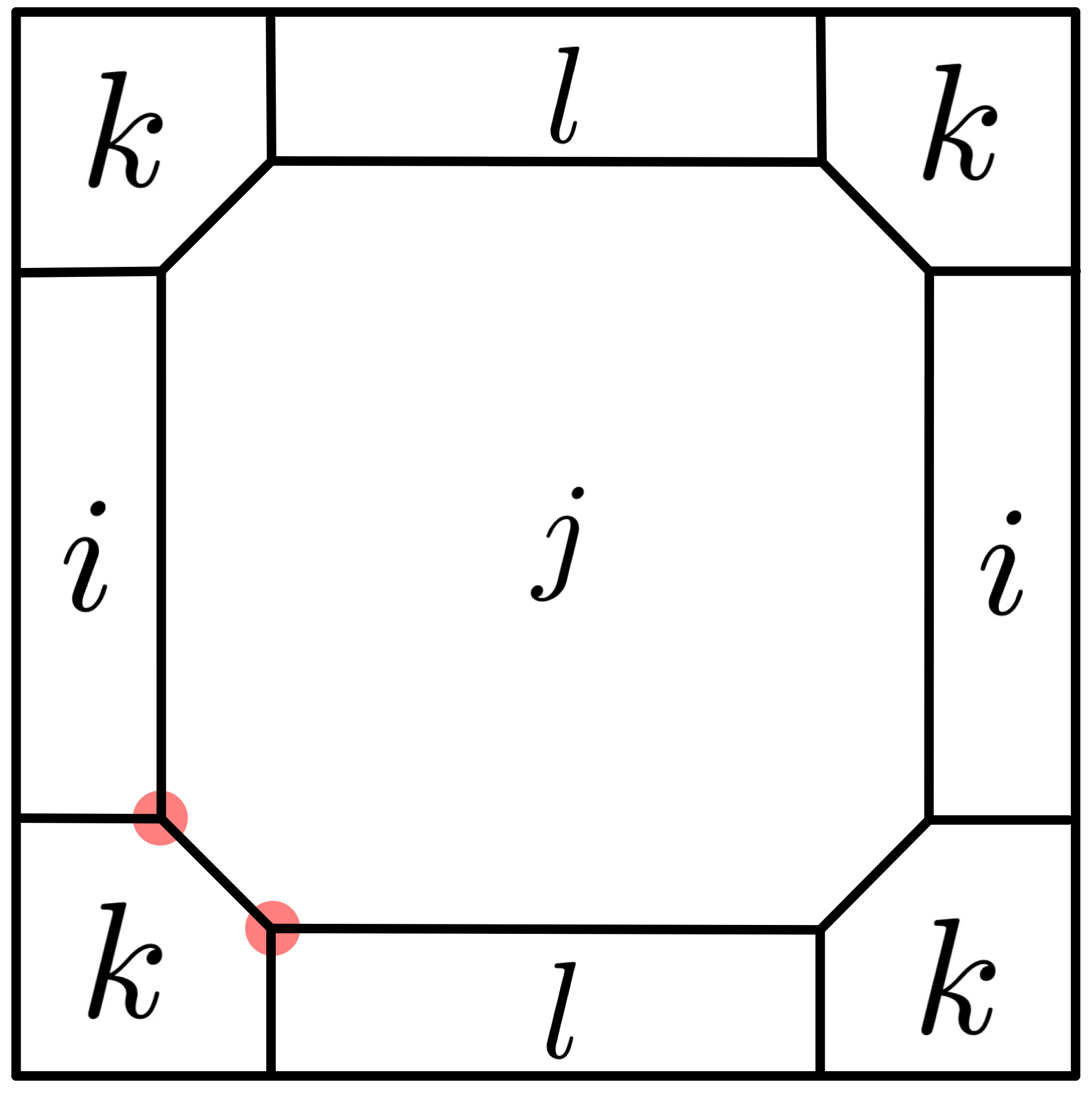}
 \end{center}
  \caption[]{The triangulation of $Y$ induced from that of $\mathrm{L}(3,1)\times S^1$, and vertices on which the Dixmier-Douady class has a value $\omega^2$.
}
\label{fig:nontrivial cocycle on Y Z3}
\end{figure}

\begin{figure}[H]
 \begin{center}
  \includegraphics[width=40mm]{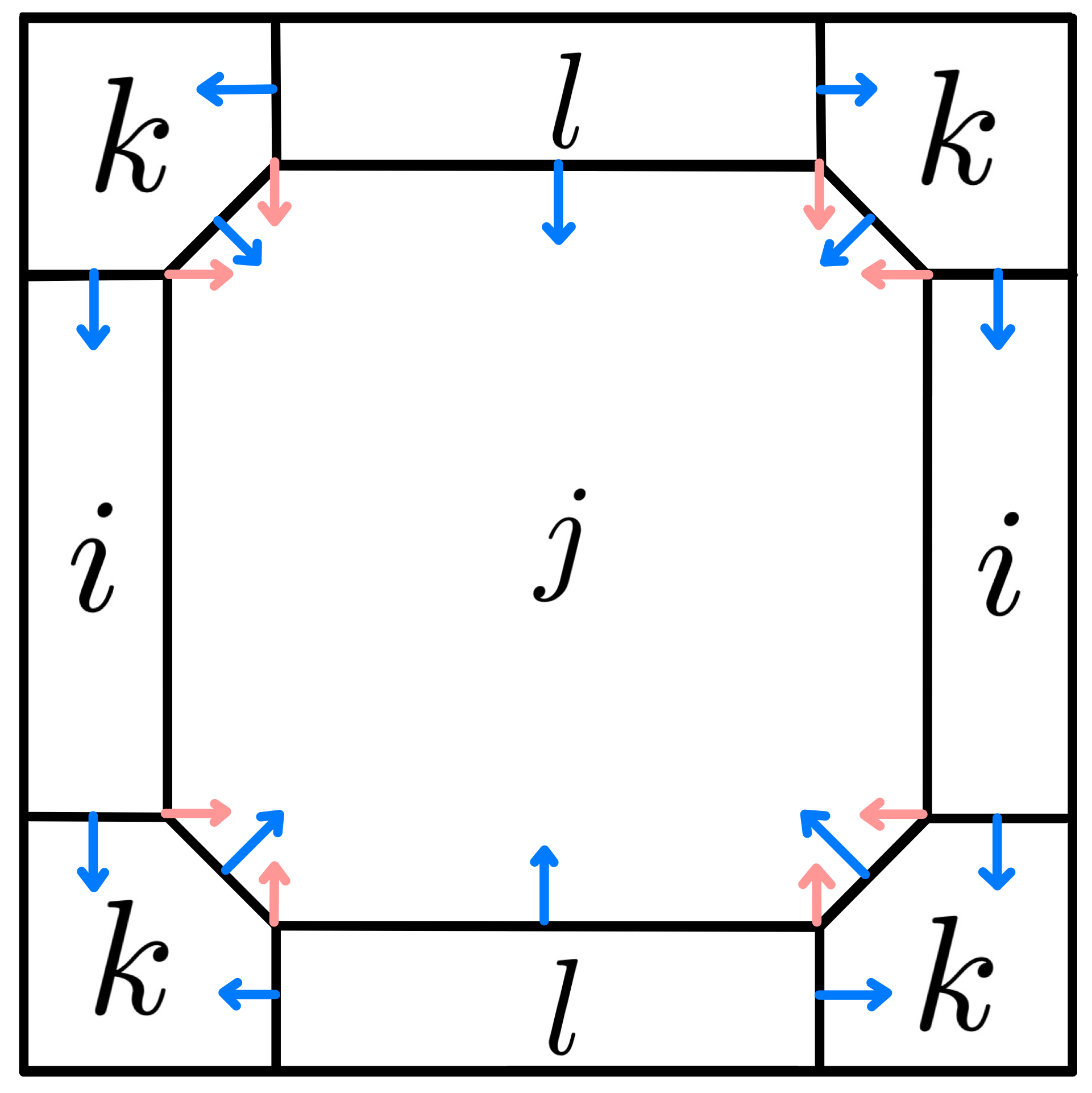}
 \end{center}
  \caption[]{An index map of the triangulation of $Y$. Blue arrows are index maps for edges, and the direction of it indicates the patch to which the edge belongs. Similarly, orange arrows are index maps for vertices, and the direction of it indicates the patch to which the vertex belongs.
}
\label{fig:index map on Y Z3}
\end{figure}

\begin{figure}[H]
 \begin{center}
  \includegraphics[width=40mm]{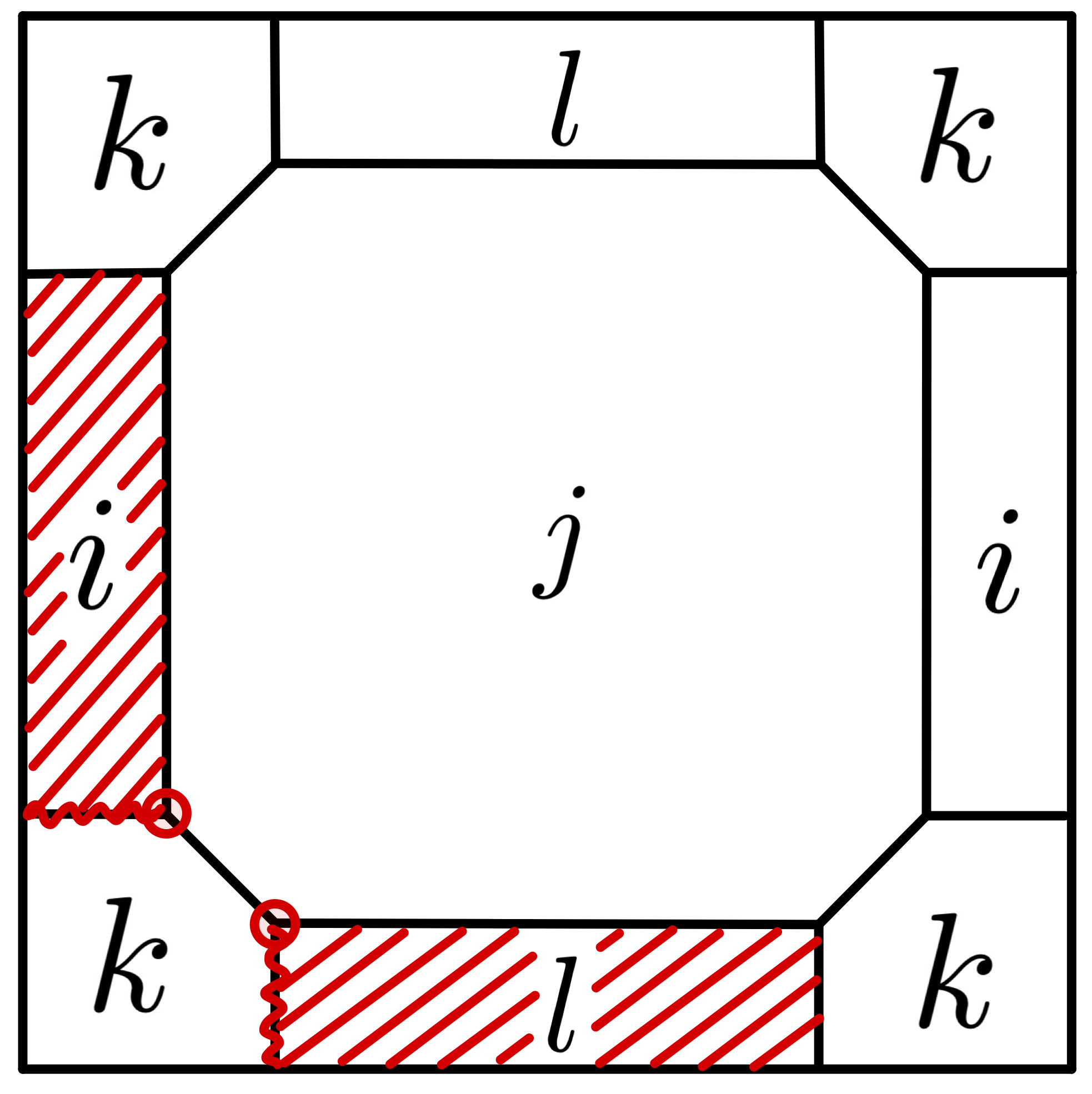}
 \end{center}
  \caption[]{
  Flags contributing to integration. In the bottom flag, the red circle represents $s^0$, the red wavy line represents $s^1$, and the red shaded face represents $s^2$. In the left flag, the red circle represents $\tilde{s}^0$, the red wavy line represents $\tilde{s}^1$, and the red shaded face represents $\tilde{s}^2$.
}
\label{fig:nontrivial flag Z3}
\end{figure}

Let's compute the higher holonomy of $c=(c_{\alpha\beta\gamma},0,0)$ along this surface $Y=\gamma\times S^1$. We take the front edge of the top triangle of the prism as a nontrivial path $\gamma$ in $\mathrm{L}(3,1)$ and take a polyhedral decomposition\footnote{Although what is shown in Fig.\ref{fig:nontrivial cocycle on Y Z3} is a polyhedral decomposition, we theoretically consider a triangulation that subdivides it and assume that the index map takes the same value for all simplices in each polyhedron.} of $Y$ induced from that of $\mathrm{L}(3,1)\times S^1$. We label the open sets of $Y$ with Roman letters ($i,j,k,...$) instead of Greek letters ($\alpha,\beta,\gamma,...$). We show the nontrivial cocycle on $Y$ in the Fig.\ref{fig:nontrivial cocycle on Y Z3}.
It is necessary to take an index map $\phi$ to perform an integration. However, to compute the invariant on $Y$, it is sufficient to determine the index map on $Y$. We take an index map as in the Fig.\ref{fig:index map on Y Z3}.
Then the higher holonomy is given by the following  formula:
\begin{eqnarray}
\hol_Y(c)=\prod_{\si=(\si^0 \subset \si^1 \subset \si^2) \in F(2)} c_{\phi_{\si^{2}} \phi_{\si^{1}} \phi_{\si^{0}}} (\si^0).
\end{eqnarray}
As we saw in Fig.\ref{fig:nontrivial cocycle on Y Z3}, the Dixmier-Douady class takes nontrivial values only on $U_{jkl}$ and $U_{ijk}$. Moreover, a full flag $\si=(\si^0 \subset \si^1 \subset \si^2)$ satisfying $\{ i,k,j \}=\{\phi_{\si^{2}},  \phi_{\si^{1}}, \phi_{\si^{0}}\}$ is only one $(s^0 \subset s^1 \subset s^2)$ and $\{ l,k,j \}=\{\phi_{\si^{2}},  \phi_{\si^{1}}, \phi_{\si^{0}}\}$ is only one $(\tilde{s}^0 \subset \tilde{s}^1 \subset \tilde{s}^2)$. We show these flags in Fig.\ref{fig:nontrivial flag Z3}. Therefore, we have 
\begin{align}
\hol_Y(c)&=\prod_{\si=(\si^0 \subset \si^1 \subset \si^2) \in F(2)} c_{\phi_{\si^{2}} \phi_{\si^{1}} \phi_{\si^{0}}} (\si^0),\\
&=c_{\phi_{s^{2}} \phi_{s^{1}} \phi_{s^{0}}} (s^0) c_{\phi_{\tilde{s}^{2}} \phi_{\tilde{s}^{1}} \phi_{\tilde{s}^{0}}} (\tilde{s}^0),\\
&=c_{ikj}c_{lkj}, \\
&=\omega^{2}\omega^{2},\\
&=\omega.
\end{align}
Since the cocycle $c$ is flat, there is no correction by the $3$-form curvature. As a result, we have that the higher pump invariant is 
\begin{eqnarray}
n_{\rm top.}(Y)=\hol_Y(c)=\omega\in\zmod{3}.
\end{eqnarray}
Therefore, the higher pump invariant is nontrivial.

\section{Discussions and Future Directions}

In this paper, we investigated a higher pumping phenomenon by constructing two models: the model parametrized by $\rp{2}\times S^1$ and the model parametrized by $\mathrm{L}(3,1)\times S^1$. We obtain these models by deforming models in a nontrivial SPT phase with $\zmod{2}\times\zmod{2}$-symmetry and $\zmod{3}\times\zmod{3}$-symmetry, respectively. As a generalization, it is expected to be possible to construct a model parametrized by (a subspace of) $BG$ based on a model in the nontrivial SPT phase with $G$-symmetry. It is an interesting problem to develop such a model construction method. 

Also, the boundary condition obstacle discussed in Sec.\ref{sec:obstacle} and Sec.\ref{sec:Z/3 obstacle} seems to be related to an anomaly of the edge theory with parameter\cite{CFLS20-1,CFLS20-2}. It is an interesting problem to consider the bulk-anomaly correspondence from an MPS perspective.

\begin{acknowledgments}
We would like to thank K. Gomi, Y. Kubota, S. Terashima and K. Yonekura for valuable discussions. S.O. acknowledge helpful discussions with T. Ando, M. Furuta, H. Kanno, R. Kobayashi,  S. Ryu, M. Sato, A. Turzillo and M. Yamashita. S.O. was supported by the establishment of university fellowships towards the creation of science technology innovation, Grant Number JPMJFS2123. K.S. was supported by The Kyoto University Foundation.
This work was supported by JST CREST Grant No. JPMJCR19T2 and JSPS KAKENHI Grant Number JP21K03240, JP22H01117.
\end{acknowledgments}

\appendix
\def\thesection{\Alph{section}}

\section{Other Boundary Conditions}\label{sec:boundary condition}

In our model, general boundary term is given by $\tau^{\delta}_{1/2}(\vec{n})\sigma^{z}_{1}$, where $\tau^{\delta}_{1/2}(\vec{n}):=\cos(\delta)\tau^{x}_{1/2}(\vec{n})+\sin(\delta)\tau^{y}_{1/2}(\vec{n})$. Let's consider the following initial and final Hamiltonians
\begin{eqnarray}\label{eq:model_with_boundary_0 delta}
 H^{\delta}_{\rm in.}(\Vec{n})=-\tau^{\delta}_{\frac{1}{2}}(\vec{n})\sigma^{z}_{1}-\sum_{j=1,2,...}\tau^{z}_{j-\frac{1}{2}}(\Vec{n})\sigma^{x}_{j}\tau^{z}_{j+\frac{1}{2}}(\Vec{n})-\sum_{j=1,2,...}\sigma^{z}_{j}\tau^{x}_{j+\frac{1}{2}}(\Vec{n})\sigma^{z}_{j+1},
\end{eqnarray}
and 
\begin{eqnarray}\label{eq:model_with_boundary_1 delta}
 H^{\delta}_{\rm fin.}(\Vec{n})=\tau^{\delta}_{\frac{1}{2}}(\vec{n})\sigma^{z}_{1}-\sum_{j=1,2,...}\tau^{z}_{j-\frac{1}{2}}(\Vec{n})\sigma^{x}_{j}\tau^{z}_{j+\frac{1}{2}}(\Vec{n})-\sum_{j=1,2,...}\sigma^{z}_{j}\tau^{x}_{j+\frac{1}{2}}(\Vec{n})\sigma^{z}_{j+1}.
\end{eqnarray}
Under this boundary condition, we can check that the ratio $r^{\delta}$ of the holonomy defined by
\begin{eqnarray}\label{eq:ratio delta}
r^{\delta}=\frac{n^{\delta}_{\rm in.}(\gamma)}{n^{\delta}_{\rm fin.}(\gamma)}=\exp(\int_{\gamma}(A^{\delta}_{\rm in.}-A^{\delta}_{\rm fin.})-\frac{1}{2}\int_{\Sigma}(dA^{\delta}_{\rm in.}-dA^{\delta}_{\rm fin.}))\frac{\braket{\mathrm{G.S.}^{\delta}_{\rm in.}(\gamma_{0})|\mathrm{G.S.}^{\delta}_{\rm in.}(\gamma_{1})}}{\braket{\mathrm{G.S.}^{\delta}_{\rm fin.}(\gamma_{0})|\mathrm{G.S.}^{\delta}_{\rm fin.}(\gamma_{1})}},
\end{eqnarray}
is equal to $-1$. Here, $\Sigma$ is a bounding manifold of $2\cdot\gamma$. Let $A^{\delta}_{\rm in.}(\vec{n})$ and $A^{\delta}_{\rm fin.}(\vec{n})$ be the Berry connections of the above Hamiltonians. Then the difference between these connections is
\begin{eqnarray}
A^{\delta}_{\rm in.}(\vec{n})-A^{\delta}_{\rm fin.}(\vec{n})=\frac{1}{4}(\bra{\mathrm{Ref}^{\delta}_{\rm in.}}f_{1}h_{\frac{1}{2}}(\vec{n})f_{1}\ket{\mathrm{Ref}^{\delta}_{\rm in.}}
-\bra{\mathrm{Ref}^{\delta}_{\rm fin.}}f_{1}h_{\frac{1}{2}}(\vec{n})f_{1}\ket{\mathrm{Ref}^{\delta}_{\rm fin.}}),
\end{eqnarray}
where $\ket{\mathrm{Ref}^{\delta}_{\rm in.}}$ is a simultaneous eigenstate of $\tau^{\delta}_{\frac{1}{2}}\sigma^{z}_{1}$ and $\sigma^{z}_{j}\tau^{x}_{j+\frac{1}{2}}\sigma^{z}_{j+1}$ with eigenvalue $1$, and $\ket{\mathrm{Ref}^{\delta}_{\rm fin.}}$ is a simultaneous eigenstate of $-\tau^{\delta}_{\frac{1}{2}}\sigma^{z}_{1}$ and $\sigma^{z}_{j}\tau^{x}_{j+\frac{1}{2}}\sigma^{z}_{j+1}$ with eigenvalue $1$. By doing the same computation as in Sec.\ref{sec:chern number pump}, we obtain that
\begin{eqnarray}
A^{\delta}_{\rm in.}(\vec{n})-A^{\delta}_{\rm fin.}(\vec{n})=\frac{1}{4}(\bra{\mathrm{Ref}^{\delta}_{\rm in.}}(1+\tau^{z}_{\frac{1}{2}})h_{\frac{1}{2}}(\vec{n})(1+\tau^{z}_{\frac{1}{2}})\ket{\mathrm{Ref}^{\delta}_{\rm in.}}
-\bra{\mathrm{Ref}^{\delta}_{\rm fin.}}(1+\tau^{z}_{\frac{1}{2}})h_{\frac{1}{2}}(\vec{n})(1+\tau^{z}_{\frac{1}{2}})\ket{\mathrm{Ref}^{\delta}_{\rm fin.}})=0.\nonumber
\end{eqnarray}
Let $\ket{{\rm G.S.}^{\delta}_{\rm in.}(\vec{n})}$ and $\ket{{\rm G.S.}^{\delta}_{\rm fin.}(\vec{n})}$ be the ground state of the Hamiltonians Eq.(\ref{eq:model_with_boundary_0 delta}) and Eq.(\ref{eq:model_with_boundary_1 delta}). Then, they meet $\ket{{\rm G.S.}^{\delta}_{\rm fin.}(\vec{n})}\propto\tau^{z}_{\frac{1}{2}}(\vec{n})\ket{{\rm G.S.}^{\delta}_{\rm in.}(\vec{n})}$, since $\tau^{z}_{\frac{1}{2}}(\vec{n})$ is anti-commute with $\tau^{\delta}_{\frac{1}{2}}(\vec{n})$. Therefore, 
\begin{eqnarray}
\braket{\mathrm{G.S.}^{\delta}_{\rm fin.}(\vec{n})|\mathrm{G.S.}^{\delta}_{\rm fin.}(-\vec{n})}&=&\bra{\mathrm{G.S.}^{\delta}_{\rm in.}(\vec{n})}\tau^{z}_{\frac{1}{2}}(\vec{n})\tau^{z}_{\frac{1}{2}}(-\vec{n})\ket{\mathrm{G.S.}^{\delta}_{\rm in.}(-\vec{n})},\\
&=&-\braket{\mathrm{G.S.}^{\delta}_{\rm in.}(\vec{n})|\mathrm{G.S.}^{\delta}_{\rm in.}(-\vec{n})},
\end{eqnarray}
and, consequently, the ratio of the holonomy is 
\begin{eqnarray}
r^{\delta}=\frac{n^{\delta}_{\rm in.}(\gamma)}{n^{\delta}_{\rm fin.}(\gamma)}=-1.
\end{eqnarray}

\section{A Comment on Unitary Matrices $\tilde{V}_{\tau}(\vec{z})$ and $\tilde{V}_{\sigma}(t)$}\label{sec:comment on uni}

In Sec.\ref{sec:def of Z/3 cluster model}, we introduced unitary matrices $\tilde{V}_{\tau}(\vec{z})$ and $\tilde{V}_{\sigma}(t)$ defined in Eq.(\ref{eq:Vtauz}) and Eq.(\ref{eq:Vsigmat}) without explanation. Here we comment on the background of its construction.

First, we would like to give $\mathrm{L}(3,1)$ dependence to $\tau$-sites. To this end, we define a unitary matrix
\begin{eqnarray}
V_{x}&:=&\frac{1}{\sqrt{3}}\begin{pmatrix}
1&\omega^2&\omega\\
1&1&1\\
1&\omega&\omega^2
\end{pmatrix},
\end{eqnarray}
that diagonalizes $\tilde{\tau}^{x}$:
\begin{eqnarray}
\ket{\bar{u}_{0}}:=V_{x}\ket{u_{0}}=\begin{pmatrix}
1\\
0\\
0
\end{pmatrix},
\ket{\bar{u}_{1}}:=V_{x}\ket{u_{1}}=\begin{pmatrix}
0\\
1\\
0
\end{pmatrix},
\ket{\bar{u}_{2}}:=V_{x}\ket{u_{2}}=\begin{pmatrix}
0\\
0\\
1
\end{pmatrix}.
\end{eqnarray}
Under this basis, we mix $\ket{\bar{u}_{1}}$ and $\ket{\bar{u}_{2}}$ by $\mathrm{SU}(2)$ transformation. Let $\vec{z}:=(z_1,z_2)$ be complex numbers such that $\left|z_1\right|^2+\left|z_2\right|^2=1$, which is coordinates of $\mathrm{SU}(2)$. We also define a unitary matrix
\begin{eqnarray}
U(\vec{z})=\begin{pmatrix}
1&0&0\\
0&z_1&-z_2^{\ast}\\
0&z_2&z_1^{\ast}
\end{pmatrix},
\end{eqnarray}
and 
\begin{eqnarray}
\ket{\bar{u}_{0}(\vec{z})}:=U(\vec{z})\ket{\bar{u}_{0}}=\begin{pmatrix}
1\\
0\\
0
\end{pmatrix},
\ket{\bar{u}_{1}(\vec{z})}:=U(\vec{z})\ket{\bar{u}_{1}}=\begin{pmatrix}
0\\
z_1\\
z_2
\end{pmatrix},
\ket{\bar{u}_{2}(\vec{z})}:=U(\vec{z})\ket{\bar{u}_{2}}=\begin{pmatrix}
0\\
-z_2^{\ast}\\
z_1^{\ast}
\end{pmatrix}.
\end{eqnarray}
Finally, let's get back to the $z$-basis:
\begin{eqnarray}
\ket{u_{i}(\vec{z})}&:=&V_{x}^{\dagger}\ket{\bar{u}_{i}(\vec{z})}=V_{x}^{\dagger}U(\vec{z})V_{x}\ket{u_{i}}.
\end{eqnarray}
In fact, 
\begin{eqnarray}
\tilde{V}_{\tau}(\vec{z})=V_{x}^{\dagger}U(\vec{z})V_{x}.    
\end{eqnarray}
This is the origin of the unitary matrix $\tilde{V}_{\tau}(\vec{z})$. 

Next, we give $S^{1}$ dependence to $\sigma$-sites. To this end, we interpolate $1_{3}$ and $\tilde{\sigma}_{x}$. For a unitary matrix 
\begin{eqnarray}
W={\frac{1}{\sqrt{3}}}\begin{pmatrix}
1&\omega^{2}&\omega\\
1&\omega&\omega^2\\
1&1&1
\end{pmatrix},
\end{eqnarray}
$\tilde{\sigma}^{x}$ satisfies
\begin{eqnarray}
\tilde{\sigma}^{x}=W\begin{pmatrix}
1&&\\
&\omega&\\
&&\omega^{2}
\end{pmatrix}W^{\dagger}.
\end{eqnarray}
$\tilde{V}_{\sigma}(t)$ is a path connecting $1_2$ and $\tilde{\sigma}^x$ as following way:
\begin{eqnarray}
\tilde{V}_{\sigma}(t)=W\begin{pmatrix}
1&&\\
&\exp(i\frac{t}{3})&\\
&&\exp(i\frac{2t}{3})
\end{pmatrix}W^{\dagger}.
\end{eqnarray}
This is the origin of the unitary matrix $\tilde{V}_{\sigma}(\vec{z})$.

\section{Complex Line Bundle}\label{sec:complex line bundle}

Let $X$ be a parameter space\footnote{Strictly speaking, we assume that $X$ is compact Hausdorff space.} and let $L\to X$ be a complex line bundle over $X$. It is well known that a complex line bundle over $X$ is classified by $\cohoZ{2}{X}$. Here, $\cohoZ{2}{X}$ is the $2$nd cohomology group with coefficient $\mathbb{Z}$. Since $\cohoZ{2}{X}$ is a finitely generated abelian group, there are integers $k,l\in \mathbb{N}$ so that
\begin{eqnarray}\label{eq:cpx line bdl isom}
\cohoZ{2}{X}\simeq\mathbb{Z}^{\oplus k}\oplus\zmod{p_1}\oplus\cdots\zmod{p_l},
\end{eqnarray}
where $\{p_{i}\}_{i=1}^{l}$ is a set of prime numbers. We define
\begin{eqnarray}
\cohoZ{2}{X}_{\rm free}&:=&\mathbb{Z}^{\oplus k},\\
\cohoZ{2}{X}_{\rm tor.}&:=&\zmod{p_1}\oplus\cdots\zmod{p_l}.
\end{eqnarray}
$\cohoZ{2}{X}_{\rm free}$ is called the free part of $\cohoZ{2}{X}$ and $\cohoZ{2}{X}_{\rm tor.}$ is called the torsion part of $\cohoZ{2}{X}$.
In this section, we review the way to extract this data for a given complex line bundle over $X$ numerically, i.e., the way to identify the image of $L\to X$ under the isomorphism  Eq.(\ref{eq:cpx line bdl isom}). 

Fix an open covering $\{U_{\alpha}\}_{\alpha\in I}$ of $X$. The topological class $\left[L\right]$ is determined by the transition function $\{g_{\alpha\beta}:U_{\alpha\beta}\to\mathrm{U}(1)\}$ which satisfies the cocycle condition
\begin{eqnarray}
g_{\alpha\beta}g_{\beta\gamma}=g_{\alpha\gamma}.
\end{eqnarray}
In fact, $\{g_{\alpha\beta}\}$ determine an element of the $1$st sheaf cohomology group with coefficient $\mathrm{U}(1)$
\begin{eqnarray}
\left[g_{\alpha\beta}\right]\in\coho{1}{X}{\underline{\mathrm{U}(1)}},
\end{eqnarray}
and since $\coho{1}{X}{\underline{\mathrm{U}(1)}}\simeq\cohoZ{2}{X}$, we obtain the element of $\cohoZ{2}{X}$ for given $L\to X$ mathematically. However, this construction is a little abstract and it is not clear to which number of the right-hand side of Eq.(\ref{eq:cpx line bdl isom}) the given $L$ corresponds. A connection and a curvature are useful tools to compute this number numerically. 

A connection on a line bundle $L\to X$ is a set of $1$-form $\{A_{\alpha}\}_{\alpha\in I}$ such that $A_{\beta}=A_{\alpha}-g_{\alpha\beta}^{\dagger} dg_{\alpha\beta}$ on nonempty intersection $U_{\alpha\beta}:=U_{\alpha}\cap U_{\beta}$. Then $\{F_{\alpha}:=dA_{\alpha}\}$ is called a curvature form of the connection $\{A_{\alpha}\}_{\alpha\in I}$. 

\begin{itemize}
    \item The free part of $\cohoZ{2}{X}$:
    
    Since $F_{\alpha}-F_{\beta}=d(g_{\alpha\beta}^{\dagger} dg_{\alpha\beta})=0$ on $U_{\alpha\beta}$, $\{F_{\alpha}\}$ define a global $2$-form $F$, which is also called a curvature form. We can also show that an integration of $F$ for any closed surface $\Sigma$ takes value in $2\pi i\mathbb{Z}$:
    \begin{eqnarray}\label{eq:free part inv}
    n(\Sigma):=\int_{\Sigma}\frac{F}{2\pi i}\in \mathbb{Z}.
    \end{eqnarray}
    
    By using the universal coefficient theorem, the free part of $\cohoZ{2}{X}$ is isomorphic to the free part of $\mathrm{H}_{2}(X;\mathbb{Z})$. If we would like to know the $i$-th component of $\cohoZ{2}{X}$, we compute the integration Eq.(\ref{eq:free part inv}) over the surface which generates the $i$-th component of $\mathrm{H}_{2}(X;\mathbb{Z})$. This is the way to compute the free part of the line bundle.
    
    \item The torsion part of $\cohoZ{2}{X}$:
    
    Let $\gamma$ be a closed path in $X$ such that $p$ copies of $\gamma$ is trivial in the homology group $\mathrm{H}_{1}(X;\mathbb{Z})$:
    \begin{eqnarray}
    p\cdot\left[\gamma\right]=0\in\mathrm{H}_{1}(X;\mathbb{Z}).
    \end{eqnarray}
    Then, we have a surface $\Sigma$ such that $\partial\Sigma=p\cdot\gamma$. Let $\{\tilde{U}_{i}\}_{i=1}^{n}$ be the open covering of $\gamma$ induced from $\{U_{\alpha}\}$. We take a point $\gamma_{ij}\in\gamma$ from each inter section $\tilde{U}_{ij}$ and let $\gamma_{i}\subset\gamma$ be a interval between $\gamma_{i-1,i}$ and $\gamma_{i,i+1}$ as in the Fig.\ref{fig:base point torsion}. Now, we consider the following quantity:
    \begin{eqnarray}\label{eq:tor part inv}
    n(\gamma)&=&\exp(\sum_{i}\int_{\gamma_{i}}A_{i}-\frac{1}{p}\int_{\Sigma}F)\prod_{i}g^{\dagger}_{i,i+1}(\gamma_{i,i+1}),\\
    &=&\hol(\gamma)\exp(-\frac{1}{p}\int_{\Sigma}F).
    \end{eqnarray}
    Note that $n(\gamma)$ does not depend on the choice of the bounding manifold $\Sigma$ and point $\gamma_{ij}$.
    We can show that $n(\gamma)$ is a gauge invariant quantity and $n(\gamma)\in\zmod{p}\subset\mathrm{U}(1)$. In fact, by using the Stokes theorem,
    \begin{eqnarray}
    n(\gamma)^p=\hol(\gamma)^p\exp(-\int_{\Sigma}F)=1,
    \end{eqnarray}
    and this implies $n(\gamma)\in\zmod{p}$. In addition, under the gauge transformation with $\{g_{i}:U_{i}\to\mathrm{U}(1)\}$, each components of $n(\gamma)$ transform as 
    \begin{eqnarray}
    A_i&\mapsto& A_{i}+g_{i}^{\dagger}dg_{i},\\
    F&\mapsto& F,\\
    g_{ij}&\mapsto& g_{i}g_{ij}g_{j}^{\dagger}.
    \end{eqnarray}
    Therefore, $n(\gamma)$ transforms as 
    \begin{eqnarray}
    n(\gamma)&\mapsto&n(\gamma)\exp(\sum_{i}\int_{\gamma_{i}}g_{i}^{\dagger}dg_{i})\prod_{i}g^{\dagger}_{i}(\gamma_{i,i+1})g_{i+1}(\gamma_{i,i+1}),\\
    &=&n(\gamma)\prod_{i}\exp(\int_{\gamma_{i-1,i}}^{\gamma_{i,i+1}}\frac{d}{dt}\log(g_i(t))dt)\prod_{i}g_{i}(\gamma_{i-1,i})g^{\dagger}_{i}(\gamma_{i,i+1}),\\
    &=&n(\gamma),
    \end{eqnarray}
    and this implies the gauge invariance of $n(\gamma)$.
    
    By using the universal coefficient theorem, the torsion part of $\cohoZ{2}{X}$ is isomorphic to the torsion part of $\mathrm{H}_{1}(X;\mathbb{Z})$. If we would like to know the $k$-th component of the torsion part of $\cohoZ{2}{X}$, we take a generator $\left[\gamma^{(k)}\right]$ of the $k$-th component of the torsion part of $\mathrm{H}_{1}(X;\mathbb{Z})$, and compute the integration Eq.(\ref{eq:tor part inv}) over the path $\gamma^(k)$. This is the way to compute the torsion part of the line bundle. We call $n(\gamma)$ the discrete Berry phase.
    \begin{figure}[H]
 \begin{center}
  \includegraphics[width=50mm]{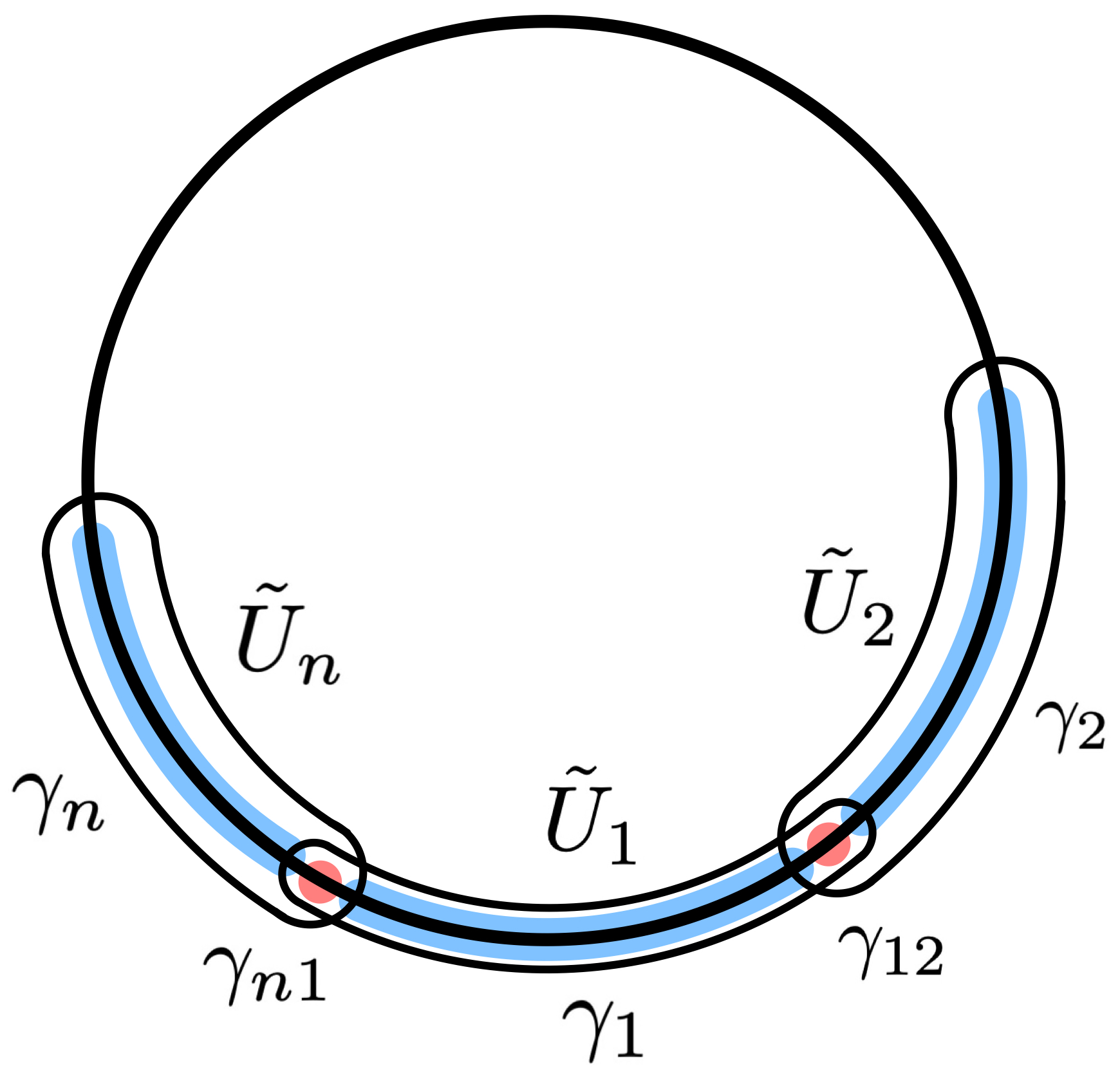}
 \end{center}
  \caption[]{The open covering of $\gamma$ and the triangulation of $\gamma$.
}
\label{fig:base point torsion}
\end{figure}

\end{itemize}

\bibliography{parentbib.bib}
\end{document}